%% file: SharkQ.tex
\documentclass[fleqn,usenatbib]{mnras}

\usepackage{newtxtext,newtxmath}

\usepackage[T1]{fontenc}
\usepackage{ae,aecompl}
\usepackage{xcolor}

\usepackage{graphicx}	
\usepackage{amsmath}	

\def\eagle{{\sc eagle}}
\def\surfs{{\sc surfs}}
\def\shark{{\sc Shark}}

\title[The SFHs of high-z massive, quenched galaxies]{The diverse star formation histories of early  massive, quenched galaxies in modern galaxy formation simulations}

\author[C.~D.~P. Lagos et al.]{
\parbox[t]{\textwidth}{
\vspace{-0.5cm}
Claudia del P. Lagos$^{1,2,3}$\thanks{E-mail: claudia.lagos@icrar.org}, Francesco Valentino$^{4}$, Ruby J. Wright$^{5,1}$, Anna de Graaff$^{6}$, Karl Glazebrook$^7$, Gabriella De Lucia$^8$, Aaron S.G. Robotham$^{1,2}$, Themiya Nanayakkara$^7$, Angel Chandro-Gomez$^{1,2}$, Mat\'ias Bravo$^9$, Carlton M. Baugh$^{10}$, Katherine E. Harborne$^{1,2}$, Michaela Hirschmann$^{11}$, Fabio Fontanot$^8$, Lizhi Xie$^{12}$, Harry Chittenden$^{7}$}
\vspace*{6pt} \\
$^{1}$International Centre for Radio Astronomy Research (ICRAR), M468, University of Western Australia, 35 Stirling Hwy, Crawley, \\WA 6009, Australia.\\
$^{2}$ARC Centre of Excellence for All Sky Astrophysics in 3 Dimensions (ASTRO 3D).\\
$^{3}$Cosmic Dawn Center (DAWN), Denmark.\\
$^4$European Southern Observatory, Karl-Schwarzschild-Str. 2, 85748 Garching, Germany.\\
$^5$Department of Physics, University of Helsinki, Gustaf Hällströmin katu 2, FI-00014 Helsinki, Finland.\\
$^6$ Max-Planck-Institut f\"ur Astronomie, K\"onigstuhl 17, D-69117, Heidelberg, Germany.\\
$^7$Centre for Astrophysics and Supercomputing, Swinburne University of Technology, PO Box 218, Hawthorn, VIC 3122, Australia.\\
$^8$INAF – Astronomical Observatory of Trieste, Via G. B. Tiepolo 11, 34143 Trieste, Italy.\\
$^{9}$Department of Physics \& Astronomy, McMaster University, 1280 Main Street W, Hamilton, ON, L8S 4M1, Canada.\\
$^{10}$Department of Physics, Institute for Computational Cosmology, Science Laboratories, Durham University, South Road, Durham, DH1 3LE, UK\\
$^{11}$Institute of Physics, Laboratory for galaxy evolution, EPFL, Observatory of Sauverny, Chemin Pegasi 51, 1290 Versoix, Switzerland.\\
$^{12}$Tianjin Normal University, Binshuixidao 393, 300387 Tianjin, PR China.
\vspace*{-0.5cm}}

\date{Accepted XXX. Received YYY; in original form ZZZ}

\pubyear{2024}

\begin{document}
\label{firstpage}
\pagerange{\pageref{firstpage}--\pageref{lastpage}}
\maketitle

\begin{abstract}
We present a comprehensive study of the star formation histories of massive-quenched galaxies at $z=3$ in three semi-analytic models (\shark, {\sc GAEA}, {\sc Galform}) and three cosmological hydrodynamical simulations ({\sc Eagle}, {\sc IllustrisTNG}, {\sc Simba}). We study the predicted number density and stellar mass function of massive-quenched galaxies, their formation and quenching timescales and star-formation properties of their progenitors. Predictions are disparate in all these diagnostics, for instance: (i) some simulations reproduce the observed number density of very massive-quenched galaxies ($>10^{11}\,\rm M_{\odot}$) but underpredict the high density of intermediate-mass ones, while others fit well the lower masses but underpredict the higher ones; (ii) In most simulations, except for {\sc GAEA} and \eagle, most massive-quenched galaxies had starburst periods, with the most intense ones happening at $4<z<5$; however, only in \shark\ and {\sc IllustrisTNG} we do find a large number of progenitors with star formation rates $>300\rm \, M_{\odot}\,yr^{-1}$; (iii) quenching timescales are in the range $\approx 20-150$~Myr depending on the simulation; among other differences. These disparate predictions can be tied to the adopted Active Galactic Nuclei (AGN) feedback model. For instance, the explicit black-hole (BH) mass dependence to trigger the `radio mode' in {\sc IllustrisTNG} and {\sc Simba} makes it difficult to produce quenched galaxies with intermediate stellar masses, also leading to higher baryon collapse efficiencies ($\approx 15-30$~per cent); while the strong bolometric luminosity dependence of the AGN outflow rate in {\sc GAEA} leads to BHs of modest mass quenching galaxies. Current observations are unable to distinguish between these different predictions due to the small sample sizes. However, these predictions are testable with current facilities and upcoming observations, allowing a `true physics experiment’’ to be carried out.
\end{abstract}




\section{Introduction}

Over the previous years, significant effort has been placed in to studying the number density of passive galaxies across cosmic time (e.g. \citealt{Straatman14,Schreiber15,Schreiber18,Merlin19,Valentino20,Shahidi20,Carnall20,Gould23,Weaver22}). 
This field has seen an explosion of results thanks to the James Webb Space Telescope ({\it JWST}), which consistently reports high number densities of massive {($M_{\star}\gtrsim 10^{10}\,\rm M_{\odot}$, {where $M_{\star}$ is the galaxy stellar mass}), passive (specific star formation rate, sSFR$\equiv \rm SFR/M_{\star}\lesssim 10^{-10}\,yr^{-1}$, where SFR is the star formation rate)} galaxies at redshifts $z>2$, even higher than the values claimed pre-{\it JWST} \citep{Carnall23,Valentino23,Nanayakkara22,Long23,Alberts23}. 
These observations have moved beyond reporting `candidates' for massive-quenched galaxies based on colours, with many now being spectroscopically confirmed \citep{Carnall23b,Carnall24,Nanayakkara22,deGraaff24,Weibel24}, with the highest redshift one being at $z\approx 7$ \citep{Weibel24}.
These studies are posing significant constraints on modern cosmological galaxy formation simulations, which for the most part seem to have difficulties in reproducing such high number densities (e.g. \citealt{Valentino23,Hartley23,DeLucia24,Lagos24,Weller24,Vani24,deGraaff24}). There are exceptions though. For example, \citet{Remus23} show that in their simulation, the number of massive-quenched galaxies at $z\approx 3-5$ agrees reasonably well with observations within the uncertainties.  

Beyond constraining the number density of massive, passive galaxies, the {\it JWST} has been used to get exquisite spectra of these galaxies, not only confirming their high redshifts ($z\gtrsim 3$), but also allowing intrinsic properties of galaxies to be inferred (e.g. \citealt{Glazebrook23,deGraaff24,Carnall23b,Carnall24,Nanayakkara22}). These galaxies tend to have large stellar masses ($\gtrsim 10^{10.3}\,\rm M_{\odot}$), and star formation histories (SFHs) indicative of relatively short-duration starbursts (few $\sim 100$~Myr). These observations are therefore offering a unique opportunity to piece together the picture of massive galaxy formation from the early to the local Universe.

In the local Universe, a consistent picture has emerged over the previous 20 years, in which some form of feedback (usually related to Active Galactic Nuclei; AGN) is responsible for quenching star formation in massive galaxies. This form of feedback is invoked by galaxy formation models and simulations to reproduce a plethora of phenomena observed in local massive galaxies, such as `downsizing', the break in the stellar mass function (SMF), the optical colour-magnitude bimodality, among others (see e.g. \citealt{Benson03,Springel05b,Croton06,Bower06,Lagos08,Somerville08,Sijacki07,Cattaneo08} to mention a few of the early studies on the topic). AGN feedback is thus generally used   to reproduce the properties of massive galaxies locally and in fact the free-parameters of AGN feedback models are usually tuned for this exact purpose (e.g. \citealt{Schaye14,Crain15,Croton16,Weinberger17,Dave19,Kugel23,Lagos24,DeLucia24}). 
The process of tuning the free-parameters involved in the feedback models included in galaxy formation simulations has led to an overall agreement of broad statistics among simulations, such as the local Universe galaxy SMF and the SFR-$M_{\star}$ relations  \citep{Somerville15}. Thus, testing the simulations in regimes that are far from those used for the tuning process is essential to understand how well current feedback models are doing in reproducing the diversity of galaxy properties observed across a wide range of cosmic epochs. 

Although effectively all cosmological galaxy formation simulations invoke some form of AGN feedback to reproduce the properties of massive galaxies locally, in detail, the exact model employed can vary significantly between codes. For example, some models and simulations invoke only a single mode of AGN feedback \citep{Schaye14,Henriques15,Lacey15}, while others invoke two modes \citep{Weinberger17,Dave19,Lagos24,DeLucia24}. Although these models do produce similar properties for massive galaxies in the local universe, by construction, they do it for very different reasons (see for example the differences in the halo gas reservoirs and gas in/out flow rates in the {\sc Eagle}, {\sc Simba} and {\sc IllustrisTNG} simulations; \citealt{Wright24}).
The new wave of observations of massive-quenched galaxies at high redshift coming from the {\it JWST} offer a unique opportunity to stress-test galaxy formation models in a new territory for which parameters have not been tuned. Equally important is to explore whether the different AGN feedback models leave clear imprints on the properties of massive galaxies in the early universe that can now be inferred from the exquisite observations the {\it JWST} is providing. 

In this paper we set up for this task by taking $6$ modern cosmological galaxy formation simulations ($3$ semi-analytic models, and $3$ cosmological hydrodynamical simulations) and analysing the high-z massive-quenched galaxies they predict to answer the following questions: (i) how well can modern galaxy formation simulations reproduce the number density and mass distributions of early massive-quenched galaxies?; (ii) what is the diversity of SFHs predicted for early massive-quenched galaxies in modern simulations?; (iii) if the differences are large, can we trace them back to the different ways in which feedback is modelled in the simulations?

This paper is organised as follows. \S~\ref{simsinfo} describes the suite of galaxy formation simulations used in this work, and presents the samples and definitions employed in this paper. \S~\ref{SMFsection} studies the number densities and SMF of massive-quenched galaxies in the $6$ simulations analysed here and compares with observations. \S~\ref{SFHssection} studies the SFHs of massive-quenched galaxies, focusing first on the rise of star formation, and then on how quenching happens and its relation to AGN feedback. In this section we also compare with observations where possible. \S~\ref{conclusions} presents our main conclusions. Appendix~\ref{MS} presents an analysis of a control sample of massive, star-forming galaxies and how they sample the galaxy main sequence evolution; Appendix~\ref{Magneticum} shows the evolution of the number density of massive-quenched galaxies in the {\sc Magneticum} simulation and compare with the other $6$ simulations used throughout this work. Appendix~\ref{maxSFReff} evaluates the impact of the different time cadence of the simulations used here on the maximum SFR of galaxies. Appendix~\ref{SFHfits} presents the fits to the SFHs of galaxies in the simulations, which we use to quantify the frequency of starburst episodes.

 \section{The suite of galaxy formation simulations}\label{simsinfo}

In this section we briefly introduce the simulations used in this work. \S~\ref{sams} and \S~\ref{hydro} describe the semi-analytic models and hydrodynamical simulations, respectively. 

\subsection{Semi-analytic models of galaxy formation}\label{sams}

Semi-analytic models (SAMs) solve for the formation and evolution of galaxies using dark matter (DM) halo populations and merger trees built from DM-only $N$-body simulations. In the three models explored here, the following physical processes are modelled (i) the collapse and merging of DM haloes; (ii) the accretion of gas onto haloes; (iii) the shock heating and radiative cooling of gas inside DM haloes, leading to the formation of galactic discs; (iv) star formation in galaxy discs; (v) stellar feedback from the evolving stellar populations; (vi) chemical enrichment of stars and gas; (vii) the growth via gas accretion and merging of black holes (BHs); (viii) heating by AGN; (ix) photoionisation of the intergalactic medium; (x) galaxy mergers
 which can trigger starbursts and the formation and/or growth of spheroids; (xi) effects of disc instabilities which in some models can trigger starbursts.
 In SAMs, the timescale in which galaxy mergers happen needs to be prescribed, and generally happens after the satellite subhalo that hosts a satellite galaxy ceases to be tracked due to its low number of DM particles.
 
 The SAMs studied here differ in the way the different processes above are modelled. For the purpose of this paper, the key processes we are interested in are AGN feedback and the physical processes triggering starbursts, and hence we especially focus on describing the way these are modelled.
Note that SAMs are sensitive to the number of snapshots produced with the $N$-body simulation (in a way that hydrodynamical simulations are not), and the disparate number of snapshots of the adopted simulations in the SAMs below will impact how finely we can sample the SFHs of galaxies.

\subsubsection{\shark}\label{sharkdesign}

\shark, \citep{Lagos18c,Lagos24} is a publicly available code hosted on GitHub\footnote{\href{https://github.com/ICRAR/shark}{\url{https://github.com/ICRAR/shark}}}. We use the version of the code presented in \citet{Lagos24}, which was shown to produce massive galaxies whose properties agreed better with observations across a wide redshift range. 

One of the novelties of \shark\ v2.0 presented in \citet{Lagos24} is the AGN feedback model, which consists of two modes: a jet and a wind mode. The jet mode depends on the jet power produced by the AGN, $Q_{\rm jet}$, which in itself depends on the BH spin, accretion rate and mass (see Equation~(31) in \citealt{Lagos24}).
For galaxies with an AGN, the cooling luminosity of the hot gas in the halo is offset by $Q_{\rm jet}$ (modulated by an efficiency parameter $\kappa_{\rm jet}$), which can completely shut off the cooling flow for sufficiently strong jets. 
The jet mode only acts on haloes that have already formed a hot halo, following the model of \citet{Correa18}. The wind mode on the other side consists of a radiation pressure-driven outflow whose velocity depends on the dust opacity, gas content of the bulge and AGN bolometric luminosity (following the model of \citealt{ishibashi2015} - see Equation~(36)). If the internal energy of the outflow exceeds the binding energy of the halo, then the outflow (or part of it) is allowed to escape the halo. This wind mode can only happen during starburst episodes, which in \shark\ can be triggered by galaxy mergers or disc instabilities. In the latter a stability parameter is computed with the properties of the galaxy disc, and if this is below a given threshold, then the disc of that galaxy is assumed to be fully destroyed, with all the gas and stellar content being moved to the bulge. The newly acquired gas of the bulge then is the fuel for the triggered starburst.

The free parameters of \shark\ were calibrated using an automatic optimiser that fits the $z<0.1$ SMF. Although not used in the calibration, other results of
the model were visually inspected to ensure their agreement with observations was not seriously compromised by the $z <0.1$ SMF only. Those included 
$z\approx 0$ gas scaling relations, the mass-metallicity relation and the $z < 2$ cosmic specific star formation rate density (CSFRD).

Here we use a version of \shark\ running over the P-Millennium $N$-body simulation (P-MILL hereafter). 
P-MILL was introduced in \citet{Baugh19} and the cosmological parameters correspond to a total matter, baryon and $\Lambda$ densities of $\Omega_{\rm m}=0.307$, $\Omega_{\rm b}=0.04825$ and $\Omega_{\Lambda}=0.693$, respectively, with a Hubble parameter of $H_{\rm 0}=h \, 100\,\rm Mpc\, km\,s^{-1}$ with $h=0.6777$, scalar spectral index of $n_{\rm s}=0.9611$ and a power spectrum normalisation of $\sigma_{\rm 8}=0.8288$. P-MILL's volume and particle mass are  $(542.16\,\rm cMpc\,h^{-1})^3$ (where cMpc refers to comoving megaparsec) and $1.06\times 10^8\,\rm M_{\odot}\,h^{-1}$, respectively. P-Mill has $271$ snapshots.  
Halo catalogues for P-MILL were constructed using {\sc Subfind} \citep{Springel01} and merger trees were built using D-halos \citep{Jiang14}. Haloes with $\ge 20$ particles are included in the catalogues, giving a minimum halo mass of $2.12\times 10^9\,\rm M_{\odot}\,h^{-1}$.

\subsubsection{{\sc GAEA}}

GAlaxy Evolution and Assembly (GAEA) was  introduced in \citet{DeLucia14} and \citet{Hirschmann16}. Here, we use the latest {\sc GAEA} version, introduced in \citet{DeLucia24}, which combined a new treatment of AGN feedback with an updated model of environmental processes affecting satellite galaxies. A key aspect highlighted in \citet{DeLucia24} is that the new version of GAEA is able to reproduce the observed number densities of massive-quenched galaxies up to $z\approx 4$ (including some of the {\it JWST} ones). 

The AGN feedback model of GAEA was first introduced in \citet{Fontanot20}, and consists of (i) a model for the inflow
of cold gas towards the central BH driven
by star formation in the central regions of galaxies; (ii) a BH
accretion rate that is determined by the viscous accretion
timescale; and (iii) an empirical scaling relation between the mass loading of quasar winds and the AGN bolometric luminosity. Note that the quasar outflow rate does not depend on the available interstellar medium reservoir, but only on the AGN bolometric luminosity. The treatment of disc instabilities in {\sc GAEA} is markedly different from that in \shark\ and {\sc Galform}. This is described in \citet{DeLucia11} and in short the model computes the same instability parameter described in \S~\ref{sharkdesign} and if a disc is unstable, enough stellar mass is transferred from the disc to the bulge as to restore stability. Hence, there is no associated starburst with disc instabilities, but they can still contribute to the mass growth of bulges. Galaxy mergers are therefore the only route to produce starbursts in {\sc GAEA}.

The free parameters in {\sc GAEA} were tuned to reproduce the $z=0-3$ SMF, the atomic and molecular hydrogen mass functions in the local Universe, and the evolution of the AGN luminosity function up to $z=4$.

The version of {\sc GAEA} used here was run over merger trees from the Millennium Simulation \citep{Springel05} (hereafter MILL), which adopts the following cosmological parameters, $\Omega_{\rm m} = 0.25$, $\Omega_{\rm b} = 0.045$, $\Omega_{\Lambda}= 0.75$, $h=0.73$, $n_{\rm s} =1$,  and $\sigma_{8}=0.8$. MILL has a volume of $(500\,\rm cMpc\,h^{-1})^3$, a particle mass of $8.625\times 10^8\,\rm M_{\odot}\,h^{-1}$, and $64$ snapshots. 
Merger trees and halo catalogues for MILL were constructed using {\sc Sub-Link} and {\sc Subfind} \citep{Springel01}. Haloes
with $\ge 20$ particles are included in the catalogues, giving a minimum halo mass of $1.73\times 10^{10}\,\rm M_{\odot}\,h^{-1}$.

\subsubsection{Galform}\label{galform}

We use the {\sc Galform} version introduced in \citet{Lacey15}, which combined several flavours of previous {\sc Galform} model versions that focused on a variety of issues, such as reproducing the properties of massive galaxies and high-redshift star-forming galaxies. 

The AGN feedback model adopted in {\sc Galform} was introduced by \citet{Bower06} and is assumed to be effective only in haloes undergoing quasi-hydrostatic cooling. In this situation, mechanical energy input by the AGN is expected to stabilise the flow and regulate the rate at which the gas cools. Whether or not a halo is undergoing quasi-hydrostatic cooling depends on the cooling and free-fall times:
the halo is in this regime if $t_{\rm cool}(r_{\rm cool})>\alpha^{-1}_{\rm cool}\, t_{\rm ff}(r_{\rm cool})$, where
$t_{\rm ff}$ is the free fall time at $r_{\rm cool}$, {$r_{\rm cool}$ is the cooling radius (within which the cooling time equals the time since halo formation)} and $\alpha_{\rm cool}\sim 1$ is an adjustable parameter close to unity.
AGNs are assumed to be able to quench gas cooling only if the available AGN power is comparable to the cooling luminosity,
$L_{\rm cool}<\epsilon_{\rm SMBH}\, L_{\rm Edd}$, where $L_{\rm Edd}$ is the BH's Eddington luminosity and 
$\epsilon_{\rm SMBH}\sim 1$ is a free parameter.
Note that {\sc Galform} models disc instabilities in the same way \shark\ does, and in fact they play a key role in the growth of BHs and galaxies in the early universe (see \S~5.3 in \citealt{Lacey15} for details).

An important distinction of the {\sc Galform} model is that the initial mass function (IMF) of stars is not universal (for all the other simulations, it is assumed to be a universal \citealt{Chabrier03} IMF). Instead, in {\sc Galform} two IMFs are invoked: if star formation is triggered by gas cooling from the hot halo a \citet{Kennicutt83} IMF is assumed (which is not too dissimilar to the \citealt{Chabrier03}); if the star formation is instead triggered by a galaxy merger or a violent disc instability (i.e. `starburst mode'), the IMF is assumed to be top-heavy\footnote{For an IMF defined as $\Phi(m)\propto m^{-x}$, {\sc Galform} adopts $x=1$ for starbursts. The \citet{Chabrier03} IMF uses $x=1.35$ for $m>1\,\rm M_{\odot}$.}. For simplicity, we will ignore the fact that {\sc Galform} internally has different IMFs and will directly compare the resulting stellar masses with other simulations. We caution, however, that if a more top-heavy IMF was to be assumed in observations when deriving stellar masses and SFRs, lower values would be derived compared to what is obtained assuming a universal Milky-Way like IMF. This difference, however, is expected to be small compared to other errors, of the order of a factor of $\approx 1.5-2.5$ depending on the exact SFR tracer used.

{The {\sc Galform} version of} \citet{Lacey15} is run over the DM haloes of the 
 Millennium-WMAP7 (or MILL7) $N$-body simulation, which adopts the cosmological parameters of the Wilkinson Microwave Anisotropy Probe (WMAP-7)
data set \citep{Komatsu11}:   
$\Omega_{\rm m} = 0.272$, $\Omega_{\rm b} = 0.0455$, $\Omega_{\Lambda}= 0.728$, $h=0.704$, $n_{\rm s} = 0.967$,  and $\sigma_{8}=0.810$. 
 MILL7 has a volume of $(500\,\rm cMpc\, h^{-1})^3$, a particle mass of $9.364\times 10^8\,\rm M_{\odot}\,h^{-1}$, and $61$ snapshots. 
Merger trees and halo catalogues for MILL7 were constructed using {\sc D-halos} and {\sc Subfind}.
Haloes
with $\ge 20$ particles are included in the catalogues, giving a minimum halo mass of $1.87\times 10^{10}\,\rm M_{\odot}\,h^{-1}$. The data of {\sc Galform} is retrieved from the Virgo public database\footnote{\url{http://virgodb.dur.ac.uk:8080/MyMillennium/}}, and is only available for the redshift range $z=0-10.0735$. So, unlike other simulations, the derived star formation histories (SFHs) only span $z=10.0735-3.06042$.\newline

\begin{figure*}
\begin{center}
\includegraphics[trim=0mm 0mm 0mm 0mm, clip,width=0.59\textwidth]{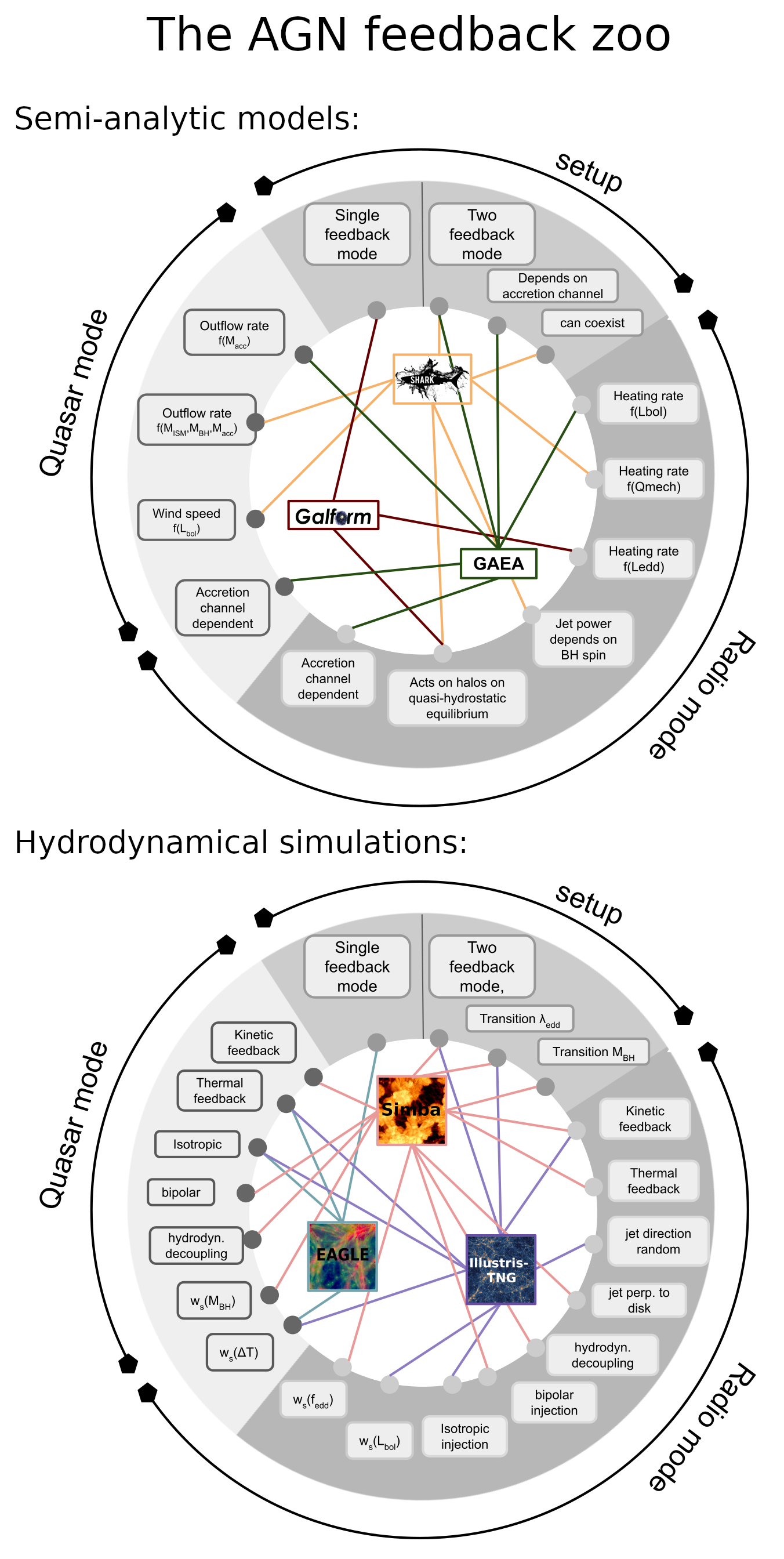}
\caption{Visualisation of the different decisions made by the SAMs (top) and cosmological hydrodynamical simulations (bottom) employed in this paper regarding their AGN feedback model. At the top, is the setup: whether one or two modes of AGN feedback are invoked. If two modes are invoked, then how one transitions  between them is shown. For simplicity we call the two modes of feedback `QSO' and `radio' mode, which generally refer to the radiatively-efficient and inefficient modes of feedback, respectively. 
For the wheel of decisions for SAMs, in both modes, $f$ represents a function, and its potential parameters are $L_{\rm bol}$, $Q_{\rm mech}$,  $L_{\rm edd}$, $M_{\rm acc}$ and $M_{\rm ISM}$, being the BH's bolometric luminosity, mechanical power, Eddington's luminosity, accretion rate and the ISM mass, respectively. 
In the case of hydrodynamical simulations, $w_{\rm s}$ refers to the wind speed and whether this is computed as depending on the BH mass, $w_{\rm s}(M_{\rm BH})$, as a change in temperature, $w_{\rm s}(\Delta\, \rm T)$, as depending on the Eddington ratio, $w_{\rm s}(f_{\rm edd})$, or the bolometric luminosity, $w_{\rm s}(L_{\rm bol})$. These figures visually show which models are more complex given their number of decisions. In the case of SAMs, {\sc Galform} has the simplest model, while \shark\ has the most complex one. For hydrodynamical simulations, {\sc Eagle} has the simplest mode, while {\sc Simba} has the most complex one. Visualisation was inspired by the method presented in \citet{Thorne21}, see their Fig.~1.} 
\label{allsims_vis}
\end{center}
\end{figure*}

The top panel of Fig.~\ref{allsims_vis} shows visually the different decisions made by the three SAMs employed in this work. For simplicity, we refer to two modes of AGN feedback as `QSO' and `radio' modes, which generally encompass the concepts of radiatively-efficient and inefficient feedback modes, respectively. Note that internally to each simulation, the two modes may not be referred to as `QSO' and `radio' modes though. {\sc Galform} has the simplest model of the three SAMs here, with a single AGN feedback mode, while \shark\ has the most complex model, with two modes that can coexist.

\subsection{Cosmological hydrodynamical simulations}\label{hydro}

For this work we analyse a suite of cosmological hydrodynamical simulations that are publicly available and that simulate large enough cosmological volumes as to contain several dozen massive-quenched galaxies at high redshift. 
All these simulations include sub-grid models that capture unresolved physics, including (i) radiative cooling and photoheating, (ii) star formation from cold, dense gas, (iii) stellar evolution and chemical enrichment, (iv) stellar feedback, and (v) BH growth and AGN feedback. Similarly to SAMs, for each simulation we describe in some detail the adopted AGN feedback model, which is modelled very differently in each of the three simulations. In general though and in hydrodynamical simulations, the energy associated with AGN feedback is coupled to the gas using `thermal' and `kinetic' approaches (in some cases, a combination of both). Thermal feedback is nominally implemented by heating the relevant gas elements, while kinetic feedback involves directly `kicking' gas elements with a prescribed velocity to generate an outflow. 
In hydrodynamical simulations no explicit treatment of galaxy mergers and disc instabilities is required, as those naturally arise as the galaxies are evolved.

\subsubsection{{\sc Eagle}}

The \eagle\ simulation suite
(described in detail in \citealt{Schaye14} and \citealt{Crain15}) consists of a large number of cosmological 
hydrodynamic simulations with different resolutions, cosmological volumes and subgrid models.  {\sc Eagle} was  performed using an extensively modified version of the parallel $N$-body smoothed particle hydrodynamics (SPH) code {\sc Gadget-3} \citep{Springel05b,Springel08}. Among those modifications are updates to the
SPH technique, which are collectively referred to as `Anarchy' (see \citealt{Schaller15} for a discussion of  the impact these changes have on the properties of simulated galaxies compared to standard SPH). {\sc Eagle} used {\sc Subfind} to identify galaxies and that is the catalogue used in this paper.

 For AGN feedback, {\sc Eagle} adopts a single model which comprises a stochastic heating model. Here, gas particles surrounding the BH are chosen randomly and heated by a temperature $\Delta T_{\rm AGN}=10^{8.5}$~K. The rate of energy injection from AGN feedback is computed from the BH accretion rate, and a fixed conversion efficiency of accreted rest mass to energy. In {\sc Eagle}, particles influenced by feedback (either stellar or AGN) are not decoupled from the hydrodynamics when they receive a temperature boost. Hence, the temperature boost is required to be large as to avoid rapid cooling and dissipation of the feedback energy (see \citealt{Crain15} for details). Because gas particles affected by AGN feedback are chosen at random, the resulting outflow has no preferential direction. 

The reference {\sc Eagle} model used in this work,  
was calibrated to ensure a
good match to the $z=0.1$ SMF, 
the sizes of present-day disc galaxies and the BH-stellar mass relation (see \citealt{Crain15} for details on the tuning of parameters).

\eagle\ adopts a \citet{Planck14} cosmology: $\Omega_{\rm m} = 0.307$, $\Omega_{\rm b} = 0.04825$, $\Omega_{\Lambda} = 0.693$, $h=0.6777$, $n_{\rm s} = 0.9611$ and $\sigma_{8} = 0.8288$. Here, we use the largest {\sc Eagle} run, which consists of a box of $(100\,\rm cMpc)^3$ with an initial number of particles $2\times 1504^3$. The initial gas particle mass is $1.81\times 10^6\,\rm M_{\odot}$ and the DM particle mass is $9.7\times 10^6\,\rm M_{\odot}$.

\subsubsection{{\sc IllustrisTNG}}

The Next Generation {\sc Illustris} simulations, known as {\sc IllustrisTNG} \citep{Springel18,Pillepich18}, are a collection of cosmological magneto-hydrodynamical simulations conducted using the moving-mesh refinement code AREPO \citep{Springel10,Pakmor11,Weinberger20}. {\sc IllustrisTNG} used {\sc Subfind} to identify galaxies and that is the catalogue used in this paper.

{\sc IllustrisTNG}'s AGN feedback model is described in detail in \citet{Weinberger17b} and in short is made up of two modes: at high BH accretion rates feedback is injected in thermal mode within a spherical region around the galaxy; at low BH accretion rates, feedback is injected in a kinetic form (as kinetic winds). In the latter, the direction of the jets is randomised at each time step. 
The transition between the two AGN feedback modes above is based on a BH-dependent accretion rate (see equation~6 in \citealt{Weinberger17b}), in a way that the BH accretion threshold below which kinetic feedback switches on increases with increasing BH mass. \citet{Terrazas20} found that given the conditions at $z=0$, the AGN feedback kinetic mode switches at a BH mass of $\approx 10^{8.2}\,\rm M_{\odot}$. In both of the accretion modes, feedback is injected into surrounding gas cells (the `feedback region') with no preferential direction, and there is no hydrodynamic decoupling of feedback-affected gas elements.

The free parameters of the {\sc IllustrisTNG} model were tuned to reproduce the cosmic SFR density evolution, the $z\approx 0$ SMF, and the present-day stellar-to-halo mass relation. 

All of the {\sc IllustrisTNG} boxes assume a \citet{PlanckCollaboration2016} cosmology, with $\Omega_{\rm m} = 0.31$, $\Omega_{\rm b} = 0.0486$, $\Omega_{\Lambda} = 0.692$, $h = 0.677$, $n_{\rm s} = 0.97$ and $\sigma_{8} = 0.8159$. 
In this paper we use the {\sc TNG100} run, a box of volume $(75.0\,\rm cMpc\,h^{-1})^3$, an initial gas cell mass of $1.40 \times 10^6\, \rm M_{\odot}$, and DM particle mass of $7.5 \times 10^6 \, \rm M_{\odot}$. 

\subsubsection{{\sc Simba}}\label{Simba}

{\sc Simba} \citep{Dave19} uses the meshless finite mass (MFM) mode of the {\sc Gizmo} hydrodynamics code \citep{Hopkins15,Hopkins17}, with a gravity solver based on {\sc Gadget-3} \citep{Springel05b}. Galaxies are identified using a friends-of-friends galaxy finder, assuming a spatial linking length of
$0.0056$ times the mean inter-particle spacing (equivalent to twice the minimum softening length). The latter is the catalogue used in this paper.

{\sc Simba}'s AGN feedback model (introduced in \citealt{Angles-Alcazar17}) includes kinetic and thermal feedback. Kinetic feedback is applied to both AGN that are radiatively-efficient and inefficient (the latter is also referred to as `jet' mode). The thermal mode (referred to as `X-ray heating' in \citealt{Dave19}) only acts in radiatively-inefficient AGN and in galaxies that are gas poor. 
In the radiatively-efficient mode,
winds with velocities $500-1500\, \rm km\,s^{-1}$ are produced, with the exact wind velocity depending on BH mass (where $v_{\rm wind}=1000\, \rm km\,s^{-1}$ corresponds to a BH mass of $10^{9}\, \rm M_{\odot}$). The jet mode occurs for BHs with masses $\ge 10^{7.5}\,\rm M_{\odot}$ and accretion rates that are $<0.2\,M_{\rm Edd}$, with $M_{\rm Edd}$ is the Eddington rate. Note, however, that the jet-mode builds-up as the Eddington ratio becomes smaller, and peaks in efficiency at $0.02\,M_{\rm Edd}$. The jet-mode can add a velocity boost of up to $7000\, \rm km\,s^{-1}$ additional to the regular wind velocity above. Thus, outflows can be as fast as $8000\, \rm km\,s^{-1}$. Gas is ejected by the jets in a bipolar fashion but with an opening angle of $0$, with the jets being perpendicular to the inner disc. 
Note that this is different to {\sc IllustrisTNG} in which the direction of the jets is randomised. 
The gas affected by jets remains hydrodynamically decoupled for a duration of $10^{-4}\,t_{\rm H}(z)$, where $t_{\rm H}(z)$ is the Hubble time at redshift $z$. The latter allows jets to travel up to $\approx10$~pkpc (physical kiloparsec) before the energy is deposited into the surrounding gas. Gas in the jets is injected at the halo virial temperature.

The free parameters of {\sc Simba} were tuned to provide a reasonable match to the $z\approx 0$ SMF and the BH-stellar mass relation. 

The {\sc Simba} simulations adopt cosmological parameters $\Omega_{\rm m} = 0.30$, $\Omega_{\rm b} = 0.048$, $\Omega_{\Lambda} = 0.70$, $h = 0.68$, $n_{\rm s} = 0.97$ and $\sigma_{8} = 0.82$.
We use the largest {\sc Simba} simulation box, which consists of a box of volume $(100\, h^{-1})^3 \,\rm cMpc$, a gas element mass of $1.8\times 10^{7}\, \rm M_{\odot}$, and DM particle mass of  $9.7\times10^{7}\, \rm M_{\odot}$.\newline

The bottom panel of Fig.~\ref{allsims_vis} shows visually the different decisions made by the three hydrodynamical simulations employed in this work. {\sc Eagle} has the simplest model of the three simulations here, with a single AGN feedback mode, while {\sc Simba} has the most complex model, with two modes, and a `radio' mode that uses both thermal and kinetic feedback. Note that for {\sc IllustrisTNG} we say the wind speed is implemented as a temperature increase because in practice the energy is saved until there is enough to kick the gas at high enough speeds, in a similar way to how gas particles are heated to a certain temperature in {\sc Eagle}. Table~\ref{simus} shows the cosmological volume and DM particle mass of each of the simulations analysed here. 

\subsection{Galaxy properties and selection of passive galaxies}\label{definitions}

\begin{table*}
        \setlength\tabcolsep{2pt}
        \centering\footnotesize
        \caption{The cosmological volume, DM particle mass, number of massive-quenched galaxies at $z=3$ ($N_{\rm pass}$),  the implied number density of massive-quenched galaxies, $\phi_{\rm pass}$,  and the percentage of galaxies with $M_{\star}\ge 10^{10}\rm \, M_{\odot}$ that are quenched at $z=3$  ($f_{\rm pass}$) of each simulation.
        For the SAMs, we also report in parenthesis the number of central massive-quenched galaxies and their number density. The percentage of passive galaxies in the right-most column for SAMs corresponds to that of central galaxies.}
        \begin{tabular}{@{\extracolsep{\fill}}l|cccccc|p{0.45\textwidth}}
                \hline
                \hline
            Simulation & volume [$\rm cMpc^3$] & $m_{\rm DM}\,\rm [M_{\odot}]$ & $N_{\rm pass}$ & $\phi_{\rm pass}\,\rm [cMpc^{-3}]$ & $f_{\rm pass}$\\
           \hline
   \shark\ & $(800)^3$ & $1.56\times10^8$ & $44,257$ ($18,831$) & $\,\,\,\,8.62\times 10^{-5}$ ($3.68\times 10^{-5})$ & $5.2$\%\\
   {\sc GAEA} & $(684.93)^3$ & $1.18\times10^9$ & $15,990$ ($12,792$) & $\,\,\,\,1.38 \times 10^{-4}$ ($1.1\times 10^{-4}$)& $9.3$\%\\
   {\sc Galform} & $(710.23)^3$ & $1.33\times10^9$ & $12,677$ ($8,846$) & $\,\,\,\,3.54 \times 10^{-5}$ ($2.45\times 10^{-5}$) & $3.9$\%\\
   {\sc Eagle} & $(100)^3$ & $9.7\times 10^6$ & $35$ & $3.5\times 10^{-5}$ & $5.5$\%\\
   {\sc IllustrisTNG} & $(110.78)^3$ & $7.6\times 10^6$ & $43$ & $3.17\times 10^{-5}$ & $6.1$\%\\
   {\sc Simba} & $(142.86)^3$ & $9.7\times 10^7$ & $36$ & $1.13\times10^{-5}$& $3.6$\%\\
   \hline
       \end{tabular}
       \label{simus}
\end{table*}

We focus on selecting passive galaxies at $z=3$, as this is a redshift where the simulations analysed here have enough galaxies (at least $30$) and where most of the current observational constraints of massive-quenched galaxies from the {\it JWST} are located (see Fig.~\ref{num_densities}). Unless specified, we show $z=3$ galaxies only. 

Throughout this paper we use stellar masses, SFRs and SFR histories of galaxies in the SAMs and hydrodynamic simulations. Below we briefly describe how these quantities are computed:

\begin{itemize}
    \item {\it Stellar masses ($M_{\star})$}: for the SAMs we use the total (remaining) stellar mass (disc plus bulge). For the hydrodynamical simulations we measured stellar masses as the sum of the remaining masses of all the stellar particles enclosed within a sphere of radius $50$~ckpc around the centre of potential of each subhalo. Visual inspection of massive galaxies show that this aperture is sufficient to encompass the galaxy while also avoiding potential contamination from substructure. 
    \item {\it SFRs}: for the SAMs we use the total SFR averaged over a snapshot (disc plus bulge). Because of the different time cadence in the SAMs, the SFR at $z=3$ corresponds to the average over the previous $\approx 41$~Myr for \shark, and $\approx 181$~Myr for {\rm GAEA} and {\sc Galform}. 
    For the hydrodynamical simulations we compute a recent SFR using the stellar particles that formed in the last $20$~Myr before the output time of interest ($z=3$). In practice,
    we sum the mass with which stellar particles formed during that period and divide by $20$~Myr. 
    \item {\it BH mass:} for the hydrodynamical simulations we save the mass of the most massive BH within a sphere of $10$~ckpc of the centre of potential. For the SAMs, this is well defined and each galaxy has a single value.
    \item {\it Host halo mass:} For \shark, \eagle, {\sc IllustrisTNG} and {\sc Simba}, we use $M_{\rm 200crit}$ to measure the host halo mass, where this is the mass enclosed by a sphere of mean density $200$ times the critical density of the universe. For {\sc Galform} instead we use the `D-halo' host halo mass, which is $\approx 1.26$ times higher than $M_{\rm 200crit}$, while for {\sc GAEA} we use the FOF halo mass, which tends to be $\approx 1.23$ times more massive than $M_{\rm 200crit}$. These factors come from \citet{Jiang14}, and we apply them to scale the halo masses down to an approximate $M_{\rm 200crit}$.
    \item {\it SFR and stellar mass histories}: For the SAMs, we compute the SFR histories by summing the SFRs at each snapshot of all the progenitors of the $z=3$ galaxies. Because of the different number of snapshots, the time cadence is different for the three SAMs. The SFHs are sampled with $100$, $28$ and $16$ timesteps for \shark, {\sc GAEA} and {\sc Galform}, respectively. These timesteps correspond to a typical time cadence of $30-40$~Myr for \shark{} and $100-180$~Myr for {\sc GAEA} and {\sc Galform}.  
    For the hydrodynamical simulations, we take all the stellar particles enclosed in the $50$~ckpc sphere described above {and take their formation age to} bin them in $100$ timesteps of $20$~Myr {width}. We compute the SFR at each timestep in the same way as we do the SFRs above. Similarly, we compute the stellar mass history for galaxies in SAMs by summing the stellar mass formed by each snapshot of all the progenitors of the $z=3$ galaxies. For the hydrodynamical simulations, we sum {the stellar masses of the particles that formed at different epochs, using the same $100$ bins of width $20$~Myr we adopted for the SFR histories.}   
    \item {\it Maximum SFR:} with the SFR histories above, we extract the maximum SFR each galaxy at $z=3$ had in the past (regardless of when that happens), and refer to that as $\rm SFR_{\rm max}$. For the same time, we also define the maximum sSFR, $\rm sSFR_{\rm max}$, as $\equiv \rm SFR_{\rm max}/M_{\star}$, where $M_{\star}(\rm at\,SFR_{\rm max})$ is the stellar mass the galaxy had when it reached $\rm SFR_{\rm max}$. The poorer time cadence in {\sc GAEA} and {\sc Galform} is likely to affect $\rm SFR_{\rm max}$ as it would be an average over $\approx 100-180$~Myr, while for the other simulations, $\rm SFR_{\rm max}$ is an average over $\approx 20-40$~Myr. Appendix~\ref{maxSFReff} shows that although there is an effect of the time cadence, it is not big, and  does not change our conclusions.
    \item {\it Stellar ages:} with the stellar mass histories defined, we compute $3$ stellar ages as the lookback time (from $z=3$) to when the $z=3$ galaxies had assembled $50$, $80$ and $90$~per cent of their stellar mass. We refer to these ages as $\rm age_{50}$, $\rm age_{80}$ and $\rm age_{90}$, respectively.
\end{itemize}

We test using different spherical apertures for the hydrodynamical simulations, using $10$~ckpc, $30$~ckpc and using twice the half-stellar mass radius, and find little difference in the results presented here. The main difference is that smaller apertures increase slightly the number of passive galaxies, but differences are $<10$ per cent. Note that the $10$~ckpc is similar to the aperture used to measure colours for galaxies with the {\it JWST} and {\it HST}, which are what is ultimately used to select quenched galaxies at high-z.

\begin{figure}
\begin{center}
\includegraphics[trim=4.5mm 4.5mm 2mm 4mm, clip,width=0.49\textwidth]{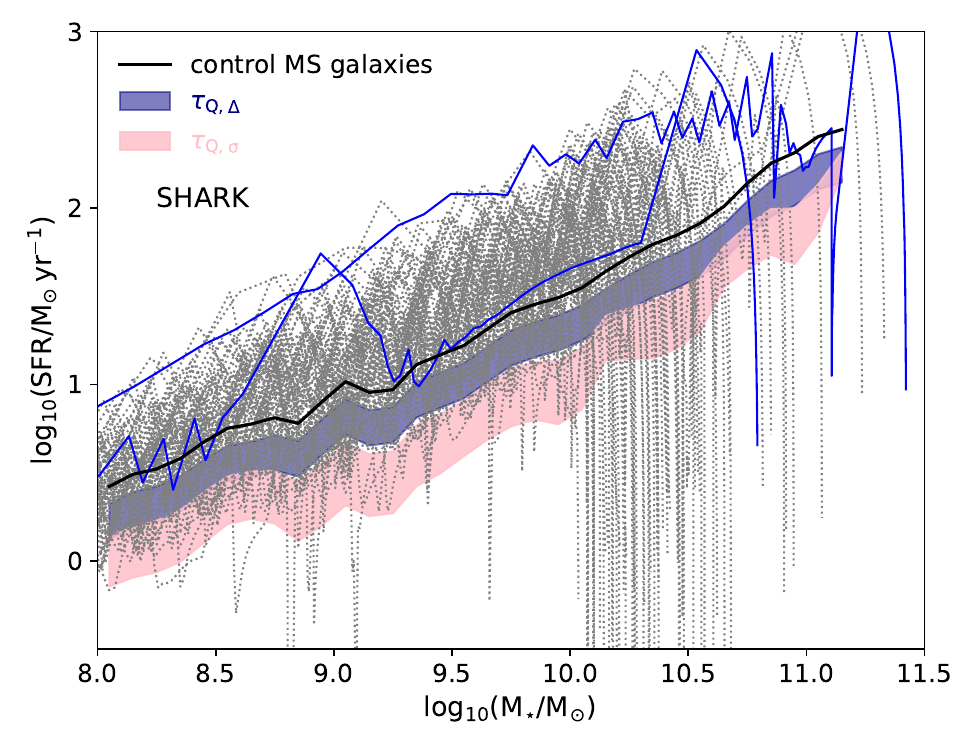}
\caption{The evolution of 70 randomly selected $z=3$ massive-quenched galaxies in \shark\ in the SFR-$M_{\star}$ plane  (individual dotted lines). We highlight two of them with blue lines: one that grows above the main sequence most of its life and that has one quenching period prior to the one that leads to it being quenched at $z=3$; and another that instead moves mostly along the main sequence, bar an early starburst, until there is a big starburst, followed by quenching close to $z=3$.
The black, solid line shows the median growth of the control MS sample. We define $\tau_{\rm Q,\Delta}$ as the time a galaxy takes to move through the blue band while quenching, while $\tau_{\rm Q,\sigma}$ is defined using the pink band instead.} 
\label{tauQex}
\end{center}
\end{figure}

With the defined properties above we select a sample of passive, massive galaxies and a control sample of star-forming galaxies following the criteria below:

\begin{itemize}
    \item {\it Massive-quenched galaxies:} are defined as those with $M_{\star}>10^{10}\,\rm M_{\odot}$ and $\rm sSFR\equiv SFR/M_{\star}<10^{-10}\rm\, yr^{-1}$. The latter is a good criterion to select passive galaxies as shown by \citet{Franx08,Schreiber18}; we discuss this threshold further in \S~\ref{SMFsection}. Table~\ref{simus} shows the number of selected galaxies in each simulation and the implied number density. For the SAMs, in addition to the selection above, we also require galaxies to be centrals. This is done with the aim of focusing the analysis around the effect of AGN feedback. Note that \shark\ predicts more passive satellite galaxies than {\sc GAEA} and {\sc Galform} due to the dynamical friction timescale employed \citep{Poulton2021}, which allows for a much longer survival timescale of satellite galaxies. {For the hydrodynamical simulations we do not enforce galaxies in the massive-quenched sample to be centrals for two reasons: the selection in stellar mass and sSFR already selects primarily centrals ($\gtrsim 90$~per cent); and massive galaxies quench similarly in hydrodynamical simulations regardless of whether they are centrals or satellites (e.g. \citealt{Wright19}). The latter is not the case in SAMs by construction and hence the need to impose galaxies to be central to isolate AGN feedback effects in SAMs.}
    \item {\it Control sample:} in order to have a reference point to compare our massive-quenched galaxy sample, we select a sample of massive-active galaxies. To do so, we first compute the median sSFR, $\rm sSFR_{\rm MS}$ (with MS referring to `main sequence') of all galaxies with $M_{\star}>10^{10}\,\rm M_{\odot}$ and sSFR$>10^{-9.9}\,\rm yr^{-1}$ at $z=3$, and then select all galaxies with $M_{\star}>5 \times 10^{10}\,\rm M_{\odot}$ and $\rm sSFR>0.63\, sSFR_{\rm MS}$ (the $0.63$ is equivalent to requiring galaxies to have a sSFR above $\rm sSFR_{\rm MS}$ minus 0.2~dex). For the control sample, we also compute SFR histories. We refer to this sample as `control MS sample'.
\end{itemize}

With a control sample defined, we can also compute a main sequence in the SFR-$M_{\star}$ plane internal to each simulation as follows. We use the growth of galaxies in the control sample in the SFR-$M_{\star}$ to measure a median SFR, $\rm \langle SFR\rangle$, and a standard deviation, $\sigma_{\rm MS}$, in bins of stellar mass (which we show later in \S~\ref{quenchingmass} in Fig.~\ref{mainseq} as black solid and dotted lines, respectively) and define this relation as our main sequence. {We show an example of the main sequence in Fig.~\ref{tauQex}.} 
We can then compute a distance to the main sequence, $\Delta_{\rm MS}=\rm log_{10}(SFR/\langle SFR(M_{\star})\rangle$, with $\rm SFR$ and $M_{\star}$ being the SFR and $M_{\star}$ {of the galaxies in our `massive-quenched' sample at any point of their history. In other words, we measure the deviation from the solid black line along the $y$-axis in Fig.~\ref{tauQex}.}
%
Because each simulation also predicts different widths of the main sequence, we also analyse the distance to the main sequence in units of the main sequence's standard deviation, $\Delta_{\rm MS}/\sigma_{\rm MS}$, where a value of $1$ indicates $1\sigma$ deviation from the main sequence. Appendix~\ref{MS} shows that this way of defining a main sequence and deviations to the main sequence is similar to measuring the main sequence at individual redshifts and using those to measure a distance to the main sequence (which is the more common way of doing it). 

With the distance to the main sequence defined at all points in the history of the $z=3$ massive-quenched galaxies, we move to quantifying quenching timescales. Defining quenching timescales from the trajectory of galaxies in the SFR-$M_{\star}$ plane involves applying some arbitrary distance to the main sequence and measure how long it takes for a galaxy to fall below some thresholds of $\Delta_{\rm MS}$ (e.g. \citealt{Wright19}). Alternatively, using colours makes for a clearer definition as the blue cloud and red sequence are quantifiable populations in the colour-magnitude diagram (e.g. \citealt{Wright19,Bravo22,Bravo23}). The drawback of the latter approach is the fact that there is no clear red sequence in place at $z=3$ yet in many of the simulations employed here (e.g. \citealt{Trayford16}). Hence, we decide to define quenching timescales based on the distance to the main sequence in the two following ways:

\begin{itemize}
    \item {$\tau_{\rm Q,\Delta}$:} We define two points of interest in $\Delta_{\rm MS}$ based on previous literature: a value of $-0.3$ to $-0.5$ indicates the point when the galaxy leaves the main sequence, and $-1$ to $-1.3$ when the galaxy is considered quenched (e.g. \citealt{Wright19,Bluck20,Mun24}). Here we measure the time it takes for the galaxies to go from $\Delta_{\rm MS}=-0.3$ to $-1$ and refer to this time as $\tau_{\rm Q,\Delta}$. If galaxies have several quenching periods, we take only the last one (i.e. the one closest to a lookback time of $0$). We tested different values within the ranges above and find that the results remain qualitatively the same.
    \item {$\tau_{\rm Q,\sigma}$:} The measurement above fails at considering the fact that each simulation predicts different widths of the main sequence. For example, in {\sc IllustrisTNG} the main sequence is extremely tight, with a standard deviation $\approx 0.1$~dex, while in \shark\ the dispersion is larger, $\approx 0.25$~dex. In this context, a $\Delta_{\rm MS}=-0.3$ in  {\sc IllustrisTNG} implies a significant deviation from the main sequence, while in \shark\ such galaxy would be just leaving the main sequence. Hence, we also measure a timescale using $\Delta_{\rm MS}/\sigma_{\rm MS}$ and thresholds of $-1$ and $-2$; i.e. the time it takes the galaxy to go from a main sequence deviation of $1\sigma$ to $2\sigma$ below the main sequence. The exact values are again arbitrary, but we find that varying these values a bit does not impact the results significantly. 
\end{itemize}

Fig.~\ref{tauQex} shows visually how we measure $\tau_{\rm Q,\Delta}$ and $\tau_{\rm Q,\sigma}$, using $70$ randomly selected $z=3$ massive-quenched galaxies. $\tau_{\rm Q,\Delta}$ is the time a galaxy takes between entering and leaving the blue band, while for $\tau_{\rm Q,\sigma}$ this corresponds to moving through the pink band. We also highlight two galaxies, with the aim of highlighting the diversity of trajectories of galaxies in this plane. 

Finally, we note that the simulations adopt slightly different cosmologies. Those adopted by \shark, {\sc Eagle}, {\sc IllustrisTNG} and {\sc Simba} are very similar, and stellar masses and SFRs are expected to differ only by $1$~per cent due to the differing cosmologies (as indicated by the differences in their $h$ parameters). This is larger for {\sc GAEA} ($\approx 9$ per cent) and {\sc Galform} ($\approx 5$ per cent). These differences are much smaller than those produced by the baryon physics included in each model, so we neglect them. 

\section{Number densities and stellar mass function of high-z massive quiescent galaxies}\label{SMFsection}

\subsection{Number density of massive-quenched galaxies at $2<z<5$}
\begin{figure*}
\begin{center}
\includegraphics[trim=5mm 4.5mm 2mm 2mm, clip,width=0.47\textwidth]{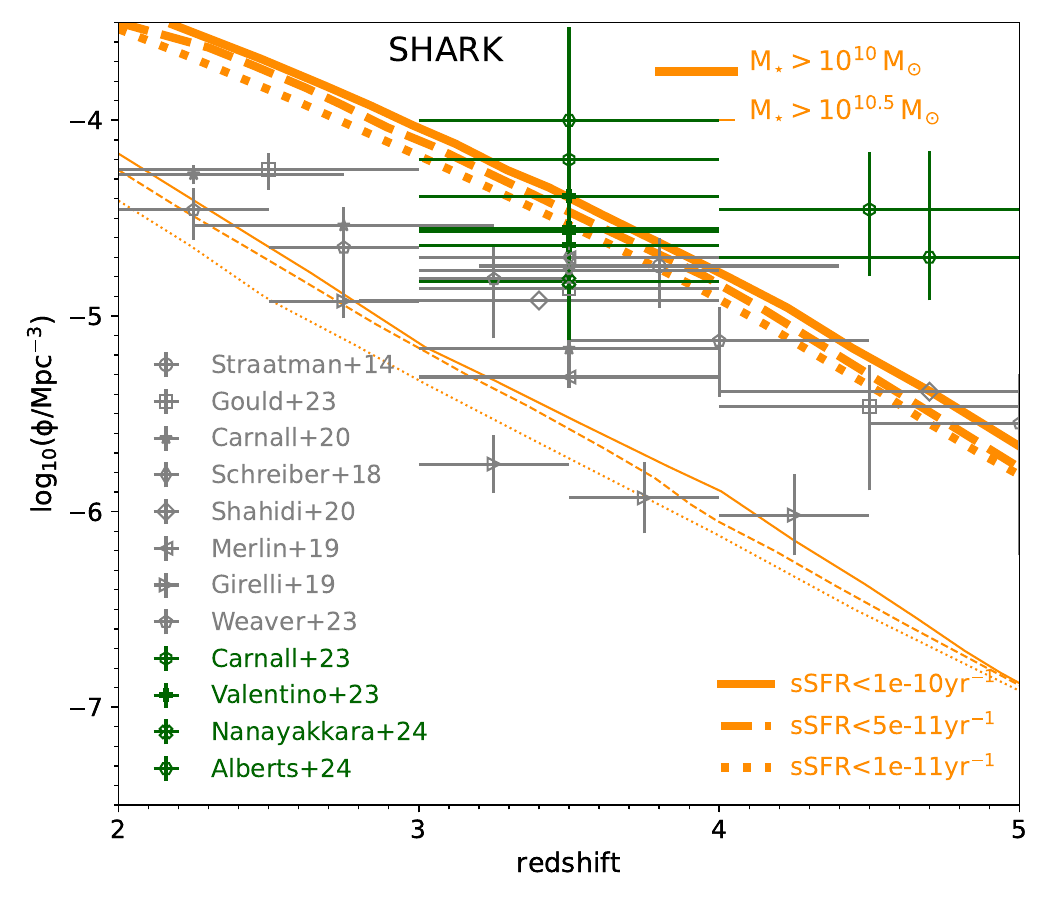}
\includegraphics[trim=5mm 4.5mm 2mm 2mm, clip,width=0.47\textwidth]{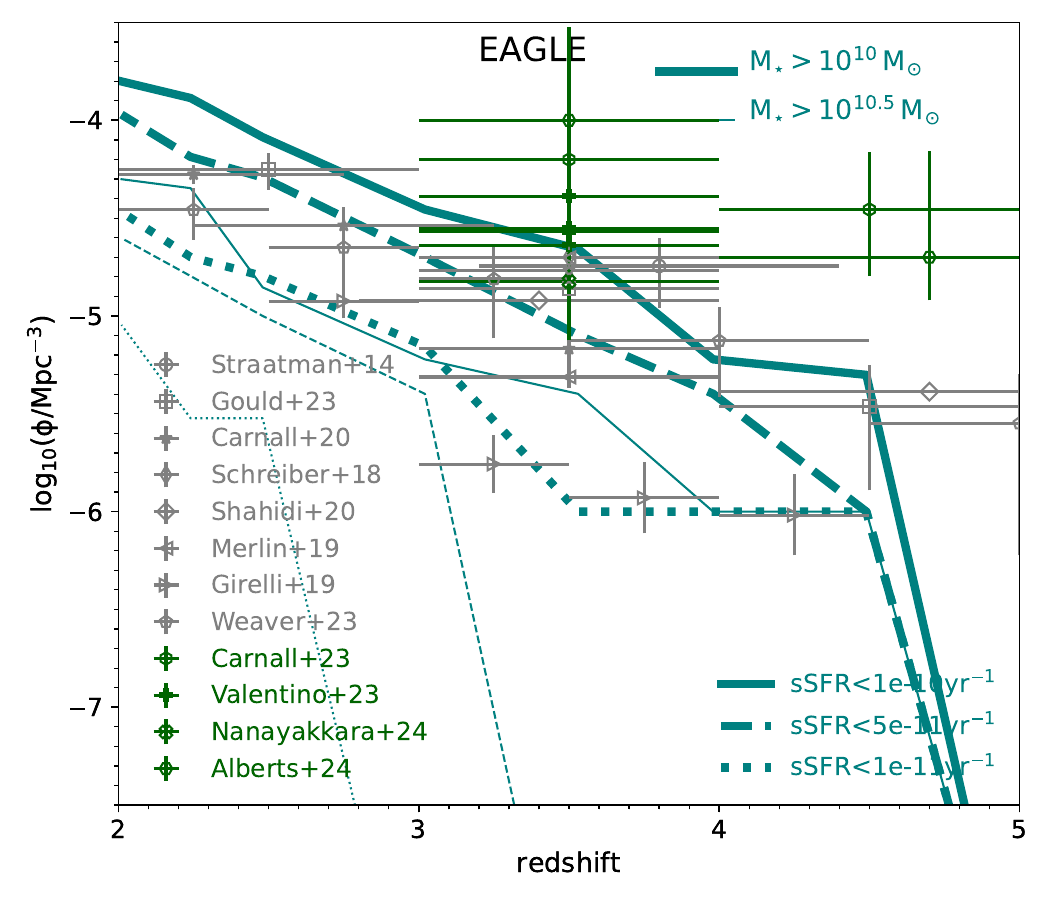}
\includegraphics[trim=5mm 4.5mm 2mm 2mm, clip,width=0.47\textwidth]{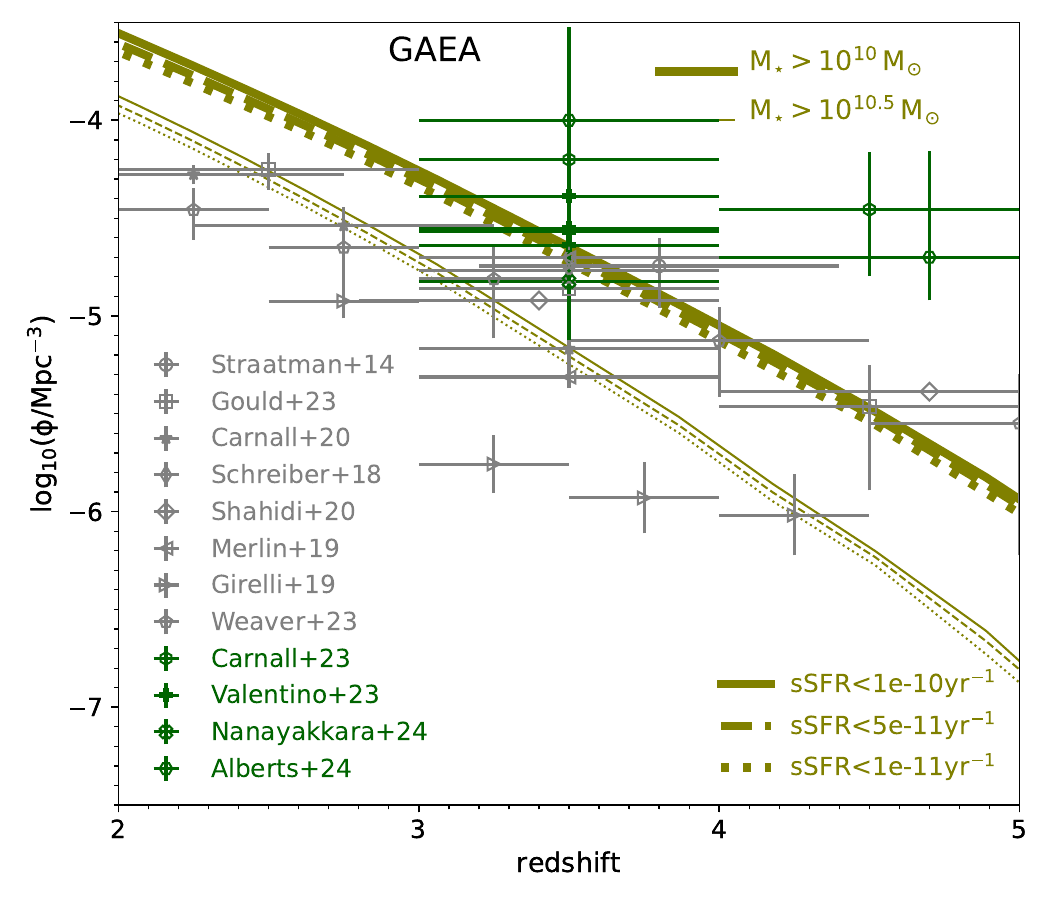}
\includegraphics[trim=5mm 4.5mm 2mm 2mm, clip,width=0.47\textwidth]{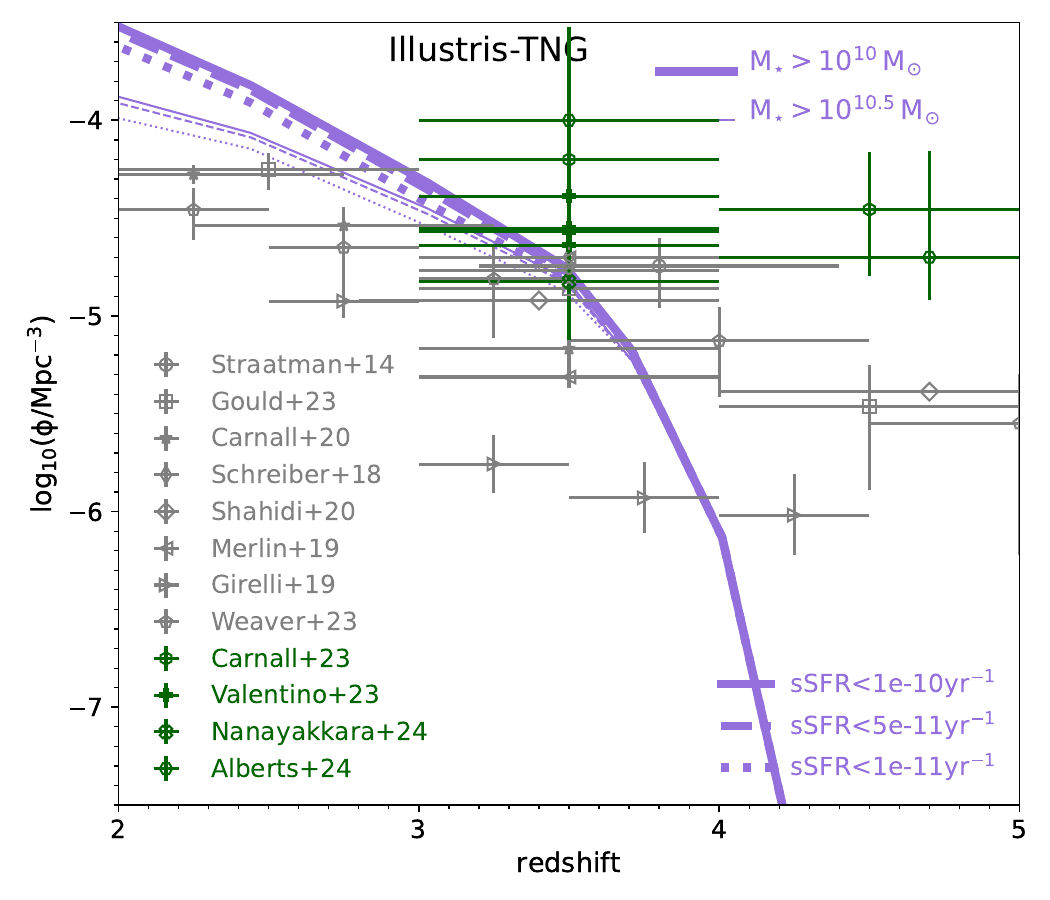}
\includegraphics[trim=5mm 4.5mm 2mm 2mm, clip,width=0.47\textwidth]{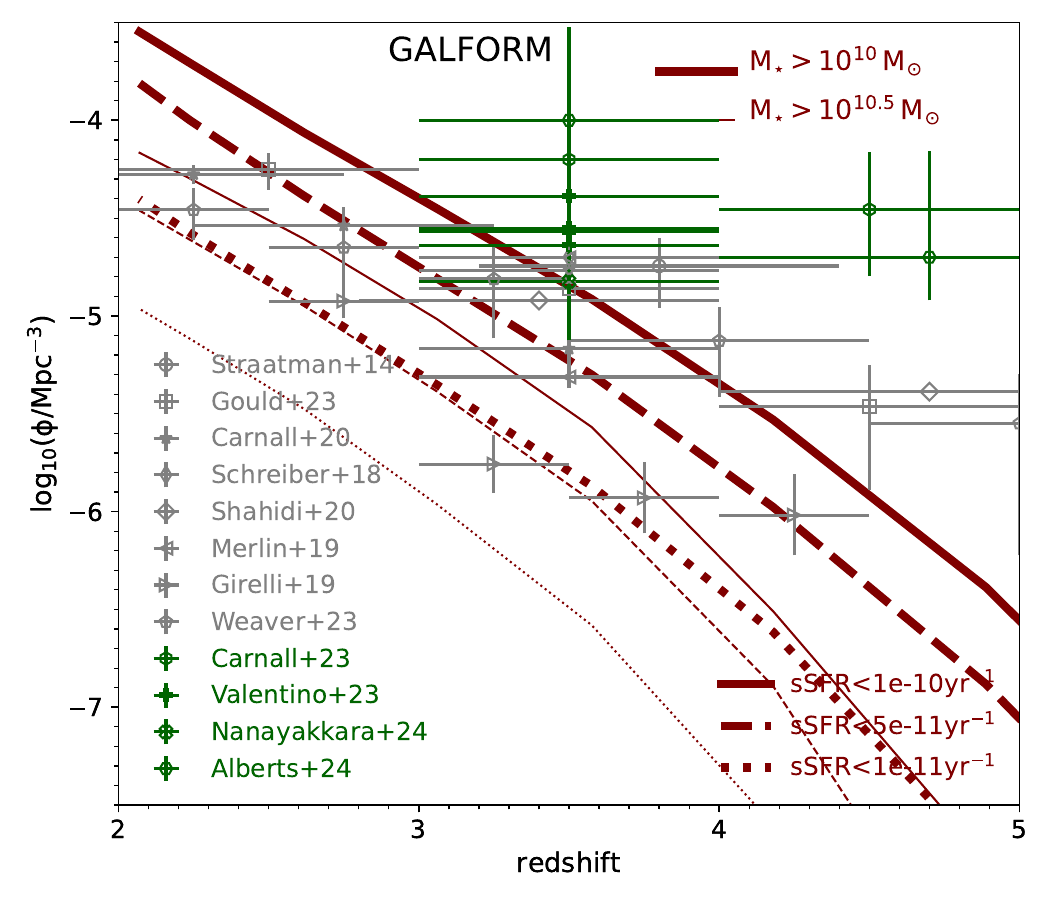}
\includegraphics[trim=5mm 4.5mm 2mm 2mm, clip,width=0.47\textwidth]{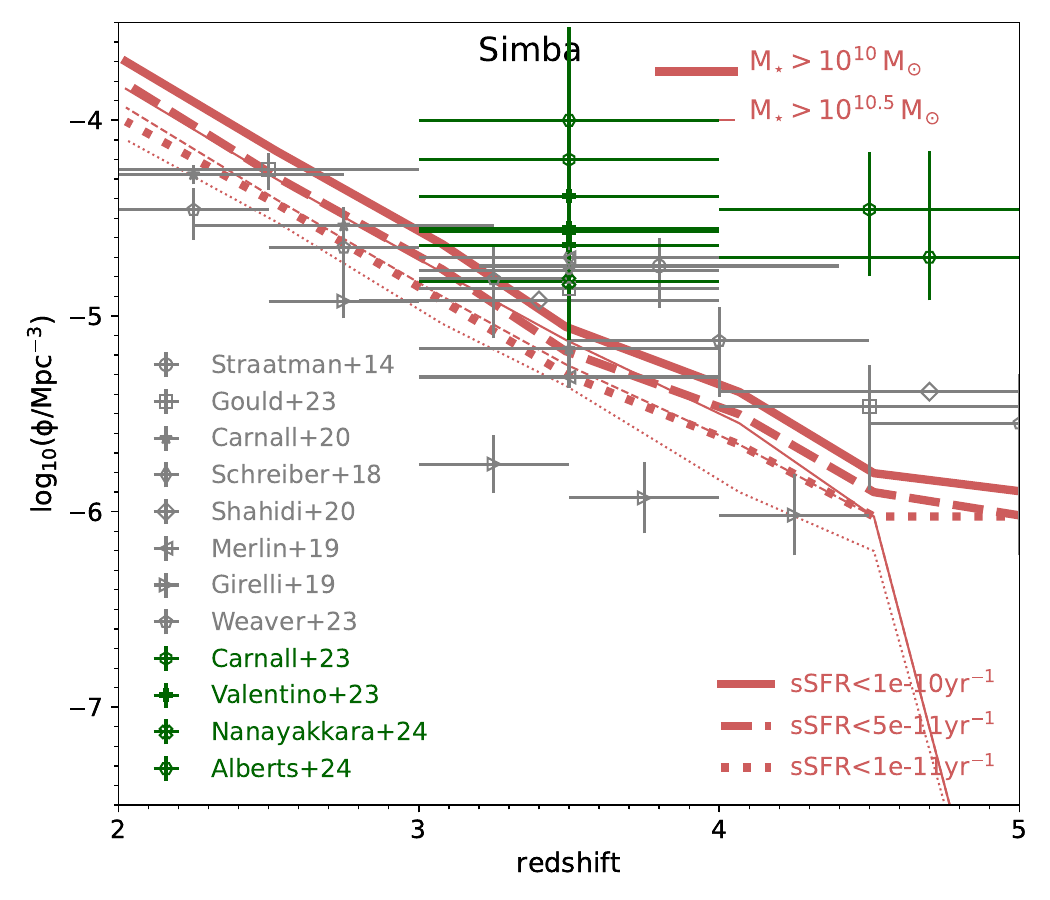}
\caption{Number density of passive massive galaxies selected based on two stellar mass thresholds, $M_{\star}>10^{10}\,\rm M_{\odot}$ (thick lines) and $M_{\star}>10^{10.5}\,\rm M_{\odot}$ (thin lines) and three sSFR thresholds, $\rm sSFR<10^{-10}\,\rm yr^{-1}$ (solid lines), $\rm sSFR<5\times 10^{-11}\,\rm yr^{-1}$ (dashed lines) and $\rm sSFR<10^{-11}\,\rm yr^{-1}$ (dotted lines), as labelled, between $2\le z\le 5$. This is shown for the $6$ simulations being analysed; the SAMs: \shark\, {\sc GAEA}, and {\sc Galform}; and the hydrodynamical simulations: {\sc Eagle}, {\sc IllustrisTNG}, and {\sc Simba}, as labelled in each panel. Masses and SFRs are measured as described in \S~\ref{definitions}. 
For the redshift bins with $0$ passive galaxies, we use an arbitrary low number density of $10^{-9}\,\rm cMpc^{-3}$. Observational estimates are also shown: in grey symbols (pre-{\it JWST} results) are from \citet{Straatman14,Schreiber18,Merlin19,Girelli19,Carnall20,Weaver22,Gould23}, and green symbols ({\it JWST} results) are from \citet{Nanayakkara22,Carnall23,Valentino23,Alberts23}, as labelled in each panel.} 
\label{num_densities}
\end{center}
\end{figure*}

\begin{figure}
\begin{center}
\includegraphics[trim=4.5mm 4.5mm 2mm 2mm, clip,width=0.49\textwidth]{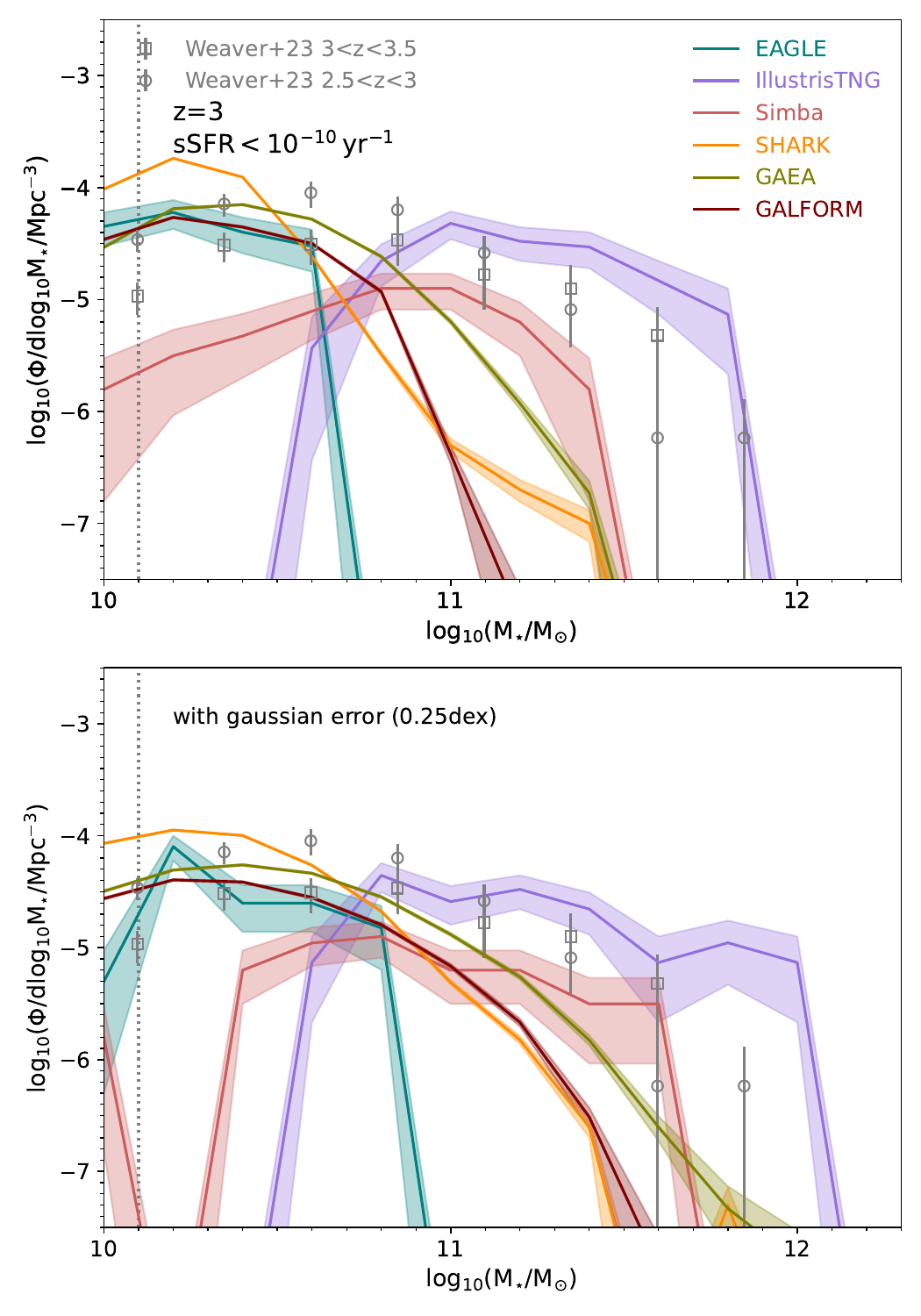}
\caption{{\it Top panel:} Stellar mass function of galaxies at $z=3$ selected to have a $\rm sSFR < 10^{-10}\,\rm yr^{-1}$ in the $6$ different simulations analysed, as labelled. We show Poisson errorbars with shaded regions to demonstrate that the differences between simulations are significant.
Observations from \citet{Weaver22} for passive galaxies are shown in symbols for redshifts around $z=3$. The latter sample was selected using colour-colour selection (see text for details). 
The vertical dotted line marks the approximate stellar mass above which observational estimates of the SMF are reliable according to \citet{Weaver22}.
{\it Bottom panel:} As in the top panel but after convolving stellar masses with Gaussian-distributed errors with a width of $0.25$~dex.}  
\label{SMFz3}
\end{center}
\end{figure}

Fig.~\ref{num_densities} shows the predicted number density of massive-quiescent galaxies at $2\le z\le 5$ in the $6$ simulations analysed. We show this using two thresholds of stellar mass, $M_{\star}>10^{10}\,\rm M_{\odot}$ (thick lines) and $M_{\star}>10^{10.5}\,\rm M_{\odot}$, and three of sSFR, $\rm sSFR<10^{-10}\,\rm yr^{-1}$,  $\rm sSFR<5\times 10^{-11}\,\rm yr^{-1}$ and $\rm sSFR<10^{-11}\,\rm yr^{-1}$. The idea of using these different thresholds is to capture the range of values adopted in the literature. Most works select passive galaxies based on their position in the UVJ colour-colour plane, but the equivalent sSFR selection is unclear. \citet{Weaver22} argue that this colour selection is roughly equivalent to requiring $\rm sSFR < 10^{-11}\,\rm yr^{-1}$ (see also \citealt{Belli19}), while \citet{Schreiber18} argue this is equivalent to a  $\rm sSFR < 10^{-10}\,\rm yr^{-1}$. Given the uncertainty, we show three selections spanning this sSFR range. As for the stellar mass, observations of passive high-z galaxies have started to push down to $10^{9.5}\,\rm M_{\odot}$ (e.g. \citealt{Valentino23}), but because the vast majority of the samples are composed of galaxies with $M_{\star}\gtrsim 10^{10}\,\rm M_{\odot}$, we focus on two bins above that. 

The predicted number densities are very different between different simulations. Focusing first on \shark, we see that the sSFR threshold has almost no effect, due to passive galaxies being very passive (i.e. being many dex below the main sequence). Conversely, the stellar mass threshold has a significant impact, with the number density decreasing by $\gtrsim 1$~dex in the sample of $M_{\star}>10^{10.5}\,\rm M_{\odot}$ galaxies. {\sc GAEA} behaves very similarly, but with a lesser difference between the two stellar mass bins, specially at $z<3$. The latter is due to {\sc GAEA} producing a less steep high-mass end of the SMF of passive galaxies compared with \shark\ (visible in Fig.~\ref{SMFz3}). 

In {\sc Eagle} and {\sc Galform}, both the stellar mass and sSFR threshold have an important effect on the number density. This shows that in these simulations there is a large population of massive galaxies that are moderately quenched (i.e. $10^{-11}\,\rm yr^{-1}\lesssim sSFR\lesssim 10^{-10}\,yr^{-1}$). This is very clearly seen later in \S~\ref{quenchingmass} in Fig.~\ref{mainseq}.
For the most massive bin in {\sc Eagle}, we see there are no galaxies with $\rm sSFR<10^{-11}\,yr^{-1}$ ($\rm sSFR<5\times 10^{-10}\,yr^{-1}$) at $z\gtrsim 2.5$ ($z\gtrsim 3$). 

{\sc Simba} shows number densities of passive galaxies that are only mildly dependent on both stellar mass and sSFR. For instance, it displays a stronger dependence on sSFR than \shark\ and {\sc GAEA} but much weaker than {\sc Eagle} and {\sc Galform}. {\sc IllustrisTNG} behaves very differently with the number density of passive galaxies of any mass and sSFR threshold falling off sharply at $z\gtrsim 3.5$ with no quenched massive galaxies at $z\ge 4$. At lower redshifts, the sSFR threshold makes a negligible difference as in \shark\ and {\sc GAEA}, while the stellar mass threshold only makes a mild difference at $z\lesssim 3.5$. This shows that in {\sc IllustrisTNG} these quenched galaxies are very massive and very quenched. 

In addition to the simulations predicting quite different number densities of massive-quenched galaxies, they also predict very different fractions of passive galaxies ($f_{\rm pass}$; see Table~\ref{simus}), with differences of a factor of $\approx 3$. {\sc GAEA} predicts the highest fraction ($9.3$ per cent) and {\sc Simba} the lowest one ($3.6$ per cent). 

The observations shown in Fig.~\ref{num_densities} display large variations and when focusing on results using {\it JWST} observations, we find that all the simulations prefer the lower number densities that are obtained using the NUVU-VJ plane, which is expected to be more robust (i.e. having fewer contaminants) than the classic UVJ selection \citep{Gould23}. This uses three colours instead of two and a probabilistic method to assign galaxies to the star-forming and passive populations. 

At $z\gtrsim 4.5$ only \shark, {\sc Simba} and {\sc GAEA} produce quenched galaxies in reasonable numbers; however they tend to have $10^{10}\,\rm M_{\odot}\lesssim M_{\star}\lesssim 10^{10.5}\,M_{\odot}$ which may be too low compared to the derived stellar masses in observations. Note that \citet{Szpila24} presented number densities of massive-quenched galaxies at high-z using {\sc Simba-C}, an improved version of {\sc Simba} with an updated chemical evolution model and tweaked feedback parameters. Those number densities are similar to the ones we obtain using {\sc Simba}, and hence we do not expect other results presented here for {\sc Simba} to be too dissimilar to what is seen in {\sc Simba-C}.

In addition to the SAMs we study here, \citet{Vani24} presented an analysis of the {\sc L-galaxies} semi-analytic model in three different flavours and found that all of them struggled to produce enough massive-quenched galaxies at $z\gtrsim 1.5$. 
\citet{Remus23} show that in the hydrodynamical simulation {\sc Magneticum}, the number density of massive-quenched galaxies matches relatively well the observations at $z\approx 3-4$. This is also shown in Fig.~\ref{num_density_magneticum} for a single stellar mass and sSFR thresholds that are comparable to the ones used in Fig.~\ref{num_densities}. {\sc Magneticum} predicts the highest  massive-quenched galaxy number density at $z=3-4$ of the simulations analysed here, but it also massively over-predicts the number density of those galaxies at slightly lower redshift $z=2-3$, likely due to a combination of overly efficient AGN feedback and resolution effects affecting the galaxies with stellar masses close to $10^{10}\,\rm M_{\odot}$ (see Appendix~\ref{Magneticum} for a discussion).  

Thus, the difficulty in reproducing the frequency of massive-quenched galaxies from cosmic dawn to noon is a very pervasive problem in modern galaxy formation simulations. 

\subsection{The $z=3$ stellar mass function of quenched galaxies}

The top panel of Fig.~\ref{SMFz3} shows the $z=3$ SMF of passive galaxies selected to have $\rm sSFR<10^{-10}\, \rm yr^{-1}$. The observations correspond to those of \citet{Weaver22}, where passive galaxies were selected from the NUVrJ plane. \shark, {\sc Eagle} and {\sc Galform} tend to produce the correct number density of passive galaxies with $10^{10}\rm \, M_{\odot} \lesssim M_{\star}\lesssim 10^{10.6}\,\rm M_{\odot}$ but predict too few passive galaxies of higher masses, while {\sc GAEA}'s agreement with observations extends to $M_{\star}\approx 10^{11}\,\rm M_{\odot}$; albeit also producing too few galaxies above that mass threshold. {\sc Simba} and {\sc IllustrisTNG} produce enough massive-quiescent galaxies with $M_{\star}\gtrsim 10^{11}\,\rm M_{\odot}$ but too few at lower stellar masses compared with observations. This shows that the suite of simulations here agree with observations in different regimes. The exact cause of this is the way AGN feedback is modelled and we discuss this in \S~\ref{AGN}.

To demonstrate the effect uncertainties in stellar masses can have in the SMF, we show in the bottom panel of Fig.~\ref{SMFz3} the mass functions after we convolve the stellar masses in each simulation with errors that are Gaussian-distributed with a width of $0.25$~dex. The tension seen between simulations that underpredict the abundance of passive galaxies with $M_{\star}\gtrsim 10^{11}\,\rm M_{\odot}$ and the observations, is greatly diminished (clearly seen for \shark, {\sc Galform} and {\sc GAEA}). The same can be said for the tension at $M_{\star}\lesssim 10^{10.5}\,\rm M_{\odot}$ seen in {\sc Simba} and {\sc IllustrisTNG}, but because the SMF at those masses is less steep, the effect of a random error is less pronounced. 

The $0.25$~dex width of error above is informed by the reported errors from spectral energy distribution (SED)-derived stellar masses at $z<0.1$ of \citet{Robotham20}. \citet{Robotham20} computed these masses using $27$ bands covering from the FUV to the FIR of galaxy SEDs and found typical errors of $0.2$~dex. At $z=3$ the stellar mass uncertainties are likely larger due to the more limited wavelength range of galaxy SEDs and redshift uncertainties, to mention a few (see \citealt{Pacifici23} and \citealt{Bellstedt24} for a comprehensive study of systematic effects in the derivation of galaxy properties from SED fitting). In fact, \citet{Wang24} show that excluding the {\it JWST} mid-IR bands in the derivation of stellar masses can bias the inferences high by even an order of magnitude (which mostly affect galaxies at $z>3$).
Hence, what is shown in the bottom panel of Fig.~\ref{SMFz3} is likely a lower limit of the effect of uncertainties in the SMF. Thus, quantifying systematic errors in the derived stellar masses and SFRs of galaxies from observations is paramount to provide stringent constraints to the simulations. 

Another source of difference between simulations and observations could be possible contamination of the colour-colour selection employed by \citet{Weaver22} in the process of constructing the SFMs of passive galaxies. \citet{Lagos24} showed, using \shark, that the NUVrJ selection can lead to significant contamination from  massive, star-forming galaxies at $z\gtrsim 2$. At lower redshifts, they found little contamination. \citet{Lagos24}, however, showed that even after applying the NUVrJ selection to \shark\ galaxies, the increased number density of passive galaxies selected by their colour was not enough to bring the simulation into agreement with \citet{Weaver22} at $z=3$. \citet{DeLucia24} using {\sc GAEA} also explored the performance of the NUVrJ selection in their simulation and found that up to $z\approx 3$ it performs well, without significant contamination. \citet{Akins22} explored the performance of the UVJ colour selection instead in {\sc Simba} up to $z=2$ and found that overall it selected the expected population of passive galaxies. The overall conclusion is that at $z=3$ is still unclear how much contamination there may be in the colour-selected samples of passive galaxies, but it is unlikely to be large enough as to bring the models into agreement with observations. 

Another potential source of difference between simulations and observations is cosmic variance. The volumes of the three hydrodynamical simulations used here are small; at $z\approx 3$ and using the cosmic variance calculator of \citet{Driver10}, we find a cosmic variance of $\approx 25$ per cent. This is still a lot smaller than the differences seen between simulations, which tend to be of factors of several.

\subsection{The stellar-halo mass relation of quenched galaxies at $z=3$}

\begin{figure}
\begin{center}
\includegraphics[trim=4mm 4.5mm 2mm 5mm, clip,width=0.49\textwidth]{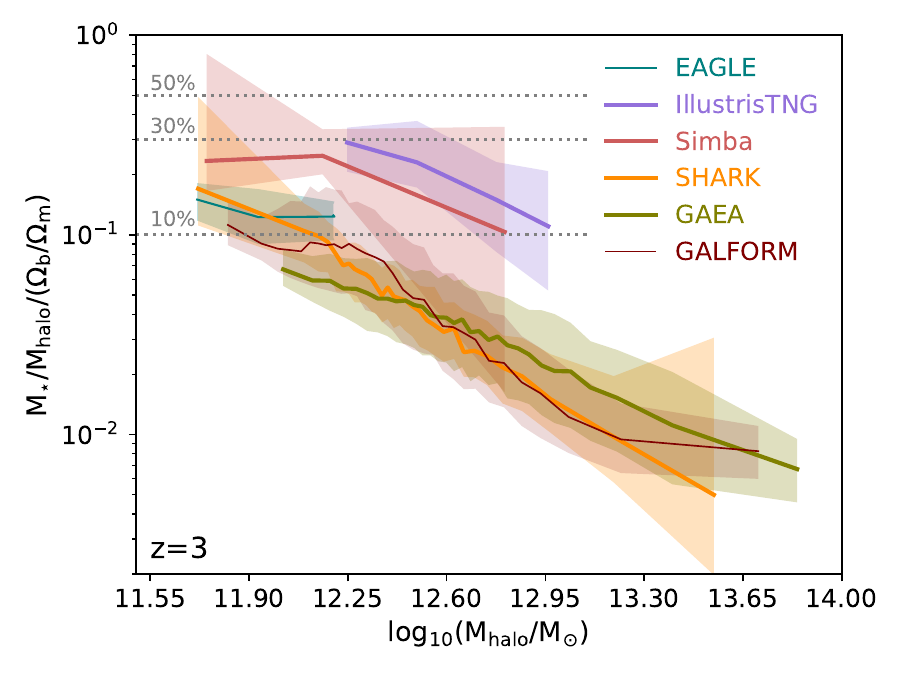}
\caption{The stellar-to-halo mass ratio in units of $\Omega_{\rm b}/\Omega_{\rm m}$ as a function of halo mass for the $z=3$ massive-quenched galaxies in each simulation, as labelled. Lines with shaded regions show the medians and $16^{\rm th}-84^{\rm th}$ percentile ranges, respectively. We use
adaptive bins to ensure a number of galaxies per bin $\ge  10$. The horizontal dotted lines show baryon collapse efficiencies of $10$, $30$ and $50$ per cent for reference.} 
\label{smhm}
\end{center}
\end{figure}

The different SMFs of quenched galaxies in the $6$ simulations of Fig.~\ref{SMFz3} naturally points to potentially different stellar-halo mass relations  between the simulations. The large stellar masses derived for massive-quenched galaxies in observations (of even of up a few $10^{11}\,\rm M_{\odot}$) have been used to argue that to form these galaxies (or at least the most massive) very efficient baryon to stellar mass conversion is required (a.k.a. `baryon collapse efficiency'; e.g. \citealt{Glazebrook17,Carnall24}). In fact, for extreme galaxies, such as ZF-UDS-7329 \citep{Glazebrook23}, baryon collapse efficiencies as high as $\approx 50-100$ per cent have been suggested \citep{Carnall24}. 

Fig.~\ref{smhm} shows the stellar-to-halo mass ratio as a function of halo mass for massive-passive galaxies at $z=3$ in each simulation. We normalise the y-axis by $\Omega_{\rm b}/\Omega_{\rm m}$ internal to each simulation to turn it into baryon collapse efficiency. We highlight three efficiencies. The $10$ per cent is the one preferred by most simulations at the lowest halo masses, while $30$ and $50$ per cent are motivated by \citet{Glazebrook17} and \citet{Carnall24} suggesting those values are required to explain some extremely massive and early-forming quenched galaxies at $z=3-4$. 

The lack of intermediate-mass galaxies in {\sc Simba} and {\sc IllustrisTNG} translate into those simulations having higher baryon collapse efficiencies compared to the other ones by a factor of $\approx 2$. For {\sc Simba} and {\sc IllustrisTNG}, the median baryon collapse efficiency is $\approx 0.22$, while for \eagle, \shark, {\sc GAEA} and {\sc Galform} is $0.13$, $0.049$, $0.037$ and $0.057$, respectively. Hence, between the extremes ({\sc IllustrisTNG} and {\sc GAEA}), there is a large difference of a factor of $\approx 5.8$ in baryon collapse efficiency.
The results here are pointing to efficiencies $\gtrsim 20$ per cent being required for 
quenched galaxies to reach stellar masses of $\gtrsim 10^{11}\rm M_{\odot}$ at $z=3$, 
as the simulations that have lower efficiencies generally fail to produce enough of those massive galaxies.
The larger range of probed halo masses in \shark, {\sc GAEA} and {\sc Galform} is likely a result of the much larger cosmological volumes compared with the three hydrodynamical simulations used here. 

In general, we find that the simulations prefer baryon collapse efficiencies below $30$ per cent, and for the most massive haloes, $M_{\rm halo}\gtrsim 10^{12.7}\,\rm M_{\odot}$, efficiencies $<10$ per cent are overall preferred. 

\section{The star formation histories of high-z massive quiescent galaxies}\label{SFHssection}

We study the SFHs of massive-quenched galaxies selected at $z\approx 3$  in the six simulations and separate the analysis between when galaxies are forming stars actively (\S~\ref{TheRiseSF}), and when quenching began (\S~\ref{quenchingmass}). In \S~\ref{AGN}, we connect the differences and similarities seen among simulations with the implementation of AGN feedback in each case.

\begin{figure*}
\begin{center}
\includegraphics[trim=4.5mm 4.5mm 2mm 4mm, clip,width=0.49\textwidth]{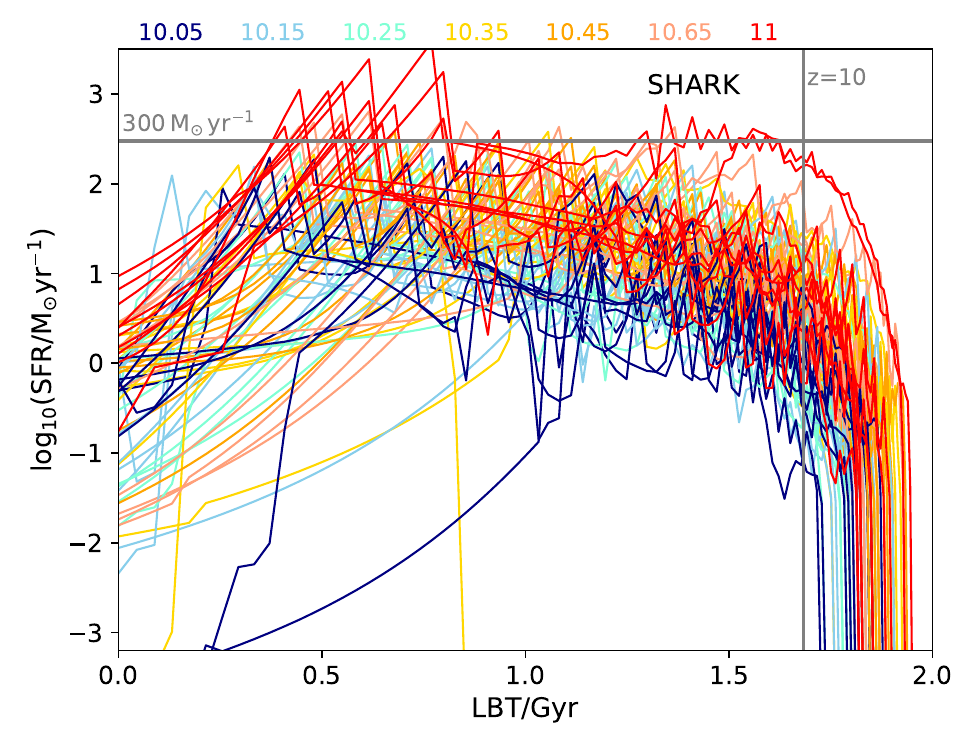}
\includegraphics[trim=4.5mm 4.5mm 2mm 4mm, clip,width=0.49\textwidth]{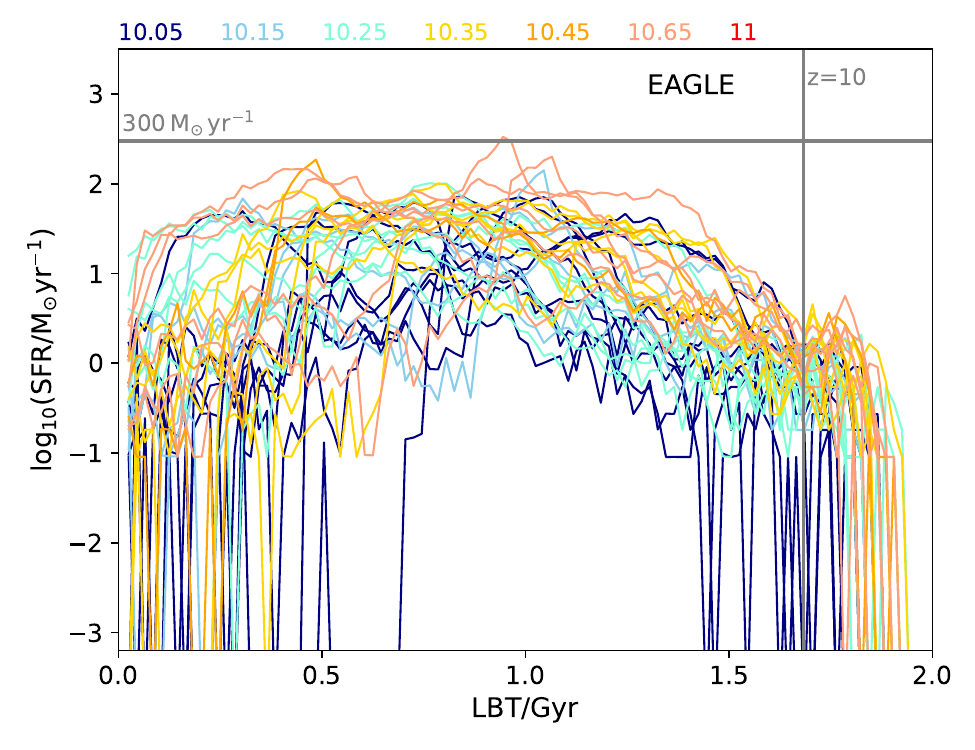}
\includegraphics[trim=4.5mm 4.5mm 2mm 3mm, clip,width=0.49\textwidth]{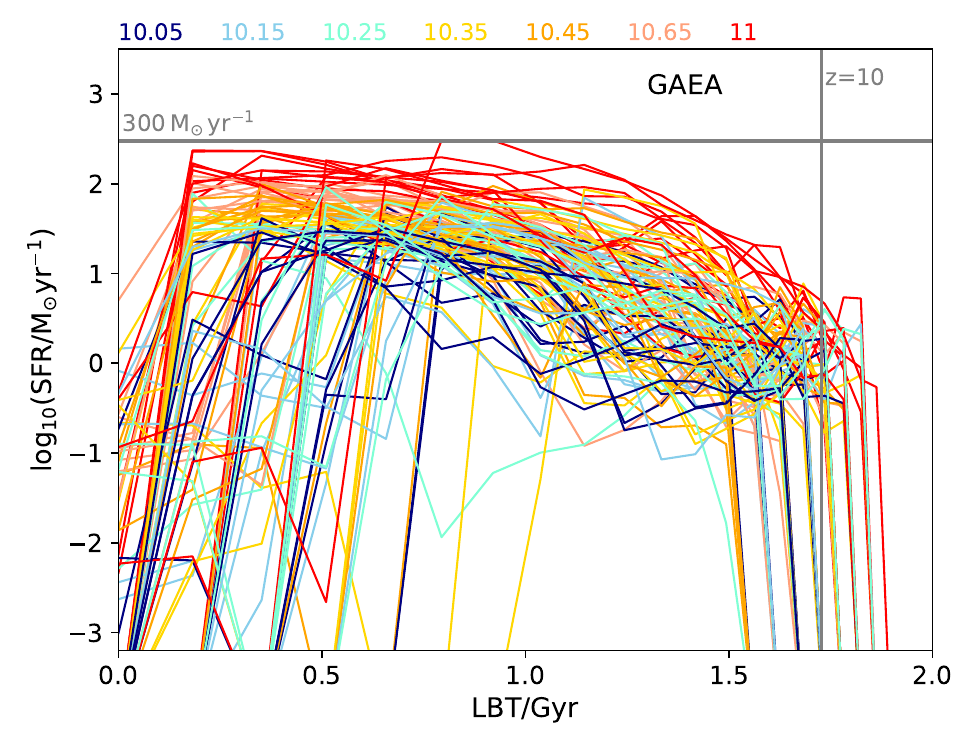}
\includegraphics[trim=4.5mm 4.5mm 2mm 4mm, clip,width=0.49\textwidth]{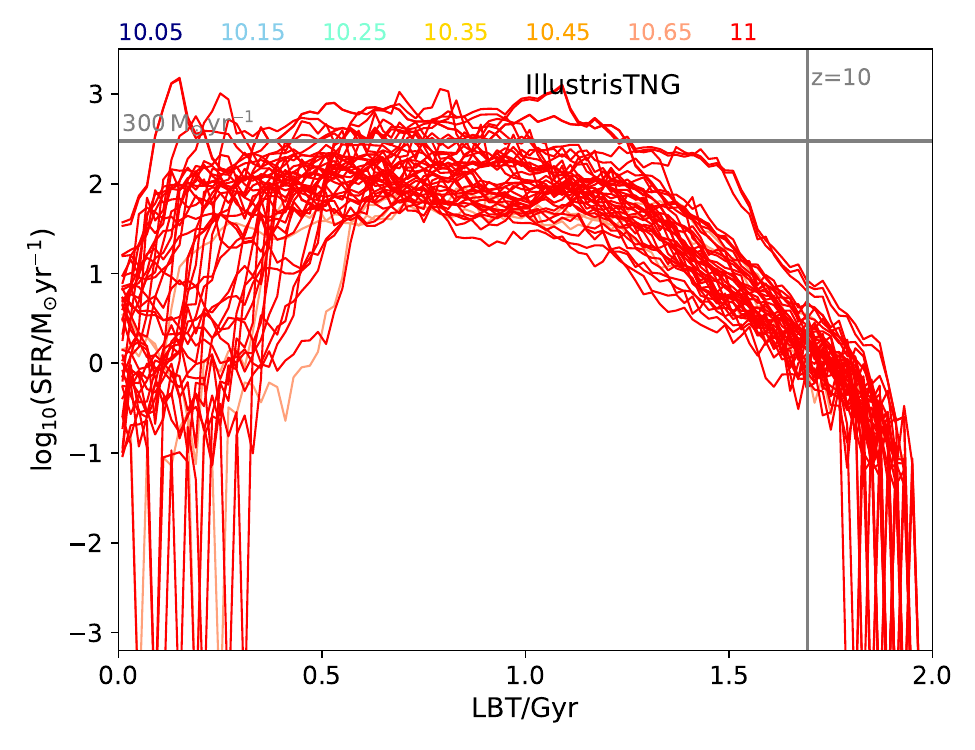}
\includegraphics[trim=4.5mm 4.5mm 2mm 4mm, clip,width=0.49\textwidth]{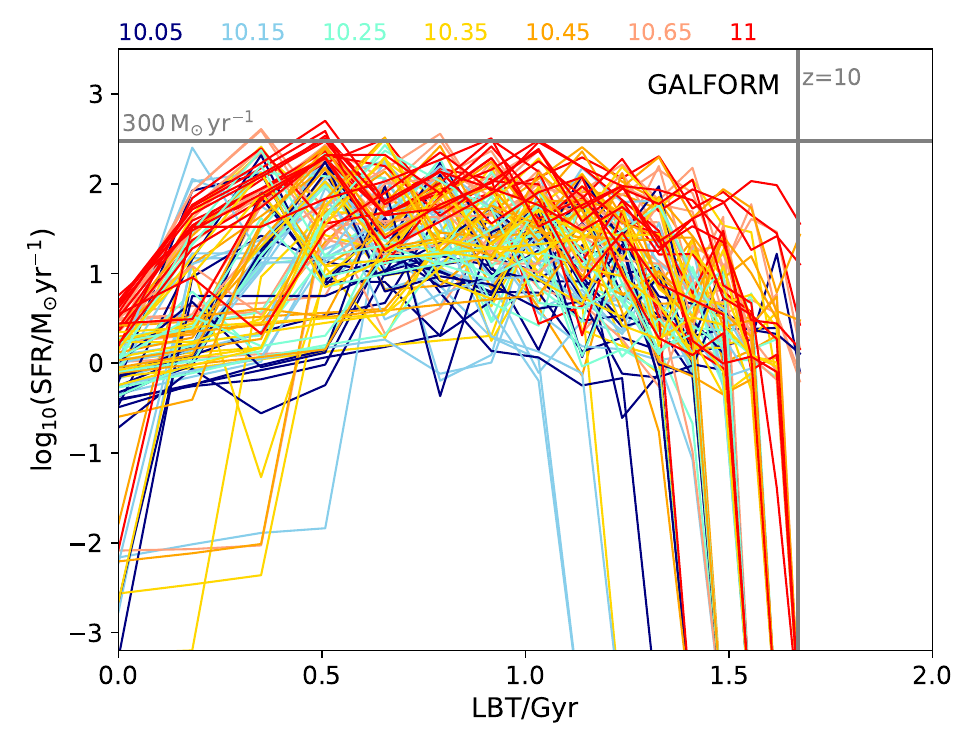}
\includegraphics[trim=4.5mm 4.5mm 2mm 4mm, clip,width=0.49\textwidth]{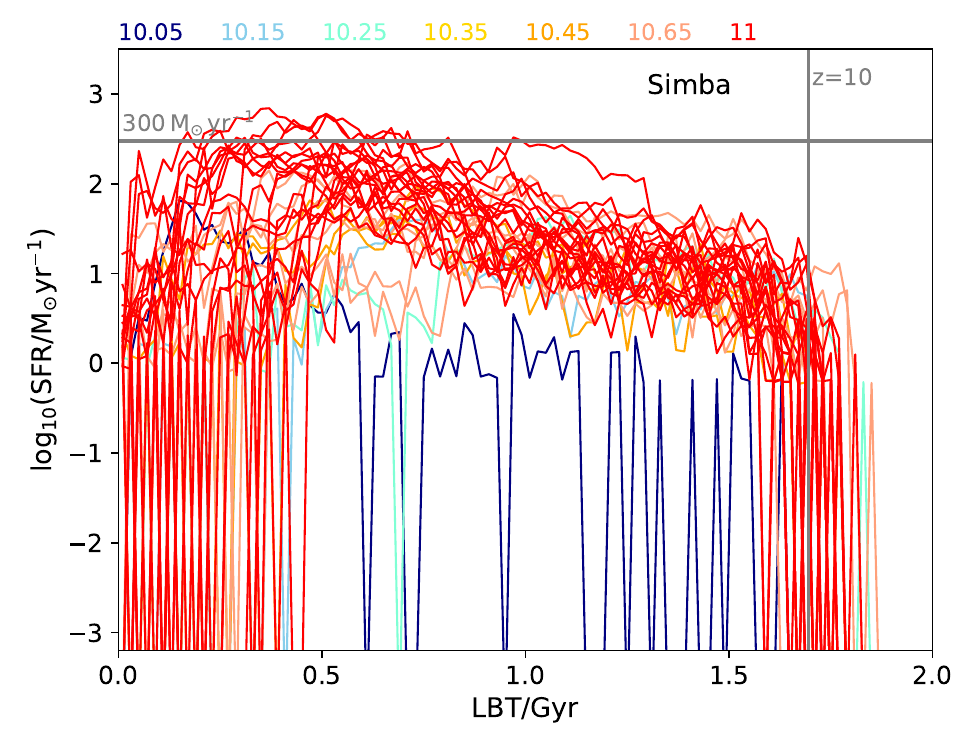}
\caption{SFR histories for galaxies in the \shark\, {\sc Eagle}, {\sc GAEA}, {\sc IllustrisTNG}, {\sc Galform}, and {\sc Simba} galaxy formation simulations, as labelled in each panel, selected at $z=3$ to be massive and passive (see \S~\ref{definitions} for details on the selection). The colour of the lines indicates the stellar mass bins (in units of $\rm M_{\odot}$): $10^{10}-10^{10.1}$, $10^{10.1}-10^{10.2}$, $10^{10.2}-10^{10.3}$, $10^{10.3}-10^{10.4}$, $10^{10.4}-10^{10.5}$, $10^{10.5}-10^{10.75}$, $>10^{10.75}$ (as indicated by the mid bin mass in each bin at the top of each panel). For reference, we show with as a horizontal line a $\rm SFR=300\,\rm M_{\odot}\,yr^{-1}$, and as a vertical line the lookback time corresponding to $z=10$. For {\sc EAGLE}, {\sc IllustrisTNG} and {\sc Simba} we show all the massive-quenched galaxies, while for \shark, {\sc GAEA} and {\sc Galform} we only show 10 randomly chosen galaxies in each mass bin, to avoid overcrowding the figures.} 
\label{SFH}
\end{center}
\end{figure*}

\subsection{The rise in the star formation rate of massive-quenched galaxies}\label{TheRiseSF}

\subsubsection{The SFHs of $z=3$ massive-quenched galaxies}

Fig.~\ref{SFH} shows the SFHs of massive-quenched galaxies selected at $z=3$ of all the simulations analysed. We remind the reader that the SFHs are computed by summing the contribution of all progenitors (see \S~\ref{definitions} for more details). To avoid overcrowding the figure, we only show $10$ randomly selected galaxies per stellar mass bin for the SAMs (\shark, {\sc GAEA} and {\sc Galform}) given their very large number of quenched galaxies (see Table~\ref{simus}), while for {\sc Eagle}, {\sc IllustrisTNG} and {\sc Simba} we show all the predicted $z=3$ massive-quenched galaxies. 
Below we pay especial attention at the predicted SFHs at $z\approx 10$, $z\approx 7$, and $z\approx 4-5$. The justification for this is that many observations of galaxies with relatively high SFRs at $z>10$ are being reported (e.g. \citealt{Bunker23}); $z\approx 7$ is considered a typical star formation epoch (e.g. \citealt{Nanayakkara22}); and because evidence suggest the peak SFR of $z=3$ massive-quenched galaxies happens at $z=4-5$ (e.g. \citealt{Valentino20,Manning22}).

Visual inspection of these SFR histories show that \shark\ and {\sc Galform} are the models producing the highest SFRs at $z\ge 10$ with values as high as $\approx 50\,\rm M_{\odot}\,yr^{-1}$ (consistent with what has been found by the {\it JWST} in \citealt{Bunker23,Castellano23,Carniani24}), while all the other simulations predict SFRs $<10\,\rm M_{\odot}\,yr^{-1}$ at $z\gtrsim 10$. The number density of these progenitors at $z=10$ with $\rm SFR>50\,\rm M_{\odot}\,yr^{-1}$ in \shark\ and {\sc Galform} is $2.8\times 10^{-7}\,\rm cMpc^{-3}$ and $7\times 10^{-8}\,\rm cMpc^{-3}$, respectively. 
Note that for {\sc Galform} the highest redshift at which we sample the SFHs is $z=10.07$ due to higher redshifts being unavailable in this database. However, from the $z=10.07$ output it is already clear that the progenitors of the $z=3$ massive-quenched galaxies in {\sc Galform} can have very high SFRs. 

Part of the difference between \shark\ and {\sc Galform} and the other simulations is the overall larger volumes simulated (although the same large volume is simulated by {\sc GAEA}), but that is not the whole story. For example, passive galaxies that at $z=3$ have stellar masses $\approx 10^{10.5}\,\rm M_{\odot}$ in \shark\ and {\sc Galform} have a median SFR at $z=10$ of $\approx 6.1\,\rm M_{\odot}\,yr^{-1}$ and $\approx 4.1\,\rm M_{\odot}\,yr^{-1}$, respectively; while in {\sc GAEA} and {\sc Eagle} (the two simulations having massive-quenched galaxies of those masses at $z=3$) the medians are $\approx 0.9\,\rm M_{\odot}\,yr^{-1}$ and $\approx 1.3\,\rm M_{\odot}\,yr^{-1}$, respectively. The reason for this difference likely resides in the higher star formation efficiency (i.e. the conversion from molecular gas to SFR surface density) \shark\ and {\sc Galform} assume for star formation that is driven by galaxy mergers or disc instabilities. In the case of \shark, this is $15$ times more efficient at converting molecular gas into stars compared to star formation happening in galaxy discs, while in {\sc Galform} this merger or disc instability-driven star formation has an efficiency that is inversely proportional to the bulge dynamical timescale, which tends to give much higher gas-to-stars conversion efficiencies than the disc star formation. {\sc Eagle} assumes instead a fixed efficiency per unit gas density, while {\sc GAEA} has very little star formation activity associated with the starburst mode (i.e. driven by galaxy mergers in their case) and hence in the practice is assuming a close to universal 
molecular-to-SFR surface density conversion efficiency. 

{\sc IllustrisTNG} and {\sc Simba} produce preferentially very massive-quenched galaxies, with $z=3$ stellar masses $\gtrsim 10^{11}\,\rm M_{\odot}$. These galaxies, however, only have modest SFRs with a median at $z=7$ of $\approx 13\rm \,M_{\odot}\,yr^{-1}$ in {\sc IllustrisTNG} and $\approx 7.5\rm \,M_{\odot}\,yr^{-1}$ in {\sc Simba}. Similarly massive galaxies in \shark\ and {\sc Galform} have 
 a median SFR at $z=7$ of $33\rm \,M_{\odot}\,yr^{-1}$ and $30\rm \,M_{\odot}\,yr^{-1}$, respectively; while {\sc GAEA} is similar to {\sc IllustrisTNG} with a median of $12\,\rm \,M_{\odot}\,yr^{-1}$, respectively. {\sc Eagle} does not produce such massive galaxies in its volume.

\begin{figure*}
\begin{center}
\includegraphics[trim=4.5mm 4.5mm 2mm 2mm, clip,width=0.99\textwidth]{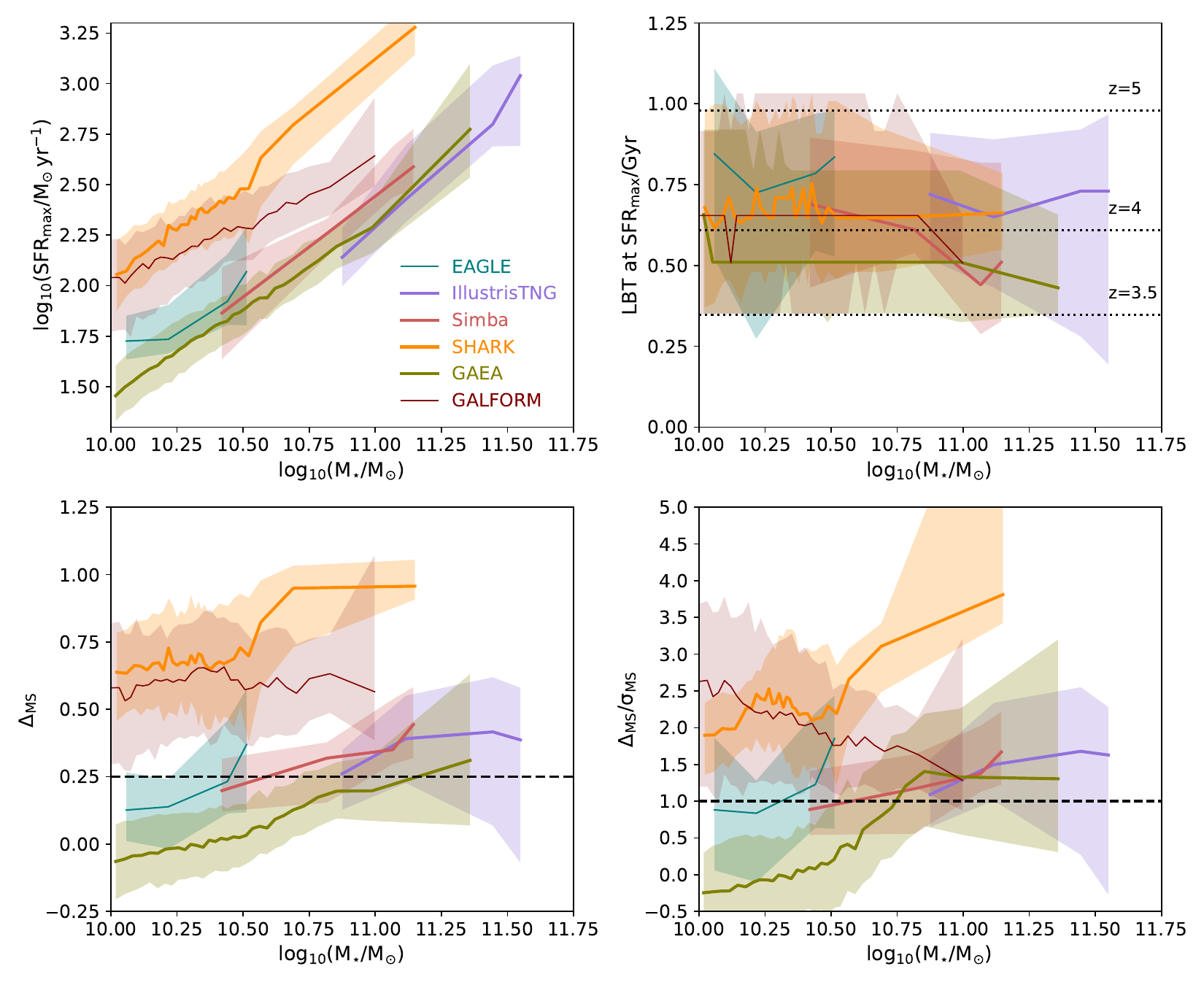}
\caption{{\it Top-left panel:} Maximum SFR achieved by the $z=3$ passive, massive galaxies (at any point of their history) as a function of the $z=3$ stellar mass in each of the $6$ simulations, as labelled. Lines with shaded regions show the medians and the $16^{\rm th}-84^{\rm th}$ percentile ranges, respectively. We use adaptive bins to ensure a reasonable number of galaxies per bin. 
{\it Top-right panel:} As in the top-left panel but for the lookback time at which SFR$_{\rm max}$ happens. Here we show with horizontal lines three different redshifts, as labelled (noting that the cosmologies of the simulations are not exactly the same, but similar enough to use these lines as guidance). {\it Bottom-left panel:} Distance to the main sequence (i.e. $\rm log_{10}(sSFR/sSFR_{\rm MS})$; see \S~\ref{definitions}), defined at the point of the maximum SFR of the $z=3$ massive-quenched galaxies as a function of the $z=3$ stellar mass. The dashed line shows the point at which galaxies are at $0.25$~dex above the main sequence. {\it Bottom-right panel:} As in the bottom-left panel but for the distance to main sequence in units of the width of the main sequence (measured as the standard deviation of the SFR-$M_{\star}$ relation at the point of the maximum SFR; see dotted lines in Fig.~\ref{mainseq}). The dashed line shows the point at which galaxies are $1\sigma$ above the main sequence.} 
\label{sfr_scaling}
\end{center}
\end{figure*}
An interesting difference between simulations is when do $z=3$ massive-quenched galaxies appear for the first time in the simulations. In \shark, {\sc Eagle}, and {\sc IllustrisTNG}, the vast majority (if not all) the massive-quenched galaxies appear for the first time at $z>10$, while in {\sc Simba}, {\sc GAEA} and {\sc Galform} there is a large fraction that start forming at $z<10$. {\sc Galform}, {\sc GAEA} and {\sc Simba} predict that $89$, $33$ and $44$ per cent of the $z=3$ massive-quenched galaxies appear for the first time in the outputs at $z<10$. In \shark, this happens $1.6$ per cent of the time, while in {\sc Eagle} and {\sc IllustrisTNG} it does not happen.

\begin{figure*}
\begin{center}
\includegraphics[trim=38mm 1.5mm 42.0mm 8mm, clip,width=0.99\textwidth]{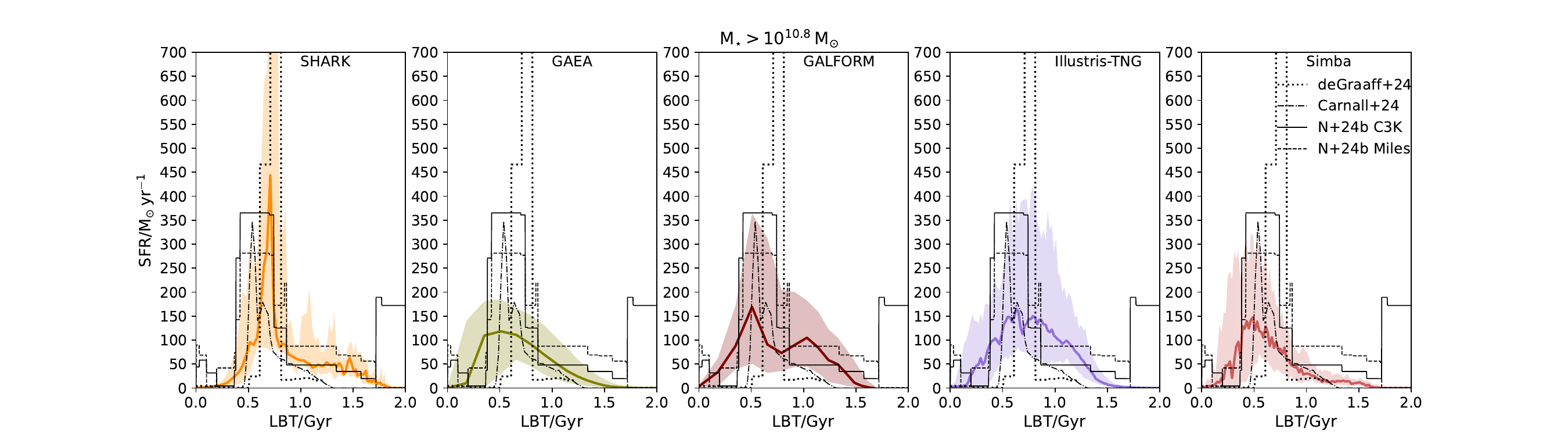}
\caption{The median and $16^{\rm th}-84^{\rm th}$ percentile range of the SFHs of $z=3$ massive-quenched galaxies are shown with solid lines and shaded regions, respectively, in each simulation, as labelled in each panel. This is shown for massive-quenched galaxies at $z=3$ with $M_{\star}> 10^{10.8}\,\rm M_{\odot}$. {\sc Eagle} has no galaxies of that mass so it is not shown. We also show the derived SFH of the galaxy presented in \citet{deGraaff24} and the medians of the $6$ galaxies from \citet{Nanayakkara24} (referred to as N+24b in the figure), and $4$ galaxies from \citet{Carnall24}, as labelled, that satisfy the same stellar mass cut. Note that for the sample in \citet{Nanayakkara24} we show the inferred SFHs using two different stellar population synthesis models (Miles and C3K), to exemplify systematic uncertainties.}
\label{SFHsAll}
\end{center}
\end{figure*}

\citet{Valentino20,Manning22} suggest that potential progenitors of $z=3-4$ massive-quenched galaxies are starburst galaxies at $z=4-5$ with SFRs from a few $\approx 100\,\rm M_{\odot}\,yr^{1-}$ to $\gtrsim 1000\,\rm M_{\odot}\,yr^{-1}$, based on the latter having a similar number density to the former. Similarly high SFRs are inferred from the spectral fitting of $z=3-4.5$ massive-quenched galaxies at their peak (e.g. \citealt{Valentino20,Glazebrook23,Carnall24,deGraaff24}). Such high SFRs are regularly seen in the progenitors of $z=3$ massive-quenched galaxies in \shark\ and {\sc IllustrisTNG}, but are rare or non-existent in the other simulations (see also the top-left panel of Fig.~\ref{sfr_scaling}). Hence, if the evolutionary link between highly star-forming galaxies at $z=4-5$ and $z=3$ passive galaxies is confirmed, it would pose a significant challenge to some of the simulations analysed here.

\subsubsection{Tracking the most intense star formation period of $z=3$ massive-quenched galaxies}

Fig.~\ref{sfr_scaling} shows the relation between $\rm SFR_{\rm max}$ and the $z=3$ stellar mass of the massive-quenched galaxies. We see that at fixed stellar mass, \shark\ is the simulation producing the highest SFR peak, followed by {\sc Galform}; while {\sc GAEA} and {\sc IllustrisTNG} produce the lowest SFR peaks. Even though the poorer time cadence of {\sc GAEA} and {\sc Galform} plays a role in the differences seen (see Appendix~\ref{maxSFReff} for details), this is not the whole story. In fact,  
 {\sc GAEA} and {\sc Galform} have a similar time cadence, but the latter produces much higher $\rm SFR_{\rm max}$ than the former. 
Overall, {\sc Eagle}, {\sc Simba}, {\sc GAEA} and {\sc IllustrisTNG} produce a similar $\rm SFR_{\rm max}$-stellar mass relation. The bottom panels of Fig.~\ref{sfr_scaling} show the median and scatter of $\Delta_{\rm MS}$ and $\Delta_{\rm MS}/\sigma_{\rm MS}$ at the time of $\rm SFR_{\rm max}$ in each simulation. We see that $\rm SFR_{\rm max}$ corresponds to starburst episodes in \shark, {\sc Galform}, {\rm Simba} and {\sc IllustrisTNG} (i.e. show a high deviation from the main sequence, with the main sequence defined internally to each simulation). In {\sc Eagle} and {\sc GAEA} only the most massive galaxies, $M_{\star}\gtrsim 10^{10.35}\,\rm M_{\odot}$ and $M_{\star}\gtrsim 10^{11}\,\rm M_{\odot}$, respectively, appear to be starbursting when they reach $\rm SFR_{\rm max}$.

The top-right panel of  Fig.~\ref{sfr_scaling} shows the lookback time to $\rm SFR_{\rm max}$. The figure shows that in \shark, {\sc Eagle}, {\sc Simba}, {\sc IllustrisTNG} and {\sc Galform} $\rm SFR_{\rm max}$ happens preferentially between $z\approx 4-5$ (highlighted by the horizontal lines), while in {\sc GAEA} $\rm SFR_{\rm max}$ happens at slightly lower redshifts, $z\approx 3.5-4$. Thus, most of the simulations predict the progenitors of the $z=3$ massive-quenched galaxies being highly star-forming at $z\approx 4-5$ in qualitative agreement with the conclusions of \citet{Valentino20,Manning22}. However, it is clear that \shark\ and {\sc Galform}  produce much more intense starbursts than the other simulations. Note that this does not mean that most of the growth of these massive-quenched galaxies happens in the starburst mode (i.e. above the main sequence), but simply that at their peak SFR, they are above the main sequence. Later, we show in Fig.~\ref{mainseq} that in fact on average galaxies grow along the main sequence in the SFR-$M_{\star}$ plane.

\subsubsection{Comparing the predicted SFHs with observations}

Fig.~\ref{SFHsAll} shows the median SFHs of massive-quenched galaxies in all the simulations that have stellar masses $>10^{10.8}\,\rm M_{\odot}$ and compares with observationally-derived SFHs. This stellar mass cut is chosen to match the range of most of the available {\it JWST} spectroscopic observations.
We do not show {\sc Eagle} here as there are no galaxies in the relevant stellar mass range. 

\shark\ produces the shortest and more pronounced {star formation episode} of all the simulations. {\sc Simba} and {\sc GAEA} produce the latest forming galaxies, with a peak at a lookback time of $\approx 0.45$~Gyr (compared with peaks at $\approx 0.75$~Gyr for \shark\ and {\sc IllustrisTNG}). {\sc Galform} produces two periods of intense star formation for these galaxies but with the latest one being more pronounced. The figure also shows observationally-derived SFHs from {\it JWST} spectroscopy of massive-quenched galaxies of the same stellar mass range as the simulation samples from \citet{deGraaff24}, \citet{Carnall24} and \citet{Nanayakkara24}. The data for \citet{Carnall24} and Nanayakkara et al. (in prep) correspond to the median of $4$ and $6$ massive-quenched galaxies selected at $z\approx 3-4$, respectively, while the \citet{deGraaff24} result represents a single galaxy at $z\approx 4.9$. 
Note that we do not include in the comparison of Fig.~\ref{SFHsAll} the galaxy ZF-UDS-7329 \citep{Glazebrook23}, as it appears to be an outlier of the 
bulk of the massive-quenched galaxies with a much earlier formation time.

Of all the simulations, 
\shark\ appears to produce the highest peak that resembles better the observationally-derived SFHs. {\sc IllustrisTNG}, {\sc Galform} and {\sc Simba} produce a lower median SFH than observations, but their $84^{\rm th}$ percentile distribution is largely consistent with observations. {\sc GAEA} on the other hand seems to produce SFHs that are too extended and do not reach the high SFRs derived from spectroscopy.
In any case, the current inferences suffer from large systematic uncertainties. \citet{deGraaff24} show that the peak of the SFH moves significantly if different metallicities are preferred, while \citet{Nanayakkara24} show that the adopted single population synthesis model impacts the recovered SFH, as shown in Fig.~\ref{SFHsAll}. 
Increasing the sample sizes in the observations is crucial to start
using the inferred SFHs as stringent constraints on the simulations. This will be possible in the near future.  

\subsubsection{Single or multiple starburst periods in the SFHs of $z=3$ massive-quenched galaxies}\label{multiSBs}

A challenge in observations has been to determine the best way to fit the SFH of massive-quenched galaxies; e.g. whether to use non-parametric SFHs or adopt some function to describe the shape of the SFH. We attempt to shed light into this problem by investigating whether the predicted SFHs can be easily fit with a simple function that displays a unique SFR peak, or whether they have complex enough SFHs that a simple function cannot describe them. In order to determine this, we fit a skewed Gaussian function to every galaxy's SFH using {\sc Highlander}\footnote{\url{https://github.com/asgr/Highlander}}, which is the preferred SFH function used to fit galaxies with {\sc ProSpect} \citep{Robotham20}. In Appendix~\ref{SFHfits} we provide details of the function, the fits and show examples of the individual SFHs and their fits. 
We calculate a goodness of fit based on the normalised model to data difference, and use those values to select galaxies that are candidates for having clear multiple episodes of starburst activity (which we define as clear unrelated peaks in the SFH). 

We then visually inspect the galaxies that were selected as candidates for multi-starbursts and find that a simple threshold in the goodness of fit works very well in {\sc IllustrisTNG}, {\sc Simba} and {\sc GAEA} to isolate galaxies with multiple starbursts. However, in \shark, {\sc Eagle}, and {\sc Galform} galaxies with short-duration starbursts are also poorly fit, so a post-processing step is applied to reclassify those as being uni-modal. 

Table~\ref{SBsFrac} shows the fraction of galaxies in each simulation displaying clear multiple starburst episodes. The fractions change dramatically between simulations. {\sc Eagle} and {\sc GAEA} have the smallest fractions, $\approx 5$ per cent.  Conversely,  \shark, {\sc Galform} and {\sc IllustrisTNG} have about half of the $z=3$ massive-quenched galaxies displaying multiple starbursts. {\sc Simba} is in between these extremes.  
Studying the stellar mass of the galaxies with multiple starbursts, we do not find any specific trend; the well-fitting vs the badly-fitting galaxies having similar masses. The only exception is {\sc Simba}, in which the best-fitting SFHs tend to be the most massive galaxies. 

The results here are not applicable to lower redshifts. In fact, using \shark, \citet{Bravo22} show that a skewed Gaussian function provides a good fit to the SFHs of $z=0$ galaxies, as most of the burstiness experienced by galaxies early on becomes irrelevant after several Gyr of evolution. A similar conclusion was reached by \citet{Diemer17} using the {\sc Illustris} simulation. 

\begin{table}
        \setlength\tabcolsep{2pt}
        \centering\footnotesize
        \caption{Fraction of $z=3$ massive-quenched galaxies exhibiting clear multiple episodes of starbursts in each simulation 
        (see Appendix~\ref{SFHfits} for details and example SFHs).}
        \begin{tabular}{@{\extracolsep{\fill}}l|cc|p{0.45\textwidth}}
                \hline
                \hline
            Simulation & $f_{\rm multiple}$\\
           \hline
   \shark\ & $0.63$\\
   {\sc GAEA} & $0.05$\\
   {\sc Galform} & $0.56$\\
   {\sc Eagle} & $0.06$\\
   {\sc IllustrisTNG} & $0.44$\\
   {\sc Simba} & $0.13$\\
   \hline
       \end{tabular}
       \label{SBsFrac}
\end{table}

The large difference between simulated SFHs warrant the exploration of an array of SFH assumptions (parametric and non-parametric) when attempting to derive these from observations.

\subsection{The quenching of massive galaxies}\label{quenchingmass}

\subsubsection{Formation timescales of $z=3$ massive-quenched galaxies}

Going back to Fig.~\ref{SFH}, we see that the quenching of the $z=3$ massive-quenched galaxies in Fig.~\ref{SFH} also happens differently in the different simulations. {\sc IllustrisTNG}'s massive-quenched galaxies display a sharp decline in their SFR, close to the target redshift. This is seen in {\sc Simba} and {\sc GAEA} as well but only for the most massive galaxies, $M_{\star}>10^{11.1}\,\rm M_{\odot}$. The decline in SFR of $z=3$ massive-quenched galaxies in \shark\ and {\sc Eagle} appears to be slower. 

\begin{figure}
\begin{center}
\includegraphics[trim=4.5mm 5mm 2mm 2mm, clip,width=0.49\textwidth]{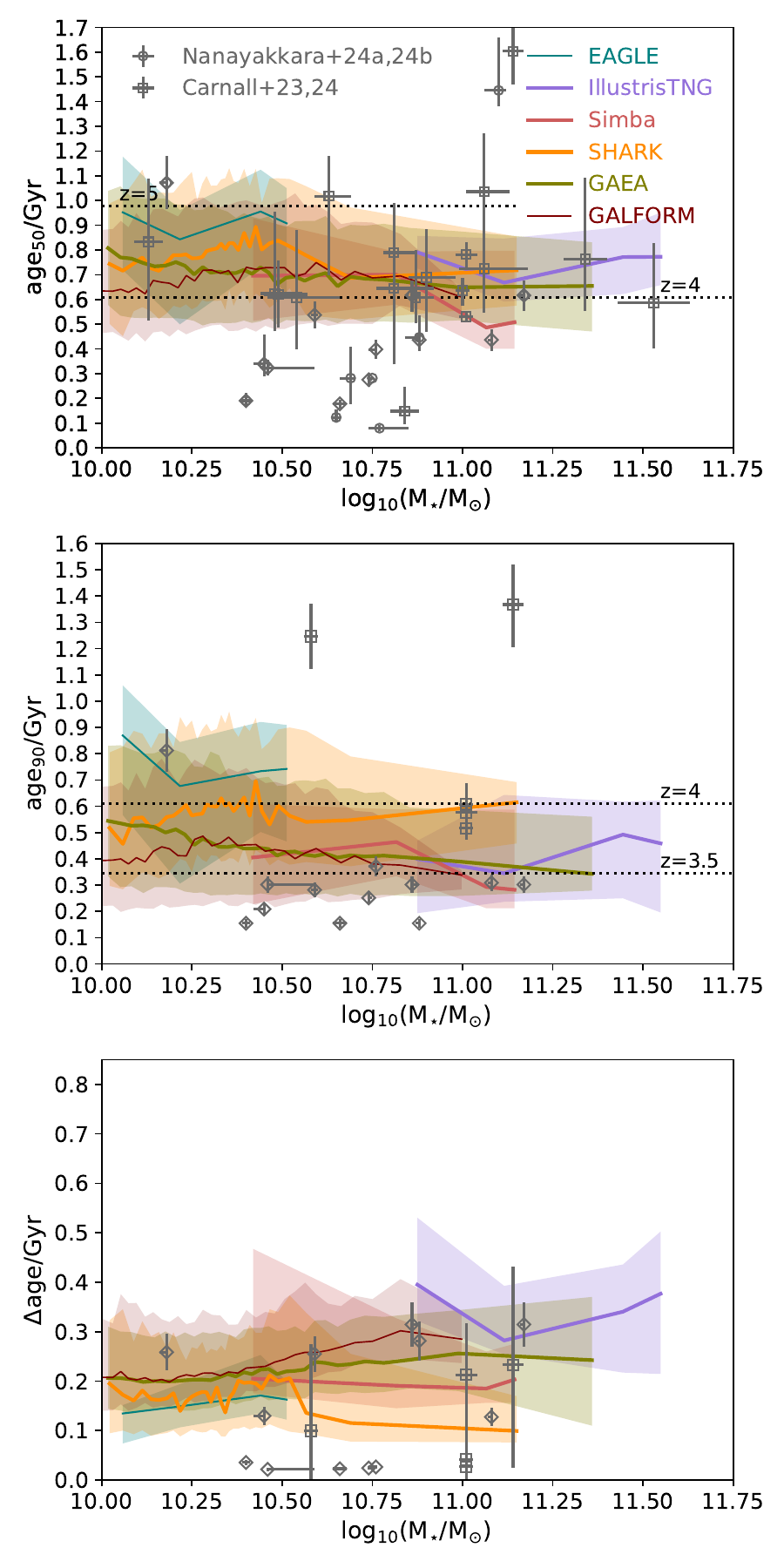}
\caption{{\it Top panel:} Lookback time to when the $z=3$ massive-quenched galaxies had formed $50$ per cent of their stellar mass as a function of the $z=3$ stellar mass, for all the simulations, as labelled. Lines with shaded regions show the medians and the $16^{\rm th}-84^{\rm th}$ percentile ranges, respectively. We use adaptive bins to ensure a number of galaxies per bin $\ge 10$. For reference, the horizontal dotted lines show $z=4$ and $z=5$. Here we also show observationally-derived measurements of $\rm age_{\rm 50}$ from \citet{Nanayakkara22,Nanayakkara24} and \citet{Carnall23,Carnall24} as open symbols, as labelled. 
{\it Middle panel:} As in the top panel but for when the galaxies had formed $90$per cent of their stellar mass. The horizontal dotted lines show $z=3.5$ and $z=4$. Here we show observations from \citet{Nanayakkara24} and \citet{Carnall24} (which is the subsample of the top panel that is spectroscopically confirmed). Although we have $\chi^2$  errors associated with the inferred SFHs for the observations, those represent lower limits, so we apply a $10$per cent error to all measurements, acknowledging that the real uncertainties are likely larger.
{\it Bottom panel:} As in the middle panel but for the time it took for the galaxies to go from $50$ to $90$ per cent of their $z=3$ stellar mass.} 
\label{age_scaling}
\end{center}
\end{figure}
To quantify this better, we show in the top and middle panels of Fig.~\ref{age_scaling} the relationship between $\rm age_{\rm 50}$ and $\rm age_{\rm 90}$, respectively, with stellar mass (see \S~\ref{definitions} for the definitions of these properties). {\sc Eagle} tends to produce the oldest massive-quenched galaxies, while those in {\sc Simba} and {\sc Galform} are the youngest (though some dependence on stellar mass emerges). The latter is especially clear with $\rm age_{\rm 90}$. {\sc IllustrisTNG} and {\sc GAEA} predicts ages that are similar or slightly older than {\sc Simba} and {\sc Galform}, while \shark\ predicts ages that are between those of {\sc Eagle} and the rest of the simulations. 
Massive-quenched galaxies in {\sc Eagle} have formed $90$per cent of their stars by $z\approx 4-4.5$, making them very old compared to the other simulations. Note that none of the simulations predict a well established age-mass relation within this population. 

The top panel of Fig.~\ref{age_scaling} shows observationally-derived $\rm age_{\rm 50}$ from \citet{Carnall23,Carnall24} and \citet{Nanayakkara22}. For \citet{Carnall23} we find that adding or not the non-robust quenched galaxies makes little difference to the median and standard deviation presented with the filled diamond. In the middle and bottom panels we show observations from \citet{Nanayakkara24} and \citet{Carnall24}. The latter are the sub-sample of galaxies in the top panel that are spectroscopically confirmed.  
The available data is currently too scarce leading to a large scatter, which makes it consistent with all the simulations within the errorbars. Increasing the sample sizes may help disentangle the predictions presented here.

\begin{figure*}
\begin{center}
\includegraphics[trim=4.5mm 4.5mm 2mm 4mm, clip,width=0.49\textwidth]{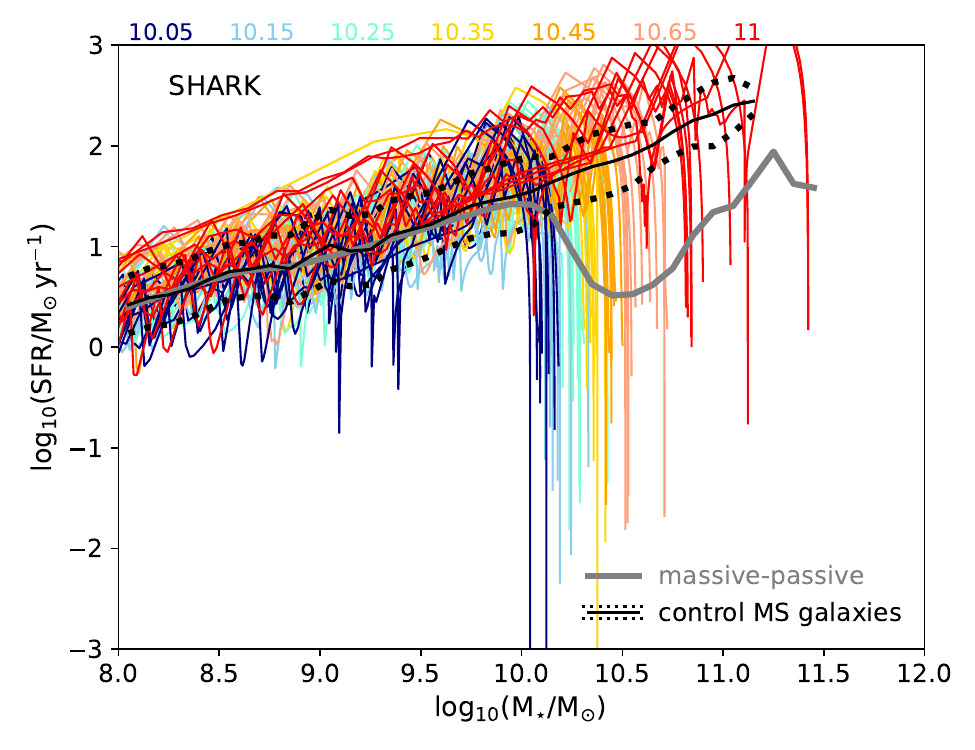}
\includegraphics[trim=4.5mm 4.5mm 2mm 4mm, clip,width=0.49\textwidth]{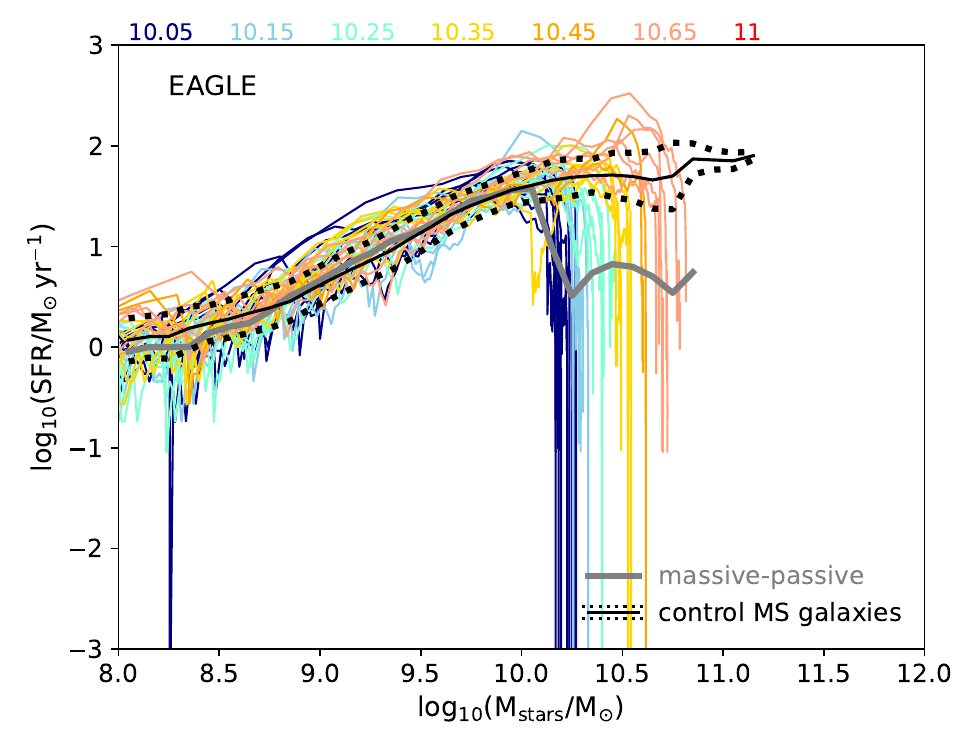}
\includegraphics[trim=4.5mm 4.5mm 2mm 4mm, clip,width=0.49\textwidth]{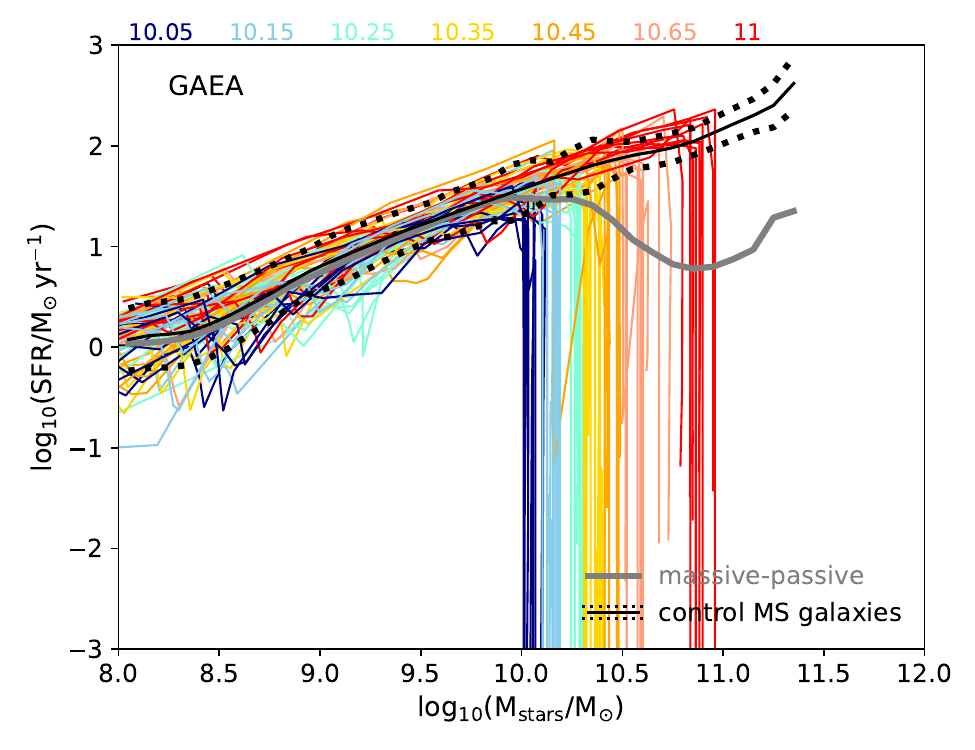}
\includegraphics[trim=4.5mm 4.5mm 2mm 4mm, clip,width=0.49\textwidth]{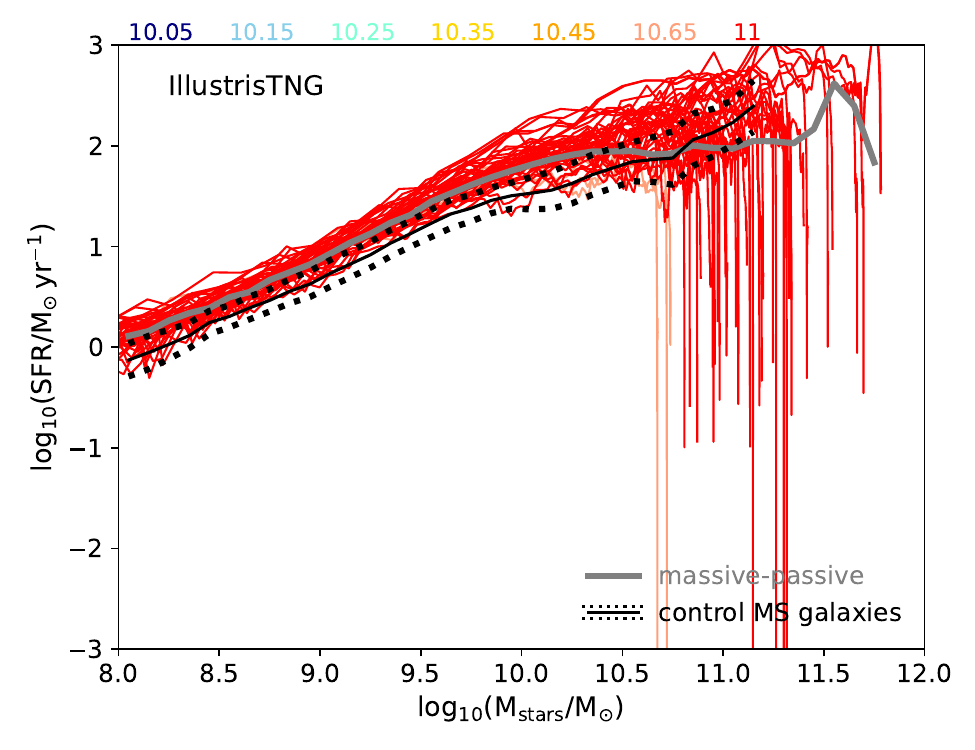}
\includegraphics[trim=4.5mm 4.5mm 2mm 4mm, clip,width=0.49\textwidth]{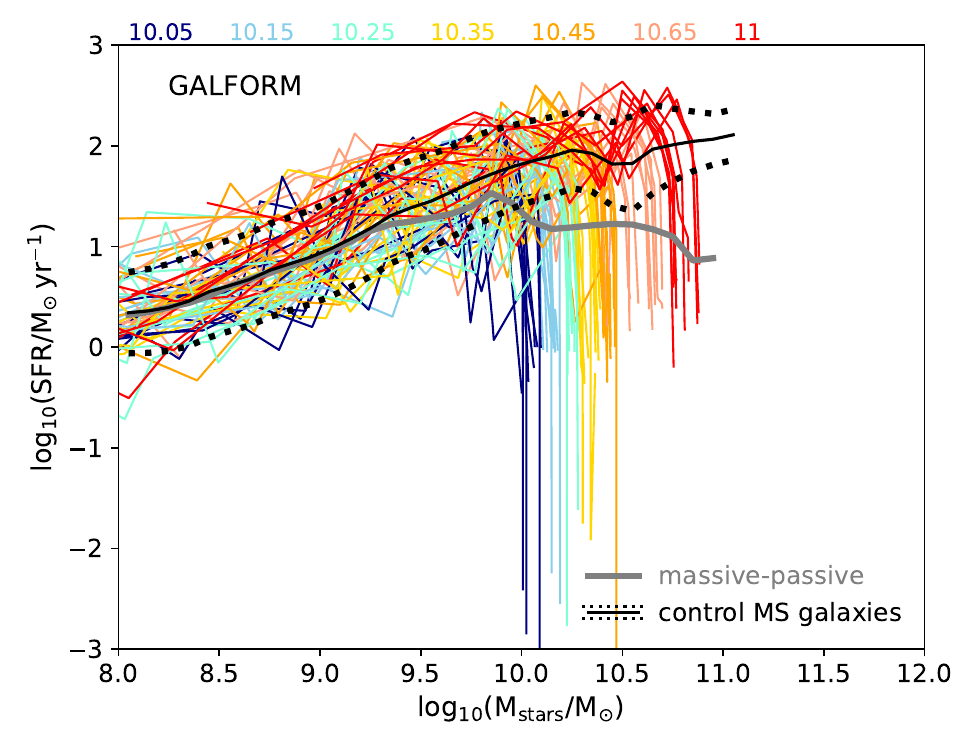}
\includegraphics[trim=4.5mm 4.5mm 2mm 4mm, clip,width=0.49\textwidth]{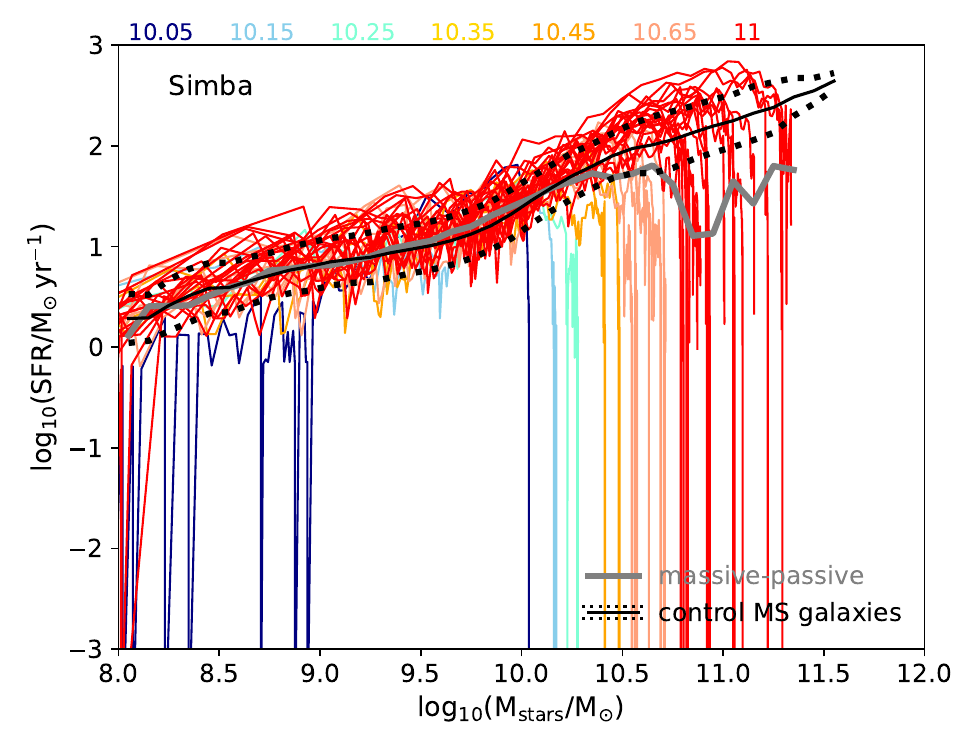}
\caption{The evolution of the $z=3$ massive-quenched galaxies shown in Fig.~\ref{SFH} in the SFR-$M_{\star}$ plane in each simulation, as labelled. 
Lines are coloured depending on their $z=3$ stellar mass using the same bins of Fig.~\ref{SFH}.
 The median SFR in bins of stellar mass of all the $z=3$ massive-quenched galaxies is shown with a thick-grey line in each panel. For comparison, in each panel we also plot the median and one standard deviation of the control MS sample (defined in \S~\ref{definitions}) with solid and dotted black lines, respectively.} 
\label{mainseq}
\end{center}
\end{figure*}

Several recent papers have used the difference between e.g. $\Delta\, \rm  age = \rm age_{\rm 90}-age_{\rm 50}$ to quantify a formation timescale (e.g. \citealt{deGraaff24}). We show this in the bottom panel of Fig.~\ref{age_scaling}. From this difference we see that {\sc IllustrisTNG} galaxies are the ones that form the slowest, while {\sc Eagle} and \shark\ galaxies form the fastest. {\sc Simba}, {\sc GAEA} and {\sc Galform} produce $\Delta\,{\rm age}$ that are in between the other simulations. The difference is quite large, with {\sc IllustrisTNG} taking a factor of $\approx 3$ ($\approx 1.7$) times longer to form compared with \shark\ ({\sc Simba}) galaxies of the same stellar mass. Note that the {\sc IllustrisTNG}'s timescales are similar to those reported by \citet{Weller24} for the same simulation using slightly different definitions.  \citet{Kimmig23} found that in {\sc Magneticum} (a cosmological hydrodynamical simulation), massive-quenched galaxies have $\Delta\, \rm  age\lesssim 170$~Myr, similar to what we find in \shark\ and {\sc Eagle}. In observations, inferred $\Delta\, \rm  age$ are mostly between $100-300$~Myr (e.g. see symbols in the figure), closer to what \shark, {\sc Eagle}, {\sc Simba}, {\sc Galform} and {\sc GAEA} predict, but the scatter is again too large to reach a robust conclusion.

\subsubsection{The growth of galaxies in the SFR-stellar mass plane}

To understand what drives the diversity of ages and $\Delta\, \rm  age$, we turn our attention to the way galaxies grow in the SFR-$M_{\star}$ plane in Fig.~\ref{mainseq}. 
Most galaxies in \shark,  {\sc Eagle} and {\sc Simba} show periods in their SFH where galaxies are clearly above main sequence (by more than $\gtrsim 0.3$~dex), in the starburst region. When not in the starburst region, galaxies grow along the main sequence. The latter is seen by the similarity between the median SFR-$M_{\star}$ relation of the massive-quenched and control MS samples before quenching happens in the former sample. In {\sc Galform} clear starbursts episodes are seen when galaxies are still relatively low mass $M_{\star}\lesssim 10^{9.5}\,\rm M_{\odot}$, while at higher mass galaxies mostly grow along the main sequence. 

Galaxies in {\sc IllustrisTNG} grow along a very tight SFR-$M_{\star}$ sequence that is slightly elevated compared with the main sequence, but showing little deviations to higher SFRs. In {\sc GAEA} we see that massive-quenched galaxies grow along the main sequence for most of their lives until quenching. In the latter, quenching is very sudden and generally happens in a single simulation snapshot. The fact that galaxies in {\sc GAEA} and {\sc IllustrisTNG} show almost no starburst episodes and that in {\sc Galform} they show clear starbursts only very early on their formation history, is why they take the longest to form (see bottom panel in Fig.~\ref{age_scaling}). Despite these differences, we find that in all simulations galaxies spend most of their time on the main sequence (between $55-99$per cent of their active lives, with the extremes of the range corresponding to {\sc Galform} and {\sc GAEA}  respectively). The large difference between the fraction of time galaxies spend on the main sequence between {\sc Galform} and {\sc GAEA} is due to the inclusion of disc instability-driven starbursts in the former and the lack of thereof in the latter. This is clear from studying the contribution of disc instabilities to the SFRs of galaxies across cosmic time in {\sc Galform} (see discussion in \S~5.3 in \citealt{Lacey15}). \shark\ also includes disc instability-driven starbursts and predicts galaxies spend $73$per cent of their lives on the main sequence, again much less than what is found in {\sc GAEA}. Despite spending most of their lives on the main sequence, at their peak SFR, galaxies in most simulations are above the main sequence as shown in Fig.~\ref{sfr_scaling}.

Fig.~\ref{mainseq} also shows that in \shark, {\sc Eagle}, {\sc Simba}, {\sc Galform} and {\sc GAEA} there is a small fraction of $z=3$ massive-quenched galaxies that have had periods of passiveness in the past. These periods of passiveness are perhaps reminiscent of what has been called `mini-quenching' in the literature \citep{Looser23}, albeit at slightly higher stellar masses, where galaxies are likely to rejuvenate after such passive periods. The galaxies here, indeed go back to the main sequence or even the starburst region after those `mini-quenching' periods. Such periods are seen in the SFHs of some $z=3$ massive-quenched galaxies in all the simulations, except for {\sc IllustrisTNG}. The latter is related to lack of massive-quenched galaxies in  {\sc IllustrisTNG} at $z\gtrsim 3.5$ shown in Fig.~\ref{num_densities}. We will go back to this in \S~\ref{AGN}. 

Another interesting difference between the simulations is the tightness of the main sequence, with {\sc Galform} predicting the least tight sequence (with a typical standard deviation of $\approx 0.45$~dex) and {\sc IllustrisTNG}, the tightest one (standard deviation of $\approx 0.14$~dex). \shark, {\sc Eagle}, {\sc Simba} and {\sc GAEA} predict a main sequence standard deviation of $\approx 0.3$, $\approx 0.2$, $\approx 0.25$, and $\approx 0.2$~dex, respectively. 
Spectroscopic observations with the {\it JWST} are starting to place constraints on the intrinsic scatter of the main sequence. \citet{Clarke24} found values of $0.4-0.5$~dex at $z\gtrsim 3$ for galaxies with stellar masses $\lesssim 10^{10}\,\rm M_{\odot}$, similar to what {\sc Galform} predicts and larger than the other simulations' predictions. One caveat is that samples are still very small (\citealt{Clarke24} employed 104 galaxies to measure the main sequence's scatter in $5$ redshift bins), and hence bigger samples are required to get more robust measurements (see \citealt{D'Silva23}).

\begin{figure}
\begin{center}
\includegraphics[trim=4.5mm 4.5mm 2mm 2mm, clip,width=0.49\textwidth]{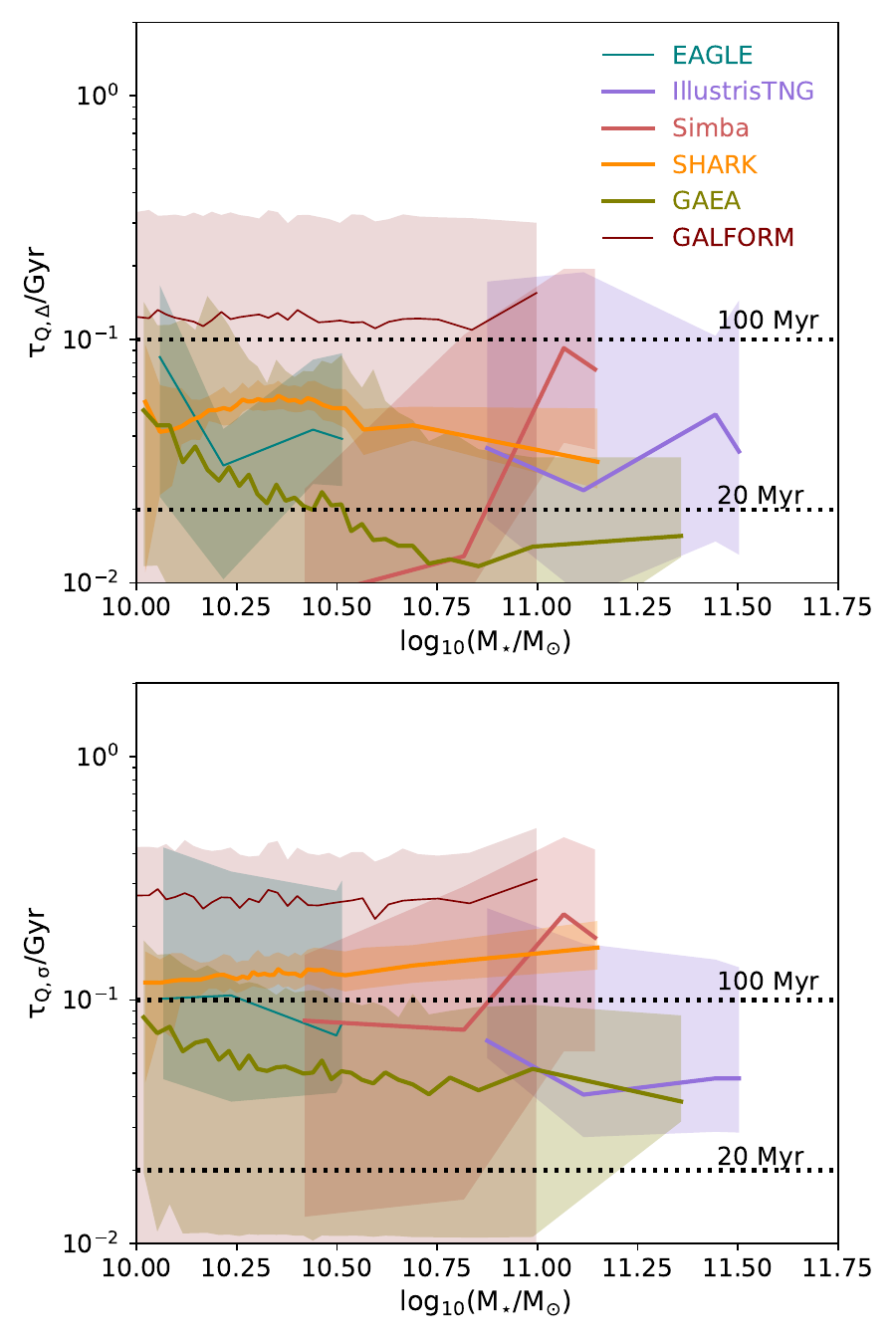}
\caption{{\it Top panel:} Quenching timescale, $\tau_{\rm Q,\Delta}$ (defined in terms of $\Delta_{\rm MS}$; see \S~\ref{definitions}) as a function of the $z=3$ stellar mass of massive-quenched galaxies in each of the $6$ simulations, as labelled. Lines with shaded regions are as in Fig.~\ref{sfr_scaling}. {\it Bottom panel:} As in the top panel but for $\tau_{\rm Q,\sigma}$ (defined in terms of $\Delta_{\rm MS}/\sigma_{\rm MS}$; see \S~\ref{definitions}).} 
\label{tauQ}
\end{center}
\end{figure}
\subsubsection{Quenching timescales of $z=3$ massive-quenched galaxies}

Finally, we study in Fig.~\ref{tauQ} the quenching timescales of the massive-quenched galaxies in each simulation. We use the two definitions introduced in \S~\ref{definitions}. When we measure the quenching timescale in terms of distance to the main sequence only ($\tau_{\rm Q,\Delta}$), we see that the quenching of the most massive-quenched galaxies happens very fast in all the simulations ($\tau_{\rm Q,\Delta}\lesssim 100$~Myr), except in {\sc Galform} which has a median of $\tau_{\rm Q,\Delta}\approx 150$~Myr, and a significant population that scatters up to $\tau_{\rm Q,\Delta}\approx 300$~Myr. 

However, when we take into consideration the tightness of the main sequence in each simulation and measure quenching in terms of $\sigma$ deviations from the main sequence (bottom panel of Fig.~\ref{tauQ}), we see that galaxies in all simulations tend to take longer: in \shark, \eagle\ and {\sc Simba} galaxies quench in $\approx 100-200$~Myr, while {\sc Galform}'s galaxies tend to take on average $\approx 300$~Myr to quench. The quenching timescales in {\sc IllustrisTNG} and {\sc GAEA} continue to be below $100$~Myr. We note, however, that such quenching timescales are much smaller than the time step of MILL, employed in {\sc GAEA}, and hence we can only assert that they are smaller or similar to $100$~Myr. 

The interpretation of quenching timescales one can draw from Fig.~\ref{tauQ} and the bottom panel of Fig.~\ref{age_scaling} can be very different in some models (see for example the change in {\sc Simba} between the top and bottom panels of Fig.~\ref{tauQ}), which emphasises the need for a convergent definition of quenching timescale in the literature.

\subsection{How AGN feedback shapes the SFHs of early massive-quenched galaxies}\label{AGN}

Below we connect the results of previous sections with how AGN feedback is implemented in each simulation. 

\begin{figure}
\begin{center}
\includegraphics[trim=5mm 4.5mm 2mm 2mm, clip,width=0.49\textwidth]{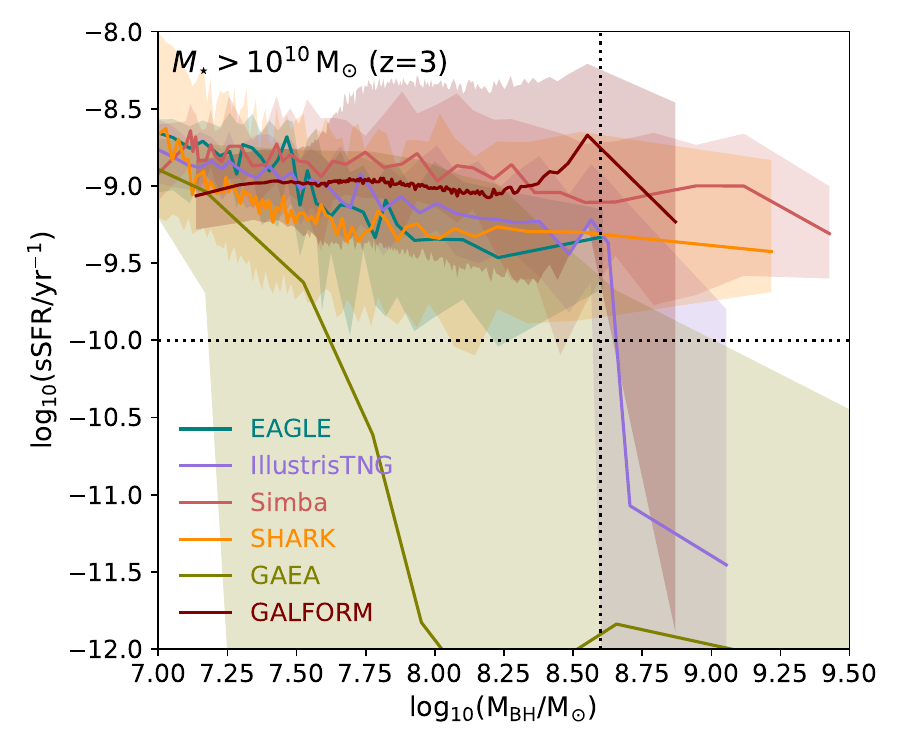}
\caption{The sSFR as a function of the central BH mass for all galaxies at $z=3$ that have a stellar mass $>10^{10}\,\rm M_{\odot}$, in all the simulations, as  labelled. Lines with shaded regions show the medians and $16^{\rm th}-84^{\rm th}$ percentile ranges, respectively. We use
adaptive bins to ensure a number of galaxies per bin $\ge  10$. The horizontal dotted line shows our passive sSFR criterion, while the vertical dotted line shows the approximate BH mass at which galaxies in {\sc IllustrisTNG} go from being star-forming to passive.} 
\label{BHSSFRz3}
\end{center}
\end{figure}

In {\sc IllustrisTNG}, AGN feedback in the radiatively-efficient mode has very little impact on the capability of galaxies to be star-forming \citep{Kurinchi-Vendhan23}. Hence, galaxies appear to grow along a tight sequence in the SFR-$M_{\star}$ plane until a milestone is reached and they are quickly quenched, in $<100\,\rm Myr$. At $z=0$, that milestone is the BH reaching a mass of $10^{8.2}\,\rm M_{\odot}$, when kinetic feedback kicks in \citep{Terrazas20}.
The BH transition mass that quenches galaxies is clearly seen in Fig.~\ref{BHSSFRz3}, where we see that {\sc IllustrisTNG} predicts a clear transition  around the mass highlighted with a vertical line; galaxies below that mass are star-forming, and above are passive. The transition is extremely sharp in {\sc IllustrisTNG}. This happens at around $10^{8.6}\,\rm M_{\odot}$ at $z=3$, which is higher than the $z=0$ transition mass of $10^{8.2}\,\rm M_{\odot}$ due to the overall accretion rates onto BHs being higher at higher redshifts in the absence of AGN feedback. 
{\sc GAEA} also displays a form of BH mass transition at $10^{7}\,\rm M_{\odot}$, with galaxies preferentially deviating from the main sequence above that mass. Note, however, that this transition is not sharp as it is in {\sc IllustrisTNG}, and the scatter around the median can be as high as $\approx 2$~dex. 

For the other simulations, \shark, {\sc Eagle}, {\sc Simba} and {\sc Galform} we see a similar behaviour, where 
massive BHs ($\gtrsim 10^{8}\,\rm M_{\odot}$) are associated with the $16^{\rm th}$ percentile being close or dropping below the sSFR threshold used to classify galaxies as passive. However, the medians remain comfortably above the sSFR threshold.
The lack of a clear BH mass threshold marking the transition from star-forming to passive in most of the simulation in Fig.~\ref{BHSSFRz3} is simply a reflection of other properties besides the BH mass being important in determining when and how AGN feedback is effective in quenching galaxies.

Going back to {\sc IllustrisTNG}, we find that the sudden effect of kinetic feedback and the lack of effective feedback below a BH mass threshold is the reason why there are almost no quenched galaxies at $z\gtrsim 3.5$ (Fig.~\ref{num_densities}), and why there are virtually no $z=3$ quenched galaxies with stellar masses $10^{10}\,\rm M_{\odot}\lesssim M_{\star}\lesssim 10^{11}\,\rm M_{\odot}$ in this simulation (Fig.~\ref{SMFz3}). The lack of effective feedback before galaxies hit a certain BH mass is also the reason why this simulation predicts massive-quenched galaxies to have higher stellar masses than the other simulations (see Fig.~\ref{SMFz3}), and a baryon collapse efficiency that is higher compared to most of the other simulations (Fig.~\ref{smhm})

{\sc Simba}'s AGN feedback model has several similarities to the one implemented in {\sc IllustrisTNG} as pointed out by \citet{Dave19} (e.g. two modes of AGN feedback, with the kinetic+thermal modes acting on slowly accreting BHs being what is effective in quenching galaxies; and a BH mass threshold above which the latter feedback can act).  
Hence, it is not surprising that there are some similarities in the AGN feedback impact: {\sc Simba} struggles to produce the density of intermediate-mass passive galaxies, and produces a similarly high baryon collapse efficiency. However, there are also several differences, with massive-quenched galaxies having progenitors that oscillate a lot more around the main sequence as they grow compared with {\sc IllustrisTNG} (Fig.~\ref{mainseq}). Another important condition for AGN feedback to be effective in {\sc Simba} besides the BH mass, is the Eddington ratio, which is required to be $<0.2$ for the jet mode to be effective, with the peak efficiency happening at an Eddington ratio of $0.02$. 
At high redshift this rarely happens \citep{Thomas21}, leading to an overall small fraction of massive-quenched galaxies (see Table~\ref{simus}).
This emphasises the condition of galaxies to be massive before they can undergo effective AGN feedback quenching. Hence, because the BH mass is not the only important parameter determining the regimes in which AGN feedback is effective, {\sc Simba} does not produce a strong correlation between sSFR and BH mass (Fig.\ref{BHSSFRz3}). 

\shark's AGN feedback model acts in a way that galaxies can get affected and fall significantly below the main sequence, but then regrow and go back to it. This can naturally happen as a jet power is always computed and can potentially produce feedback as long as there is a hot halo to work against. As soon as the BH does not produce enough jet power, then quenching becomes less efficient and BHs can go out of the maintenance mode. Hence, there is a natural stochasticity associated with the model. \citet{Lagos24} mention that even though \shark\ includes a model for AGN wind-driven feedback, in general this has a small effect compared to the jet feedback model. Because AGN feedback in general can be effective early in the formation of galaxies in \shark, galaxies then struggle to get to extremely high masses, $\gtrsim 10^{11}\,\rm M_{\odot}$, by $z=3$ and the model ends up predicting too few extremely massive galaxies.

{\sc Eagle}'s AGN feedback model becomes efficient as soon as the outflows driven by stellar feedback cease to be buoyant (and hence become less effective). The latter happens at around a halo mass of $10^{12}\,\rm M_{\odot}$ \citep{Bower17}. This was not modelled explicitly but instead is a consequence of how stellar and AGN feedback interact. AGN feedback thus can be effective quite early on in a galaxy's history, preventing them from growing too much in stellar mass. Similarly to \shark, this thus leads to the lack of extremely massive galaxies at early cosmic times.  

{\sc Galform} employs a model of AGN feedback that depends on the formation of a hot halo and the Eddington luminosity of BHs (rather than its accretion rate, which is directly or indirectly what the other simulations use). The only way this AGN feedback model acts is by stopping gas cooling, and hence the quenching necessarily happens on a scale similar to the star formation efficiency (i.e. what depletes the remaining gas reservoir in galaxies). This is a slow process and hence why {\sc Galform} tends to produce the longest quenching timescales of all simulations (Fig.~\ref{tauQ}). 

{\sc GAEA}'s latest AGN feedback implementation seems to be particularly effective at very high stellar masses. They quench extremely quickly going passive within a single snapshot of the simulation and as soon as galaxies reach modest BH masses (see Fig.~\ref{BHSSFRz3}). Because the AGN quasar wind scales with the AGN bolometric luminosity without an explicit dependence on the available gas reservoir, it makes quenching very strongly dependent on BH mass and weakly on other galaxy properties. This leads to a strong correlation between sSFR and BH mass, which is not seen in the other simulations. 

Interestingly, in {\sc GAEA} most of the quenching of massive-quenched galaxies is due to quasar winds as detailed in \citet{DeLucia24}. They show this by switching off that model and finding very few quenched galaxies at high redshift. In \shark\ and {\sc Galform} instead, quenching happens when a form of mechanical feedback is implemented (similar to what is done in {\sc Simba} and {\sc IllustrisTNG}). The {\sc Eagle} AGN feedback model, even though it injects the energy in a thermal mode, the way it quenches galaxies ends up being phenomenologically similar to the classical `radio-mode' feedback \citep{Croton06}; see \citet{Bower17} for details. This is because the quenching is not associated with gas being ejected from the galaxy but with heating a galaxy’s corona and preventing further cooling. We remark, however, that sometimes in hydrodynamical simulations it is hard to fully disentangle the role of ejective vs preventative AGN feedback \citep{Brennan18}.

\begin{figure}
\begin{center}
\includegraphics[trim=7mm 21mm 16mm 34mm, clip,width=0.49\textwidth]{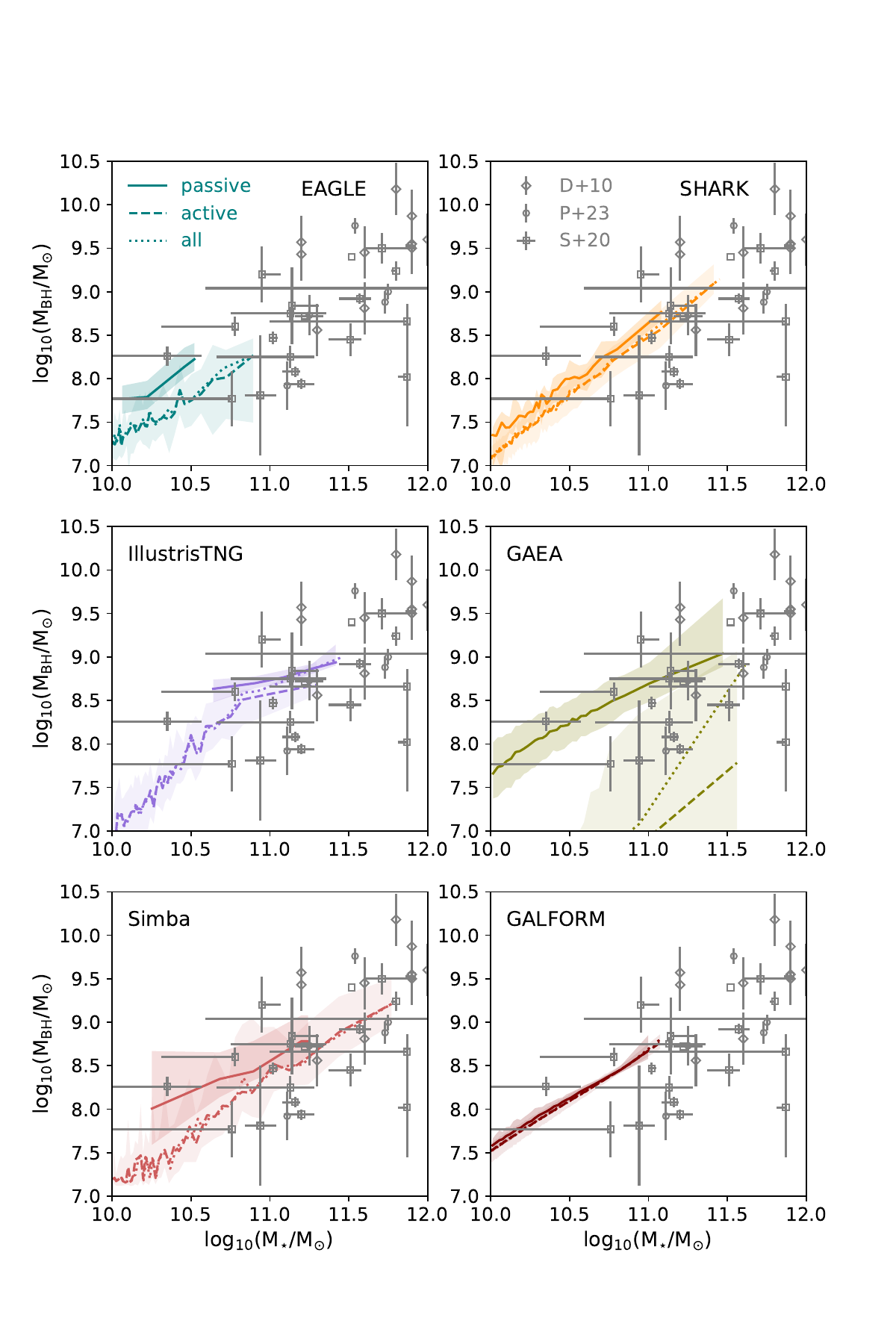}
\caption{The central BH mass a function of stellar mass for all galaxies at $z=3$ that have a stellar mass $>10^{10}\,\rm M_{\odot}$, in all the simulations as labelled (dotted lines). We also show separately passive ($\rm sSFR<10^{-10}\,\rm yr^{-1}$) and active ($\rm sSFR>10^{-10}\,\rm yr^{-1}$) galaxies, with solid and dashed lines, respectively. 
Lines show the medians, and the shaded regions (shown only for the passive and active sub-populations) the $16^{\rm th}-84^{\rm th}$ percentile ranges, respectively. We use adaptive bins to ensure a number of galaxies per bin $\ge  10$. We also show with symbols observational inferences from \citet{Decarli10a,Decarli10b} (D+10), \citet{Suh20} (S+20) and \citet{Poitevineau23} (P23), as labelled in the top-right panel, corresponding to measurements at $2\le z\le 3.7$.}  
\label{BHMstarz3}
\end{center}
\end{figure}
Fig.~\ref{BHMstarz3} shows the BH-stellar mass relation for massive galaxies in all the simulations, separating active ($\rm sSFR>10^{-10}\,\rm yr^{-1}$) and passive ($\rm sSFR\le 10^{-10}\,\rm yr^{-1}$) galaxies. \shark, {\sc Eagle}, {\sc IllustrisTNG}, {\sc Simba} and {\sc GAEA} predict a clear difference between active and passive galaxies, with passive galaxies having more massive BHs at fixed stellar mass than their active counterparts. The quantitative difference however, varies widely between the simulations: \shark, {\sc Eagle}, {\sc IllustrisTNG}, {\sc Simba} and {\sc GAEA} produce differences of $\approx 0.25$, $0.4$, $0.3-0.75$, $0.1-0.5$ and $\gtrsim 1.5$~dex, respectively, between the two populations.
At $z=0$, there is a documented difference between passive and star-forming galaxies \citep{Terrazas17}, where the former have more massive BHs than the latter at fixed stellar mass. \citet{Lagos24} showed that at $z=0$ and in \shark\ the difference can be as high as one order of magnitude, in agreement with what \citet{Terrazas17} find. At these high-z however, the samples are too small still to split between star-forming and passive galaxies, but this is likely to improve, with high-precision {\it JWST} BH-mass measurements using reverberation mapping becoming available \citep{Golubchik24}. Hundreds of such measurements are expected to be carried out with {\it Euclid} \citep{Euclid24} in the near future and thousands with the {\it Roman} space telescope\footnote{\url{https://roman.gsfc.nasa.gov}} in the longer term.

The explicit BH-mass dependence of the switch to go from the radiatively efficient to inefficient modes of AGN feedback in {\sc Simba} and {\sc IllustrisTNG} lead to a difference in BH mass between the passive and star-forming population that increases with decreasing stellar mass. This is not seen in the other simulations, where there is either only a single mode of AGN feedback (\eagle\ and {\sc Galform}), or where the transition between modes does not depend on BH mass (\shark\ and {\sc GAEA}).

\citet{Kimmig23} and \citet{Weller24} show that a similar BH mass difference is seen in the {\sc Magneticum} and {\sc Astrid} simulations, respectively.
Conversely, {\sc Galform} predicts no difference between the BH masses of active and passive galaxies at fixed stellar mass. 
The simulations in which a clear difference in BH mass is seen have AGN feedback models that depend strongly on the BH accretion rate while in {\sc Galform}, the most important condition is that the haloes are in quasi-hydrostatic equilibrium. The latter thus leads to little BH mass differences between massive and passive galaxies. It is, however, fair to say that the vast majority of simulations (including those analysed here) predict passive galaxies to have more massive BHs than active counterparts. 

On the other extreme, in {\sc GAEA}, star-forming galaxies seem to prefer very low BH masses, that deviate by orders of magnitude from the local BH-stellar mass relation, and more so considering the latest {\it JWST} results, which tend to prefer a higher normalisation at $z=3$ (e.g. \citealt{Bertemes24,Furtak24}). Current observational estimates are shown as symbols, and tend to prefer higher masses, implying that {\sc GAEA}'s predictions may be in tension with  observations unless they are biased towards passive galaxies. Note, however, that within the scatter, most of the simulations agree with the observational constraints quite well.  

Regarding the stellar populations of massive-quenched galaxies, 
\citet{Kimmig23} show that the short formation timescales predicted for massive-quenched galaxies at high-z in {\sc Magneticum} ($\Delta\,\rm age\lesssim 200$~Myr) leave clear imprints in the form of e.g. alpha-enhancements, flat age gradients and steep metallicity profiles. Similar features would be expected for the simulations here that predict similarly short formation timescales (e.g. {\sc Eagle} and \shark; see bottom panel of Fig.~\ref{age_scaling}).  

The impact of the different AGN feedback models implemented in the simulations is also seen in the gaseous haloes around galaxies \citep{Wright24} even at $z=2$ (with many of the differences becoming larger towards $z=0$). For example, \citet{Wright24} show that the gas content and gas temperature of haloes in {\sc Eagle}, {\sc IllustrisTNG} and {\sc Simba} can differ by $30$per cent and $1.5$~dex, respectively, at the scales where AGN feedback is effective. However, observing the gaseous haloes of high-z galaxies is challenging, and hence it is promising to see that AGN feedback also impacts galaxy properties of massive-quenched galaxies that can perhaps be more easily collected.

Every simulation thus have a different implementation of AGN feedback, which has a clear impact on how galaxies grow and end up quenching, including how quickly that happens. This impact is clearly seen in the SFHs of massive-quenched galaxies at $z=3$. 
This is not to say that other baryon physics processes do not affect the way in which massive galaxies grow in these simulations, but simply that AGN feedback is required to produce the quenching.
As the observational samples grow in size, studying the differences in SFHs of passive galaxies in bins of stellar mass would greatly help disentangle some of the predictions presented in this paper.  
 
\section{Conclusions}\label{conclusions}

We have presented a thorough study of the SFHs of high-z massive-quenched galaxies in $6$ modern galaxy formation simulations ($3$ SAMs, \shark, {\sc GAEA}, {\sc Galform};  and $3$ cosmological hydrodynamical simulations, {\sc Eagle}, {\sc IllustrisTNG}, {\sc Simba}). All of these have been primarily tuned to the local Universe (except for {\sc GAEA}), so the comparisons performed in this paper constitute real tests of the predictions of these simulations. 

We summarise our main findings below:
\begin{itemize}
    \item All the simulations struggle to easily reproduce the number density of massive-quenched galaxies observed by the {\it JWST} (e.g. \citealt{Carnall23,Valentino23,Nanayakkara22,Alberts23}), especially at $z\gtrsim 3$. However, there are varying degrees of tension (from $\approx 0.3$~dex to several~dex) depending on the exact stellar mass and sSFR used to select the samples in simulations (Fig.~\ref{num_densities}). 
    \item Comparing the simulations with observations of the $z\approx 3$ SMF of passive galaxies reveals that there are broadly two distinct behaviour outcomes: {\sc Simba} and {\sc IllustrisTNG} reproduce relatively well the number density of passive galaxies with $M_{\star}\gtrsim 10^{11}\,\rm M_{\odot}$ but fall short at $10^{10}\lesssim M_{\star}\lesssim 10^{11}\,\rm M_{\odot}$. The opposite is true for \shark, {\sc Galform}, {\sc Eagle} and {\sc GAEA} (Fig.~\ref{SMFz3}). We find that this is due to the explicit dependence on the BH mass of the switch to go from the radiatively-efficient to inefficient AGN feedback modes in {\sc Simba} and {\sc IllustrisTNG}, which allows galaxies to grow efficiently until their BHs hit a certain mass. In {\sc Simba} this is compounded with the slow build-up of the jet mode efficiency, which reaches its peak at a very low Eddington ratio of $0.02$. A consequence of this difference between {\sc IllustrisTNG}, {\sc Simba} and the rest of the simulations is the higher baryon collapse efficiency at fixed halo mass, $\approx 15-30$per cent, in the former simulations (Fig.~\ref{smhm}).
    \item The SFHs of the $z=3$ massive-quenched galaxies present several differences between the simulations. \shark, {\sc Eagle} and {\sc IllustrisTNG} predict their formation to start at $z\gtrsim  10$ (i.e. appear for the first time in the catalogues), while in {\sc Galform}, {\sc Simba} and {\sc GAEA} a large fraction of the star-forming progenitors start forming at $z<10$ (Fig.~\ref{SFH}). The peak SFRs of the galaxies in every case depend strongly on the $z=3$ stellar mass but with \shark\ and {\sc Galform} predicting peak SFRs $\approx 0.5-0.7$~dex higher at fixed stellar mass than the other simulations due to the higher star formation efficiency they invoke for starbursts (Fig.~\ref{sfr_scaling}). However, relative to the main sequence, we find that \shark, {\sc Galform}, {\sc IllustrisTNG} and {\sc Simba}
    predict galaxies to be starbursting at their peak SFR (i.e. well above the main sequence), while this is only true for the most massive galaxies in \eagle\ and {\sc GAEA}.
    \item Compared with observationally-inferred SFHs, we find that \shark\ and {\sc IllustrisTNG} produce the highest SFR peaks, closer to the observations (Fig.~\ref{SFHsAll}). However, the observational samples are very small and the inferences sensitive to the exact details of the spectral modelling and the spectral resolution of the available data. We also compare the simulations with the formation times derived in the literature for massive-quenched galaxies (defined as the time where half of the stellar mass has been formed) and find that within the scatter ($\approx 300-1,000$~Myr), all the simulations agree well with observations (Fig.~\ref{age_scaling}). 
    \item \shark\ and {\sc Eagle} produce the earliest and fastest-forming galaxies, while {\sc Galform} and {\sc IllustrisTNG} predict the latest and slowest-forming galaxies (Fig.~\ref{age_scaling}). \shark, {\sc Eagle} and {\sc Simba} predict $\Delta\,\rm age\lesssim 200$~Myr (with $\Delta\,\rm age$ being the difference between the time at which galaxies reach $50$ and $90$ per cent of their $z=3$ stellar mass), while the other simulations prefer $\Delta\,\rm age\gtrsim 200$~Myr, with $\Delta\rm age$ being the difference between the time the $z=3$ massive-quenched galaxies had formed $50$ and $90$ per cent of their stellar mass. By measuring the quenching timescale in terms of distance to the main sequence (which we define internally to each simulation for consistency) we find that all the simulations, except for {\sc Galform}, predict quenching timescales $\lesssim 100$~Myr (Fig.~\ref{tauQ}). However, if we consider the tightness of the main sequence, we see that the timescales for quenching become longer in a way that in \eagle, \shark, and {\sc Simba} they are $\approx 100-200$~Myr, emphasising the need for a convergent definition of quenching timescale.
    \item By comparing the sSFR with the BH mass of galaxies at $z=3$ with $M_{\star}>10^{10}\,\rm M_{\odot}$, we find that only {\sc IllustrisTNG} and {\sc GAEA} produce a strong correlation between the two (Fig.~\ref{BHSSFRz3}). In {\sc IllustrisTNG} this looks like a sharp transition from star-forming to quenched galaxies at a BH mass of $\approx 10^{8.65}\,\rm M_{\odot}$. In {\sc GAEA}, galaxies start quenching as soon as they reach a relatively modest BH mass of $10^7\,\rm M_{\odot}$, so that the BH-stellar mass relation of star-forming galaxies is $\approx 1.5$~dex lower in normalisation compared with passive galaxies (Fig.~\ref{BHMstarz3}).
    \item All the simulations, except for {\sc Galform}, predict passive galaxies to have heavier BHs than their star-forming counterparts at fixed stellar mass (Fig.~\ref{BHMstarz3}). The lack of difference in  {\sc Galform} is due to AGN feedback primarily depending on whether a halo is found to be on quasi-hydrostatic equilibrium, which is independent of BH mass. 
\end{itemize}

Our findings combined with literature ones \citep{Vani24,Weller24,Szpila24} point to the pervasive problem of reproducing the properties of massive-quenched galaxies in the early universe in modern cosmological galaxy formation simulations. 
The amount of tension with current observations is extremely sensitive to potential errors in stellar masses and SFRs, and contamination of the passive galaxy samples in observations. Significant effort is required to  understand the magnitude of errors in the derivation of these quantities in observations to have a precise understanding of the true tension with the simulations.

We also find that the exact way AGN feedback is modelled leaves clear imprints on the SFHs of massive-quenched galaxies in a way that there are clear differences in the predictions of the different simulations presented here that are testable with existing facilities and upcoming observations. 
In fact, recent observations are beginning to place constraints on the SFHs of massive-quenched galaxies across different cosmic epochs
offering a unique and novel opportunity for further testing of the models using observations. 

\section*{Data Availability}

The data for each simulation is available as follows: 

\begin{itemize}
    \item \shark: The \surfs\ halo and subhalo catalogue and corresponding merger trees used in this work can be accessed from \url{https://tinyurl.com/y6ql46d4}. \shark\ is a public code and the source and python scripts used to produce the plots in this paper can be found at \url{https://github.com/ICRAR/shark/}.
    \item {\sc Galform:} the data used in this paper is publicly available from \url{http://virgodb.dur.ac.uk:8080/MyMillennium/}.
    \item {\sc GAEA:} data and code are proprietary and access can be given upon request. Visit \url{https://sites.google.com/inaf.it/gaea/home} for details. 
    \item {\sc Eagle:} The {\sc Eagle} simulations are publicly available; see \citet{McAlpine15,EAGLE17} for how to access {\sc EAGLE} data.
    \item {\sc IllustrisTNG:} data is publicly available from \url{https://www.tng-project.org} \citep{Nelson19}.
    \item {\sc Simba:} data is publicly available from \url{http://simba.roe.ac.uk/}.

\end{itemize}

\section*{Acknowledgements}

We thank the referee for their constructive report. 
CL thanks Pascal Oesch, Robert Feldmann and Elizabeth Rembelska for organising the conference `Observing and Simulating Galaxy Evolution in the Era of {\it JWST}' (Ascona, Switzerland, August 2024) and funding her participation, as many ideas discussed in this paper arose during that meeting. CL thanks the ARC for the Discovery Project DP210101945. RW acknowledges the support of the Forrest Research Foundation. MB acknowledges the funding by McMaster University through the William and Caroline Herschel Fellowship. CMB acknowledges support from the Science Technology Facilities Council (ST/X001075/1). KH acknowledges funding from the Australian Research Council (ARC) Discovery Project DP210101945.

We thank Daniel Angles-Alcazar for his advice on how to recover initial masses for stellar particles and for sharing his thoughts on the conditions for AGN feedback quenching to be effective in {\sc Simba}. We thank Nicole Thomas for discussions about AGN properties in {\sc Simba}.
We thank Violeta Gonzalez-Perez for providing the scaling between different tracers of SFR and the intrinsic SFR for different IMFs. We thank Rhea-Silvia Remus and Adam Carnall for sharing their data. 

This work was supported by resources provided by The Pawsey Supercomputing Centre with funding from the 
Australian Government and the Government of Western Australia.

We acknowledge the Virgo Consortium for making
their simulation data available (for both {\sc Eagle} and {\sc Galform}). 
We acknowledge the {\sc IllustrisTNG} and {\sc Simba} teams for making their simulation data available. 

We acknowledge the use of INAF-OATs computational resources within the
framework of the CHIPP project and the INAF PLEIADI
program (\url{http://www.pleiadi.inaf.it}).




\bibliographystyle{mn2e_trunc8}
\bibliography{SharkQ}



\appendix
\input{AppendixMainSequence}
\input{AppendixSFHs}

\bsp	
\label{lastpage}
\end{document}

%% file: AppendixMainSequence.tex
\section{The main sequence evolution and control samples}\label{MS}

\begin{figure*}
\begin{center}
\includegraphics[trim=5mm 4.5mm 2mm 2mm, clip,width=0.49\textwidth]{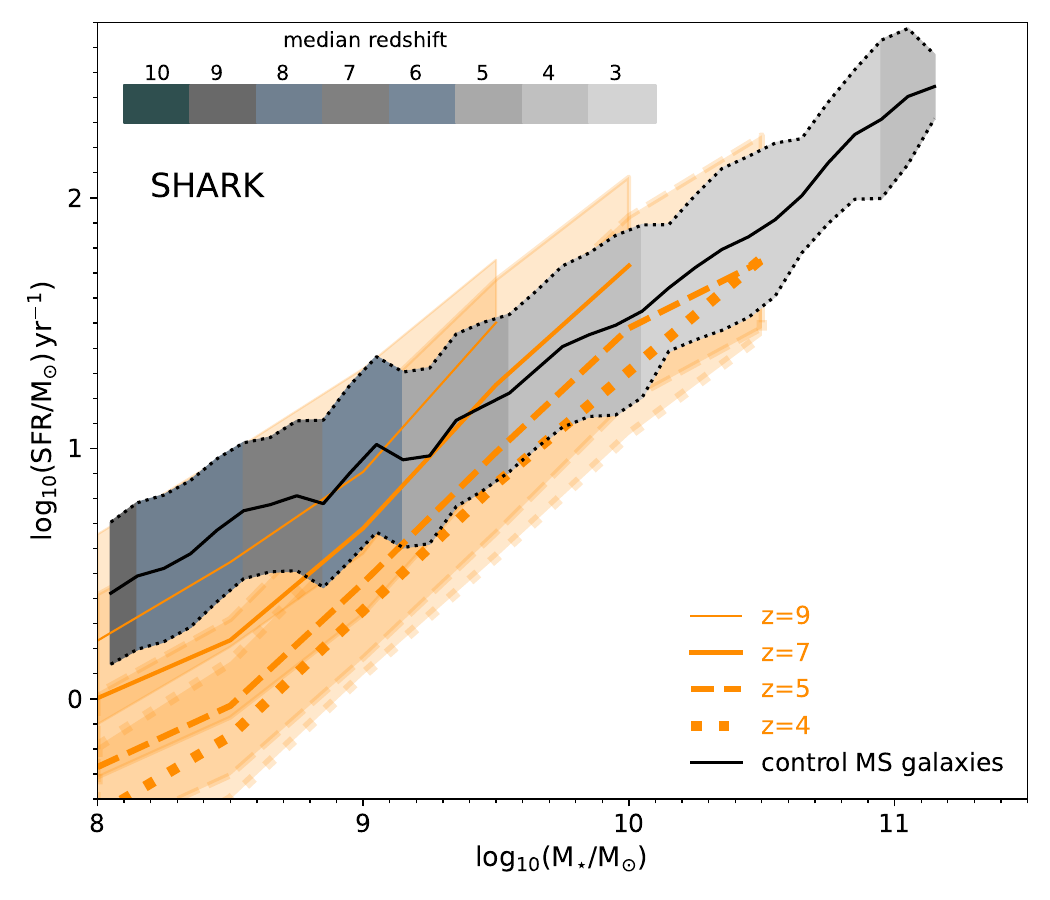}
\includegraphics[trim=5mm 4.5mm 2mm 2mm, clip,width=0.49\textwidth]{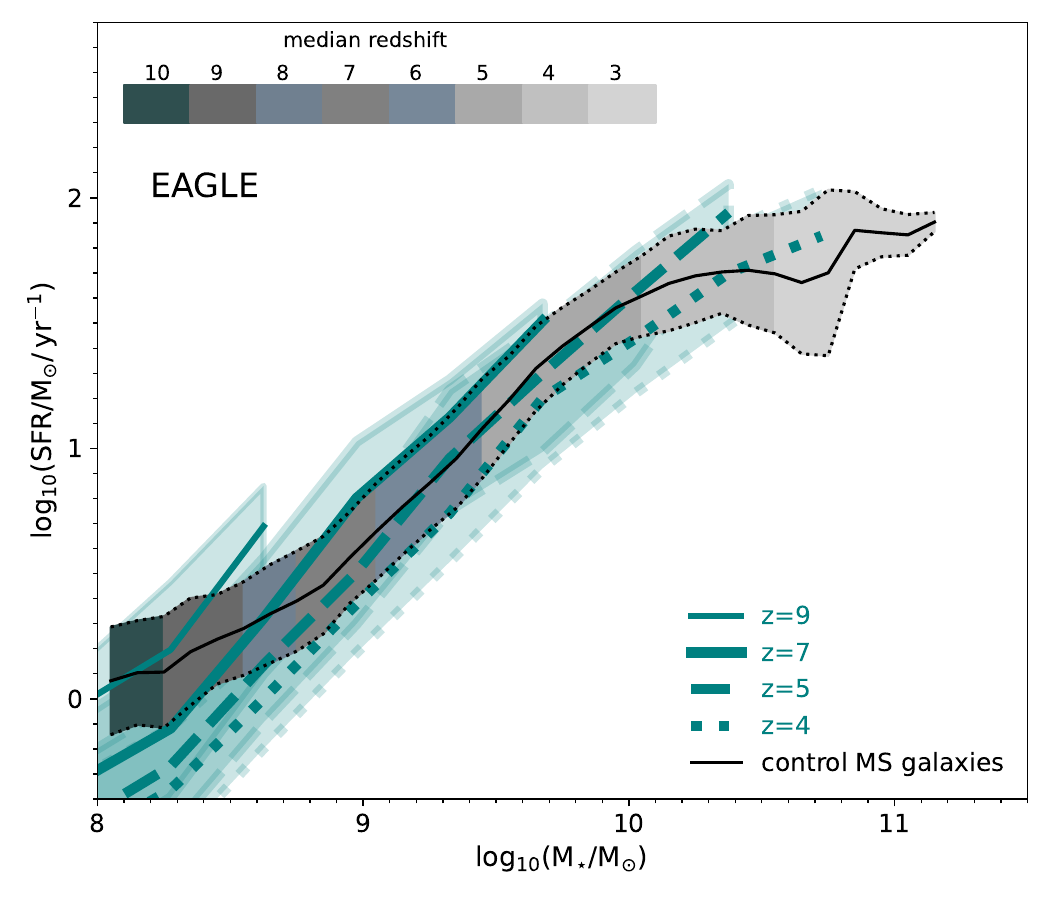}
\includegraphics[trim=5mm 4.5mm 2mm 2mm, clip,width=0.49\textwidth]{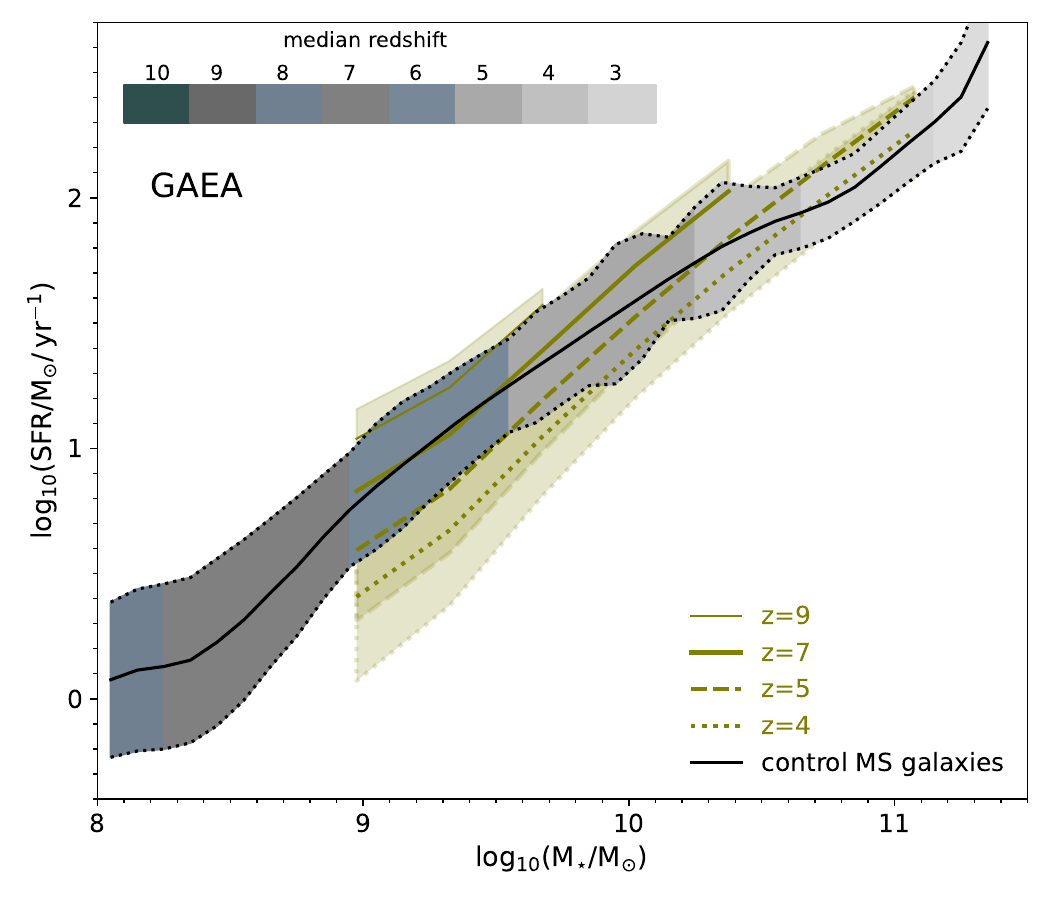}
\includegraphics[trim=5mm 4.5mm 2mm 2mm, clip,width=0.49\textwidth]{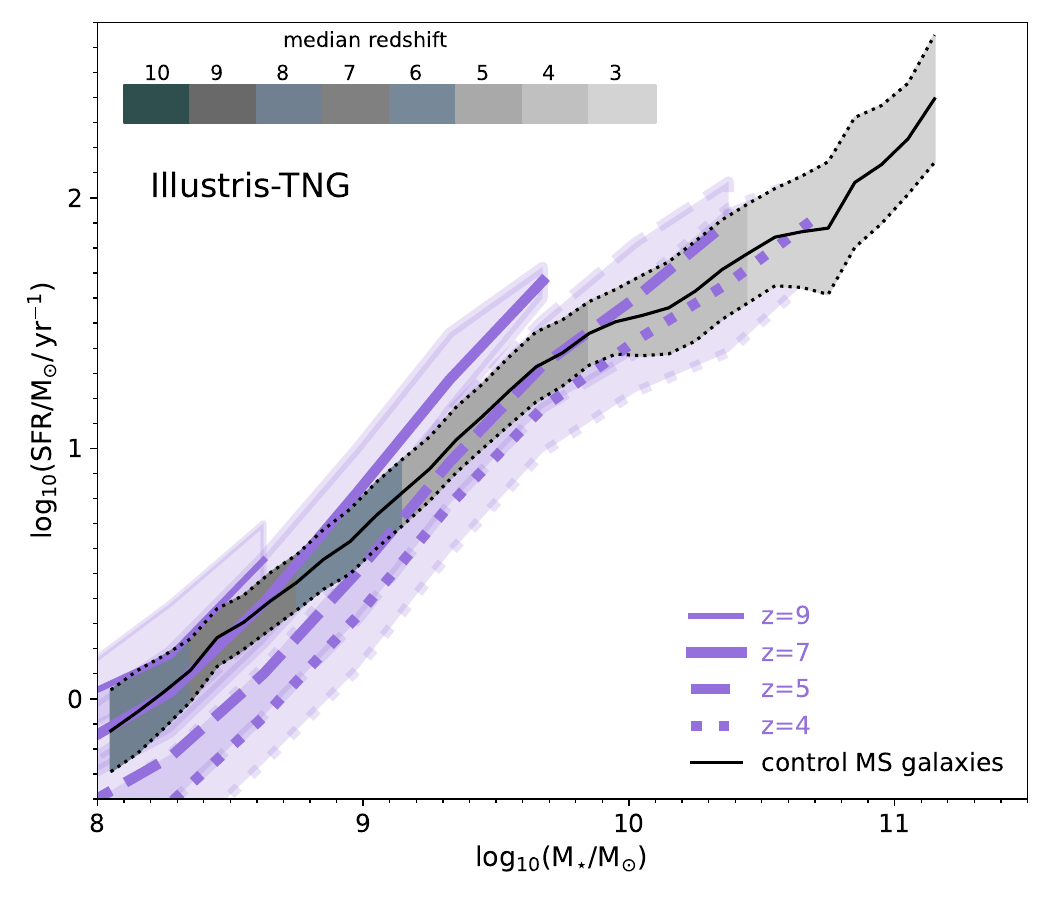}
\includegraphics[trim=5mm 4.5mm 2mm 2mm, clip,width=0.49\textwidth]{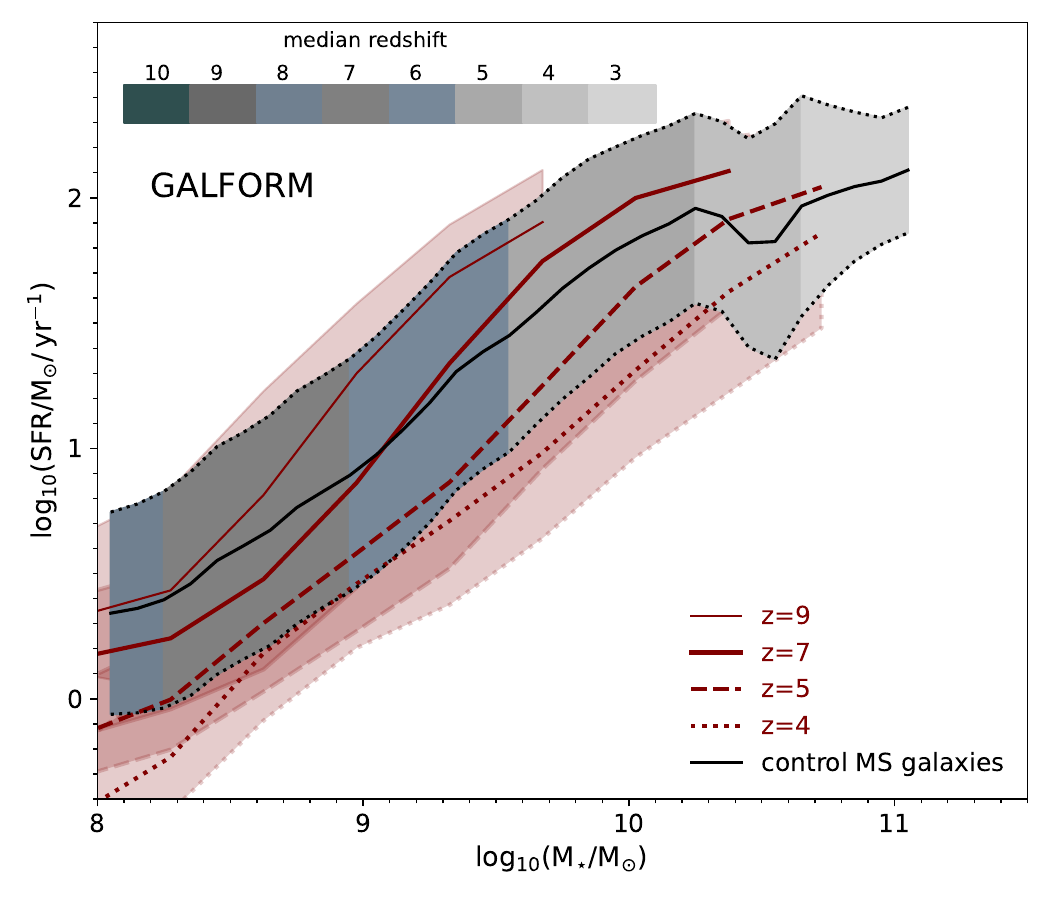}
\includegraphics[trim=5mm 4.5mm 2mm 2mm, clip,width=0.49\textwidth]{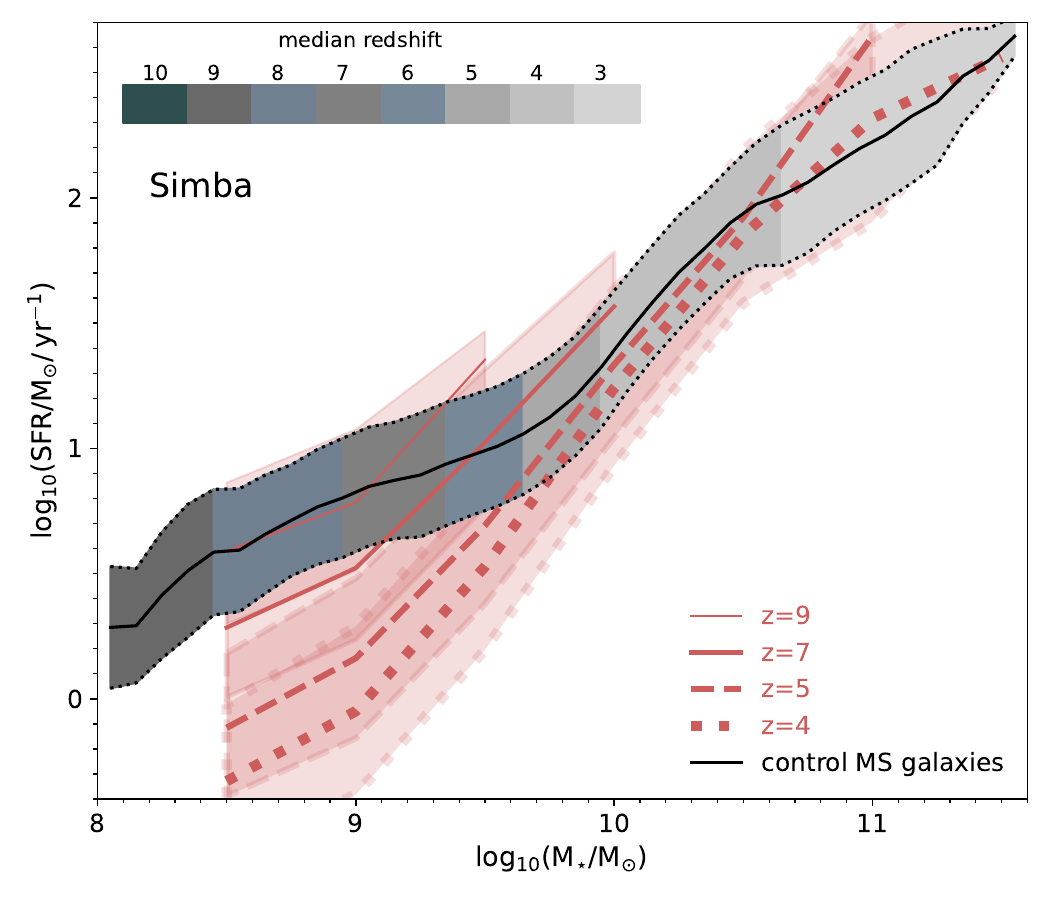}
\caption{The median relation between SFR and stellar mass with $1\,\sigma$ (standard deviation) of the control MS sample (solid and dotted lines, respectively) compared with the main sequence predicted in each simulation at $z\approx 9,\,7,\,5,\,4$ (as labelled in each panel). We colour the control MS sample by the median redshift of the data points in bins of stellar mass, as labelled in each panel. This is shown for the $6$ simulations \shark\, {\sc Eagle}, {\sc GAEA}, {\sc IllustrisTNG}, {\sc Galform} and {\sc Simba}, as labelled in each panel. Note that for {\sc GAEA} we only use galaxies with stellar masses $\ge 10^9\,\rm M_{\odot}$, while for the other simulations we cover the entire stellar mass range.} 
\label{mainseqtest}
\end{center}
\end{figure*}

\S~\ref{definitions} presents the way we construct a control sample of massive-active galaxies in each simulation, and explains how we use that control sample to measure a distance to the main sequence and define quenching timescales. Here, we show that this is a good assumption by comparing the median SFR-$M_{\star}$ relation of the control sample presented in Fig.~\ref{mainseq} (solid and dotted black lines) with the main sequence across different redshifts in each simulation. This is presented in Fig.~\ref{mainseqtest}.

Each panel of Fig.~\ref{mainseqtest} shows the median SFR in bins of stellar mass at $z=9,\,7,\,5,\,4$ using colour lines, with the coloured shaded regions showing $1$~standard deviation around the median. The latter is measured using the galaxy catalogues produced at each of these snapshots using the same SFR and stellar mass definitions introduced in \S~\ref{definitions}.

Each panel of Fig.~\ref{mainseqtest} also shows, using black lines, the median and $1\sigma$ dispersion of the SFR-$M_{\star}$ tracks of the control MS sample. The latter is also coloured by the median redshift of the points in the tracks in bins of stellar mass (with the grey-colour band at the top showing the corresponding median redshifts). Typically the lower mass part of the SFR-$M_{\star}$ tracks of the control MS sample, $M_{\star}\lesssim 10^{8.7}\,\rm M_{\odot}$, are dominated by the early growth of those galaxies, with typical median redshifts of $9$ or $8$. The control MS sample galaxies typically reach a stellar mass of $M_{\star}\approx 10^{9.5}\,\rm M_{\odot}$ by $z\approx 5$. The exact redshifts at which this happens varies slightly between simulations. 

We see that the main sequence traced by the SFR-$M_{\star}$ tracks of the control MS sample is flatter than the main sequence measured at individual redshifts in all the simulations. This is because different parts of the main sequence traced by the SFR-$M_{\star}$ tracks of the control MS sample correspond to different redshifts. The lower mass part tends to follow well the main sequence measured at $z=9$ while the higher mass parts are better described by the main sequence at $z=5$ or $z=4$. 
In principle, we could measure a distance to the main sequence at each point in the evolution of the massive-passive galaxy sample of each simulation using a main sequence defined at the same redshift of the point of interest (which is the more common approach adopted in the literature). However, the cadence of the snapshot outputs of the simulations is not as good as the cadence we use to reconstruct the SFHs of the massive-passive galaxies. To circumvent this issue, we define an alternative main sequence using the  
SFR-$M_{\star}$ tracks of the control MS sample, which in turns gives us a time dependent main sequence (due to the median redshift changing along the main sequence of the control MS sample). The fact that the median redshift of the different stellar mass portions of the control MS sample follow the measured main sequence at a redshift close to the median one implies that 
using the control MS sample to measure distance to main sequence 
would give a similar result to 
using the main sequence defined at individual redshifts. This means that the control MS sample is, on average,  growing along the main sequence.

%% file: AppendixSFHs.tex
\section{Number density of massive-quenched galaxies in the {\sc Magneticum} simulations}\label{Magneticum}

\begin{figure}
    \centering
    \includegraphics[trim=5mm 4mm 1.5mm 3mm, clip,width=0.49\textwidth]{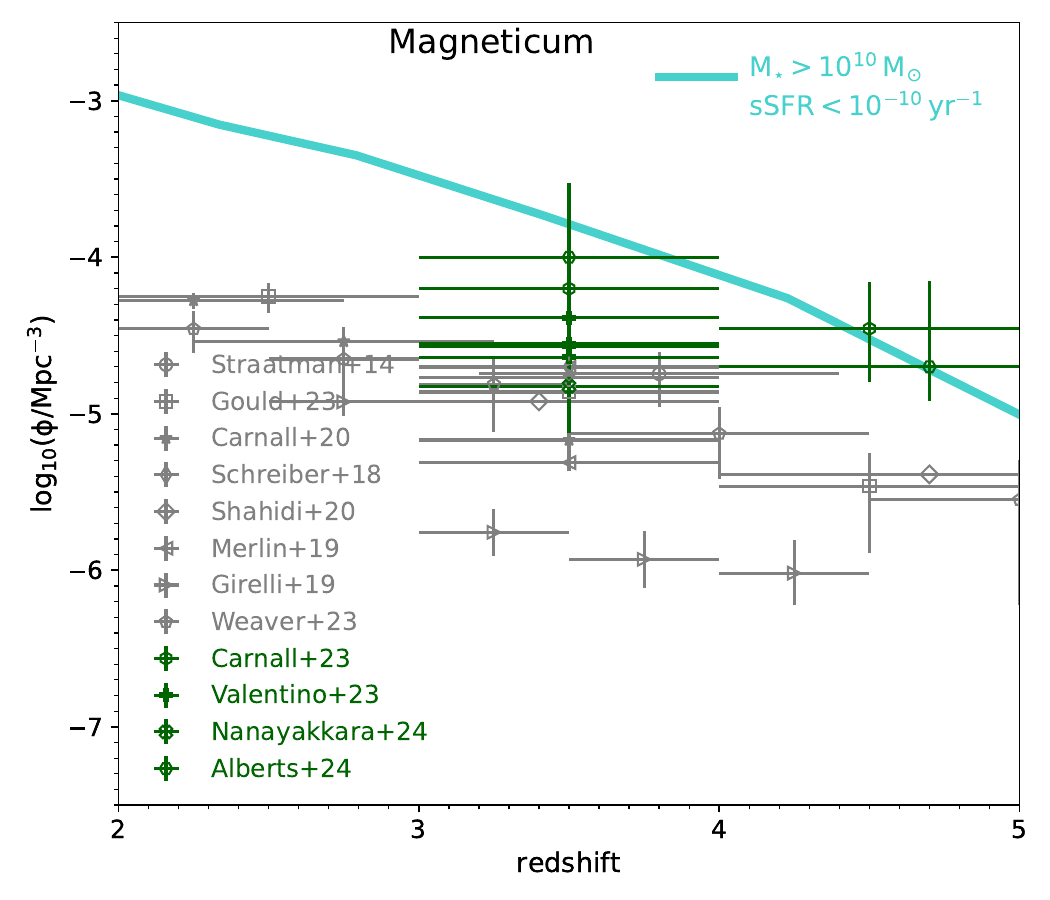}
    \caption{As in Fig.~\ref{num_densities} but for the {\sc Magneticum} simulations and using a single selection for massive-quenched galaxies: $M_{\star}\ge 10^{10}\,\rm M_{\odot}$ and $\rm sSFR\le 10^{-10}\,\rm yr^{-1}$.}
    \label{num_density_magneticum}
\end{figure}

One of the only cosmological hydrodynamical simulations to agree better with current observations of the number density of massive-quenched galaxies at $z=3-4$ is {\sc Magneticum} \citep{Remus23,Kimmig23}. Here we show the number density of galaxies in {\sc Magneticum} with $M_{\star}\ge 10^{10}\,\rm M_{\odot}$ and $\rm sSFR\le 10^{-10}\,\rm yr^{-1}$ in Fig.~\ref{num_density_magneticum}.

{\sc Magneticum} indeed produces a higher number density of massive-quenched galaxies at $z=3-4$ compared with the simulations presented in Fig.~\ref{num_densities}. However, it tends to overpredict by an order of magnitude the number of massive-quenched galaxies at slightly lower redshift, $z=2-3$. \citet{Lustig23} showed that at $z=2$ {\sc Magneticum} produced roughly the correct number density of masses $\ge 10^{11}\,\rm M_{\odot}$, indicating that the problem is that {\sc Magneticum} is producing too many passive galaxies with masses $10^{10}-10^{11}\,\rm M_{\odot}$. 

This over-production of intermediate-mass quenched galaxies could be in part due 
the AGN feedback model been overly effective in {\sc Magneticum}, and in part due to resolution. The resolution of the box used here (DM and gas particle masses of $5.1\times 10^7\,\rm M_{\odot}$ and $1.03\times 10^7\,\rm M_{\odot}$, respectively), which is the same as used in \citet{Remus23,Kimmig23}, is $10$ times worse than that used in {\sc Eagle} and {\sc IllustrisTNG}, which could lead to overly effective quenching at around a stellar mass of $10^{10}\,\rm M_{\odot}$ and below. 
Regardless of the exact causes, this shows that in the models analysed in this work and others, such as {\sc Magneticum}, the agreement with observations of massive-quenched galaxies in one redshift does not warrant the agreement over a wide redshift range.

\section{The effect of the time cadence on the maximum SFR}\label{maxSFReff}

\begin{figure}
    \centering
    \includegraphics[trim=5mm 4mm 1.5mm 3mm, clip,width=0.49\textwidth]{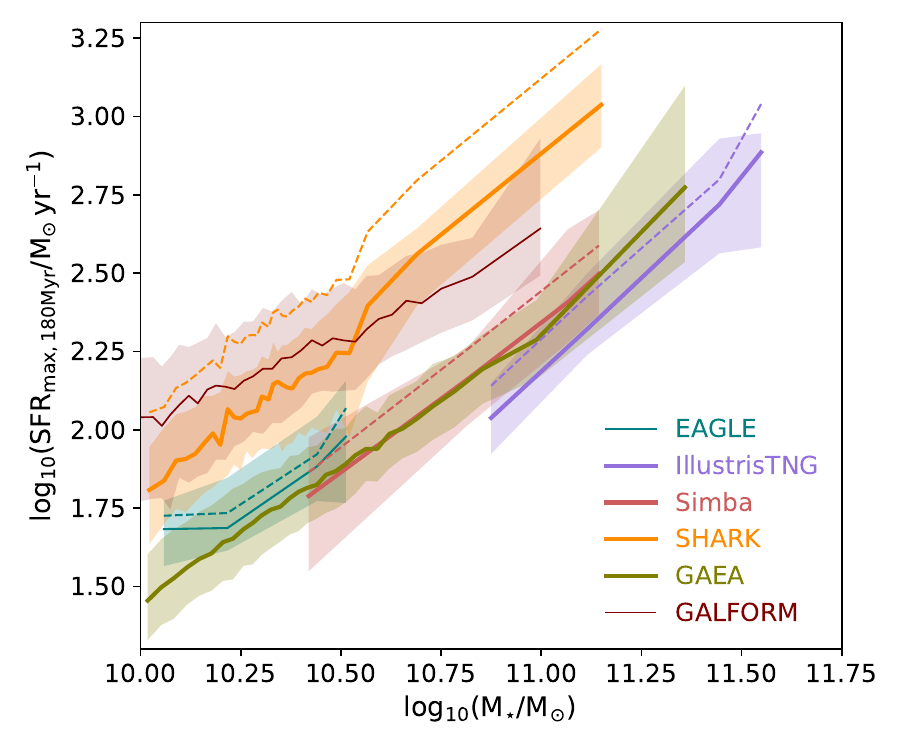}
\caption{$\rm SFR_{\rm max}$ as a function of the $z=3$ stellar mass of massive-quenched galaxies in $6$ simulations, as labelled. The solid lines and shaded regions show the medians and $16^{\rm th}-84^{\rm th}$ percentile ranges, respectively. We sample the SFHs of galaxies in \shark, {\sc Eagle}, {\sc IllustrisTNG} and {\sc Simba}  using a poorer time cadence that resembles that of {\sc GAEA} and {\sc Galform} ($\approx 180$Myr). To facilitate the comparison with 
Fig.~\ref{sfr_scaling}, we show with dashed lines the medians presented there for \shark, {\sc Eagle}, {\sc IllustrisTNG} and {\sc Simba}, with are the simulations whose SFHs are being resampled.  
Overall we see similar results to those presented in the top-left panel of Fig.~\ref{sfr_scaling}: \shark\ and {\sc Galform} produce the highest $\rm SFR_{\rm max}$, while {\sc IllustrisTNG} produces the lowest values.}
    \label{maxSFR120}
\end{figure}

Fig.~\ref{maxSFR120} shows the equivalent to the top-left panel of Fig.~\ref{sfr_scaling} but when we sample the SFHs in \shark, {\sc Eagle}, {\sc IllustrisTNG} and {\sc Simba} to have timesteps that are $\approx 180$~Myr, to make them more comparable to the time cadence of {\sc GAEA} and {\sc Galform}. We see small changes - for example $\rm SFR_{\rm max}$ in {\sc GAEA} is similar to that of {\sc Simba}, while {\sc IllustrisTNG} has the smallest values; at stellar masses $\lesssim 10^{10.5}\,\rm M_{\odot}$ {\sc Galform} produces higher $\rm SFR_{\rm max}$ than \shark, but this reverts at higher stellar masses. However, qualitatively the results and hence our conclusions remain.

\section{Fits to the star formation histories}\label{SFHfits}

We fit each SFH with a skewed Gaussian function (as presented in \citealt{Robotham20}):

\begin{eqnarray}
    X{\rm (age)} &=& \frac{{\rm age} - m_{\rm peak}}{m_{\rm period}},\\
    Y{\rm (age)} &=& X{\rm (age)} * \left(e^{m_{\rm skew}} \right)^{\rm asinh({\it X}{\rm (age)})}\\
    {\rm SFR(age)} &=& m_{\rm SFR}\, e^{-Y^2{\rm (age)}/2},\nonumber\\
\end{eqnarray}

\noindent where ``age'' is the lookback time from the redshift of interest, and $m_{\rm peak}$, $m_{\rm period}$, $m_{\rm SFR}$ and $m_{\rm skew}$ are parameters. We choose this function because its parameters are orthogonal and that makes it better from a statistical point of view than a function in which the parameters are potentially degenerate with each other. It has also been used to fit galaxies at high-z in \citet{D'Silva23b,D'Silva23c}. We acknowledge however, that other functions have more broadly been used to fit the SFHs of high-z galaxies than the one used here (e.g. \citealt{Schreiber18,Carnall20}).

\begin{figure*}
    \centering
    \includegraphics[trim=5mm 5mm 1.5mm 3mm, clip,width=0.3\textwidth]{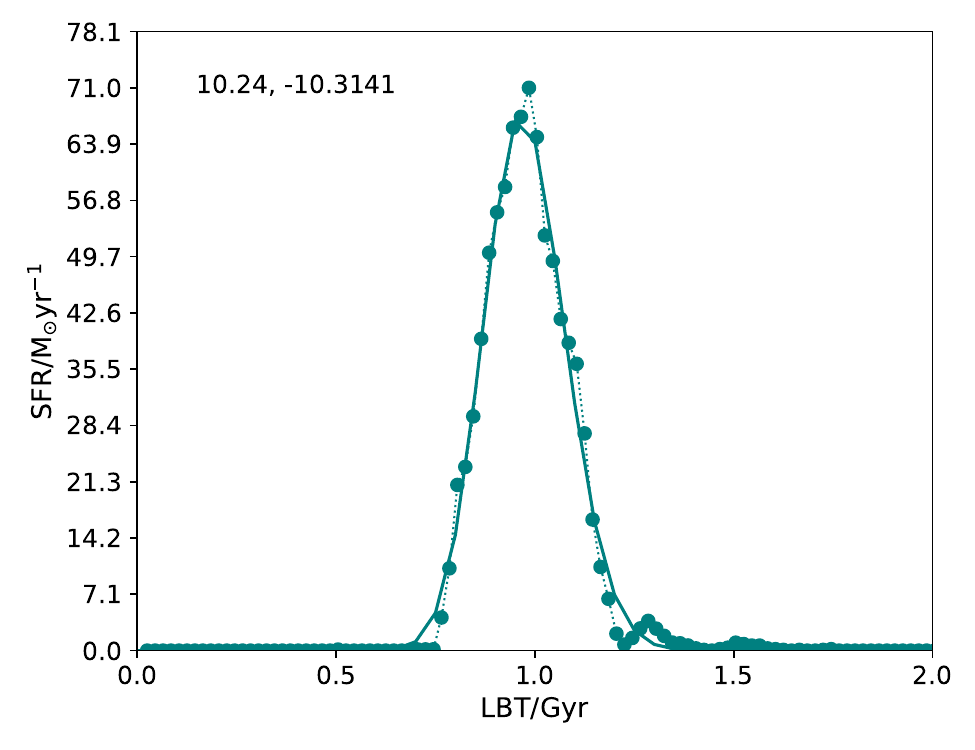}
    \includegraphics[trim=5mm 5mm 1.5mm 3mm, clip,width=0.3\textwidth]{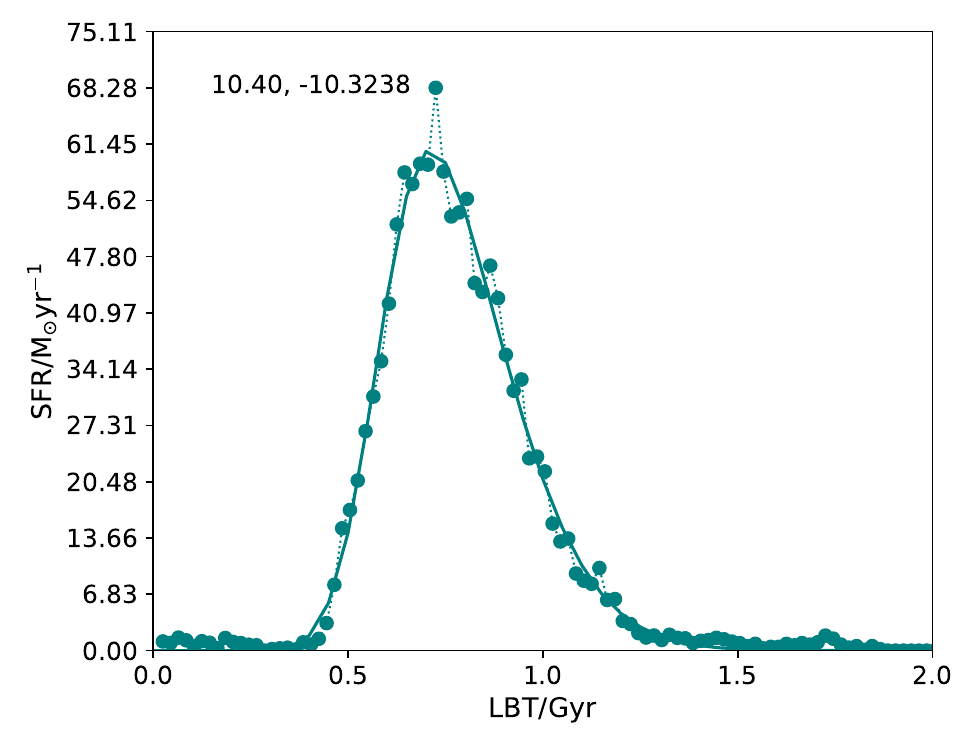}
    \includegraphics[trim=5mm 5mm 1.5mm 3mm, clip,width=0.3\textwidth]{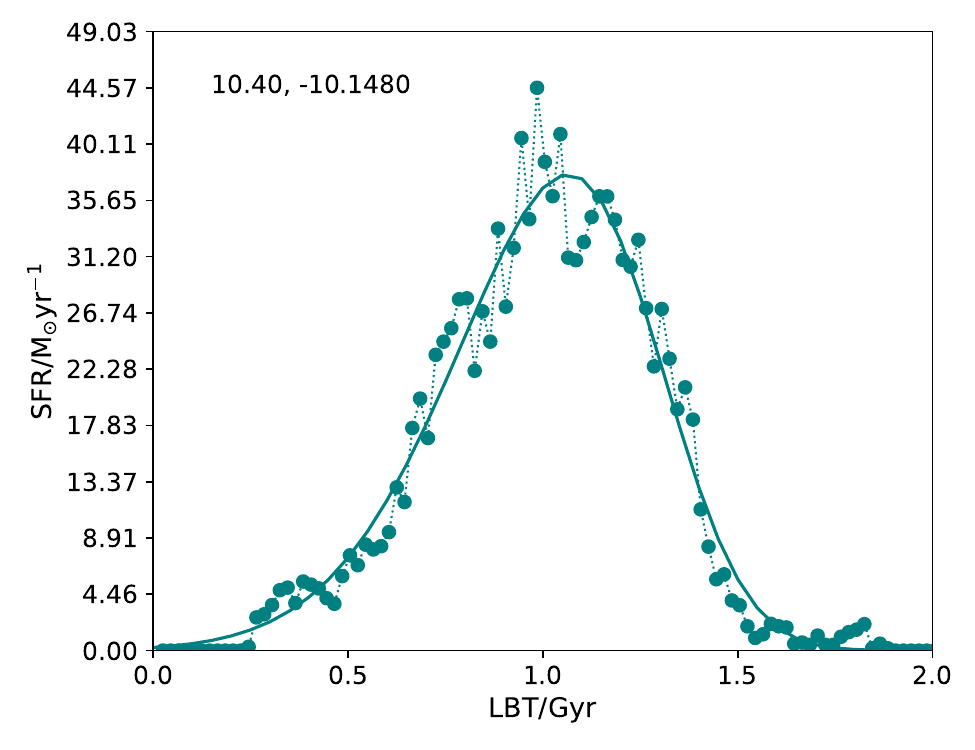}
    \includegraphics[trim=5mm 5mm 1.5mm 3mm, clip,width=0.3\textwidth]{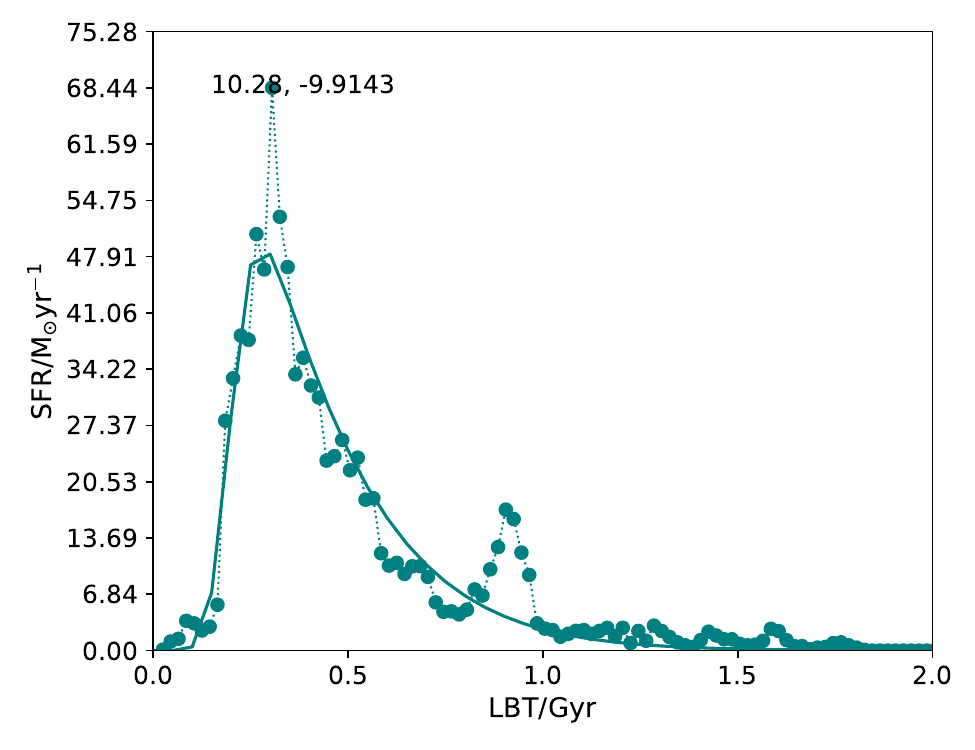}
    \includegraphics[trim=5mm 5mm 1.5mm 3mm, clip,width=0.3\textwidth]{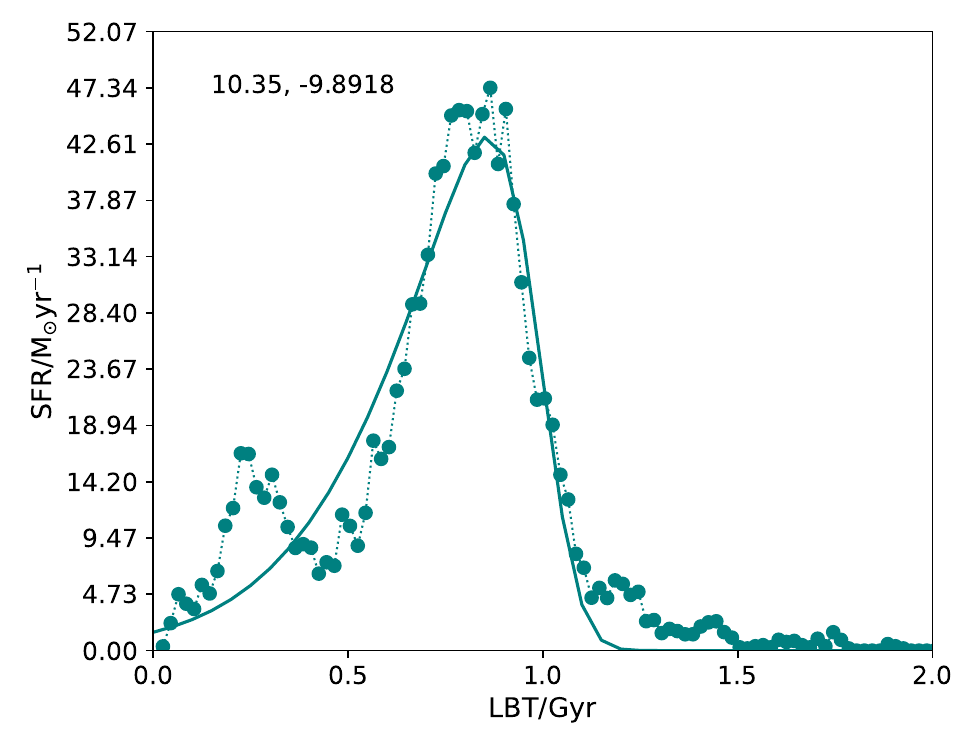}
    \includegraphics[trim=5mm 5mm 1.5mm 3mm, clip,width=0.3\textwidth]{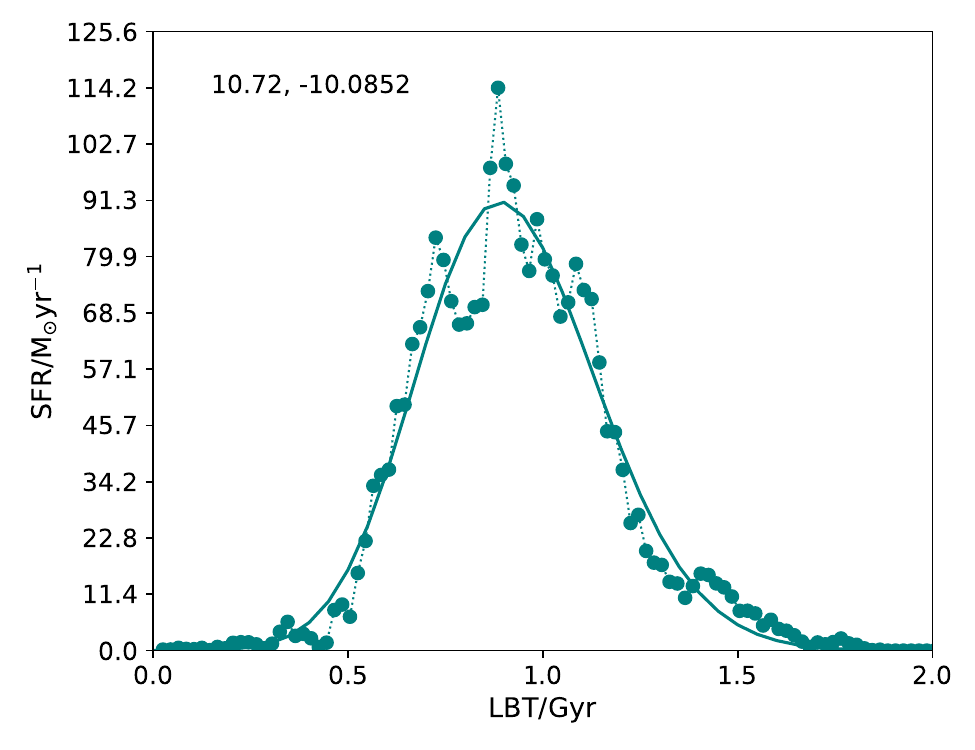}
    \includegraphics[trim=5mm 5mm 1.5mm 3mm, clip,width=0.3\textwidth]{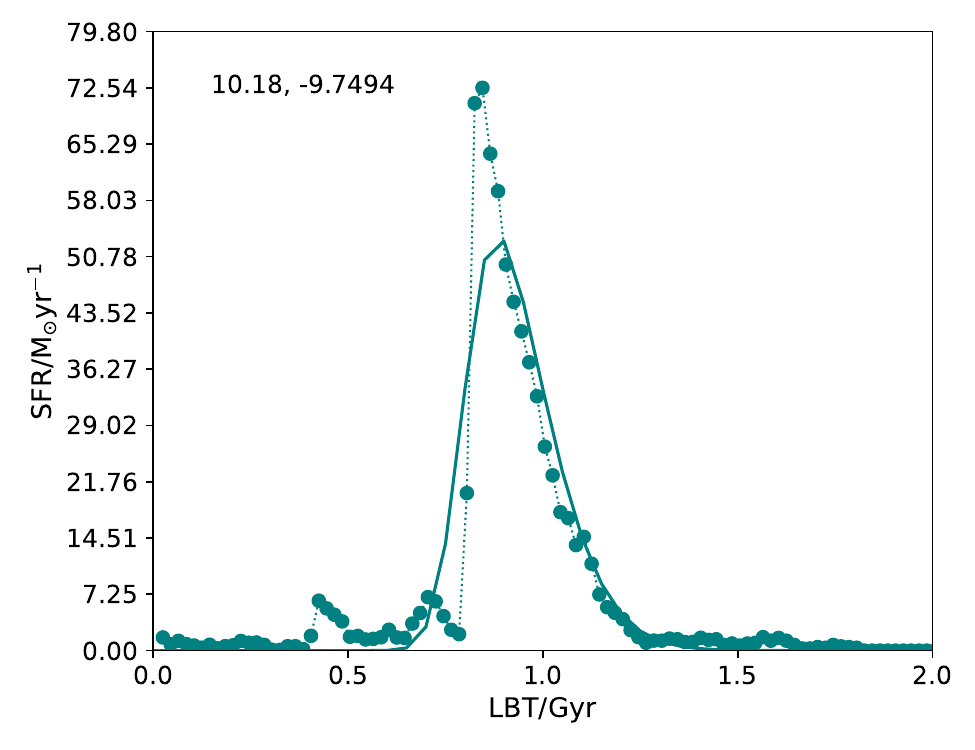}
    \includegraphics[trim=5mm 5mm 1.5mm 3mm, clip,width=0.3\textwidth]{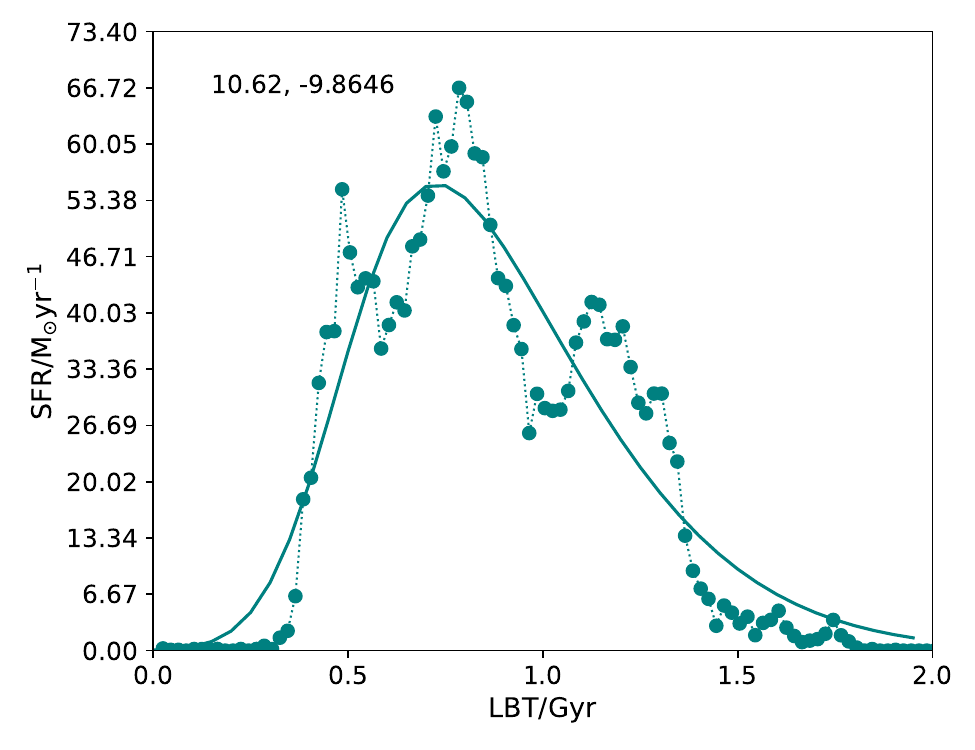}
    \includegraphics[trim=5mm 5mm 1.5mm 3mm, clip,width=0.3\textwidth]{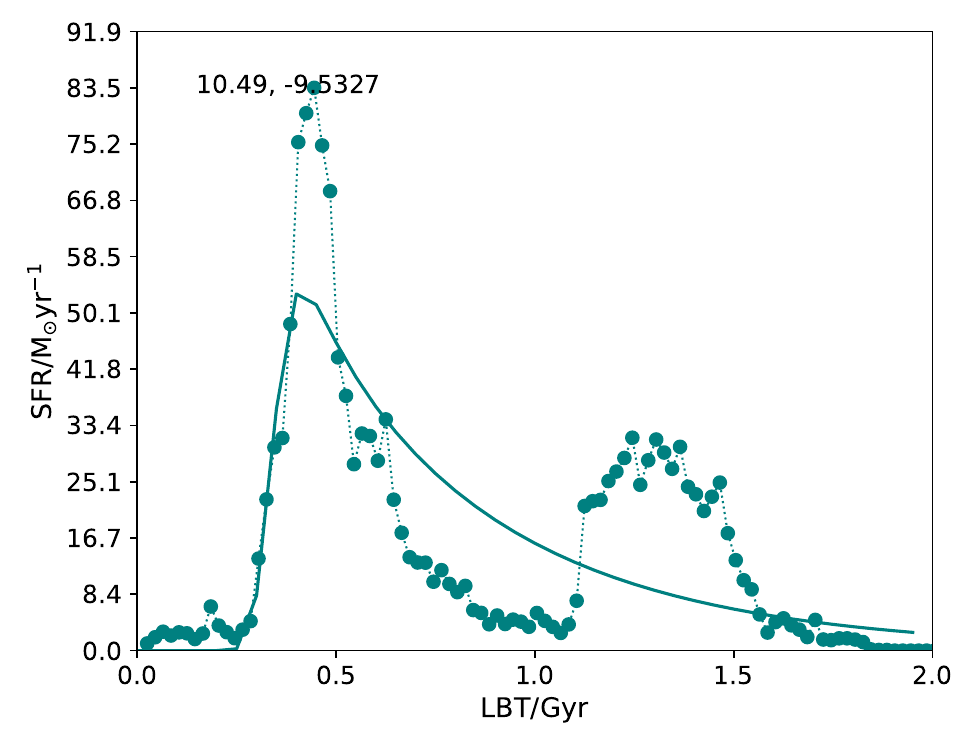}
    \caption{Examples of the star formation histories of $z=3$ massive quiescent galaxies in {\sc Eagle} (symbols connected with dotted lines) and the best skewed Gaussian fit (solid line). In each panel we show the $z=3$ stellar mass and ``goodness'' of fit, as defined in this Appendix. The top three panels show very good fits, while in the middle panel we show galaxies with reasonably good fits, which have small bumps, but which we do not consider as multi-starbursts. The bottom-left panel shows an example of a uni-modal SFHs that is poorly fit by a skewed Gaussian due to the very shark decrease after the peak. We reclassify those as being uni-modal despite the poor goodness of fit. The bottom middle and right panels show galaxies that have a poor fit due to them displaying clear multi-starbursts. The goodness threshold used in {\sc Eagle} to separate uni-model vs multiple starbursts is $-9.8$.}
    \label{SFHfitEAGLE}
\end{figure*}

\begin{figure*}
    \centering
    \includegraphics[trim=5mm 5mm 1.5mm 3mm, clip,width=0.3\textwidth]{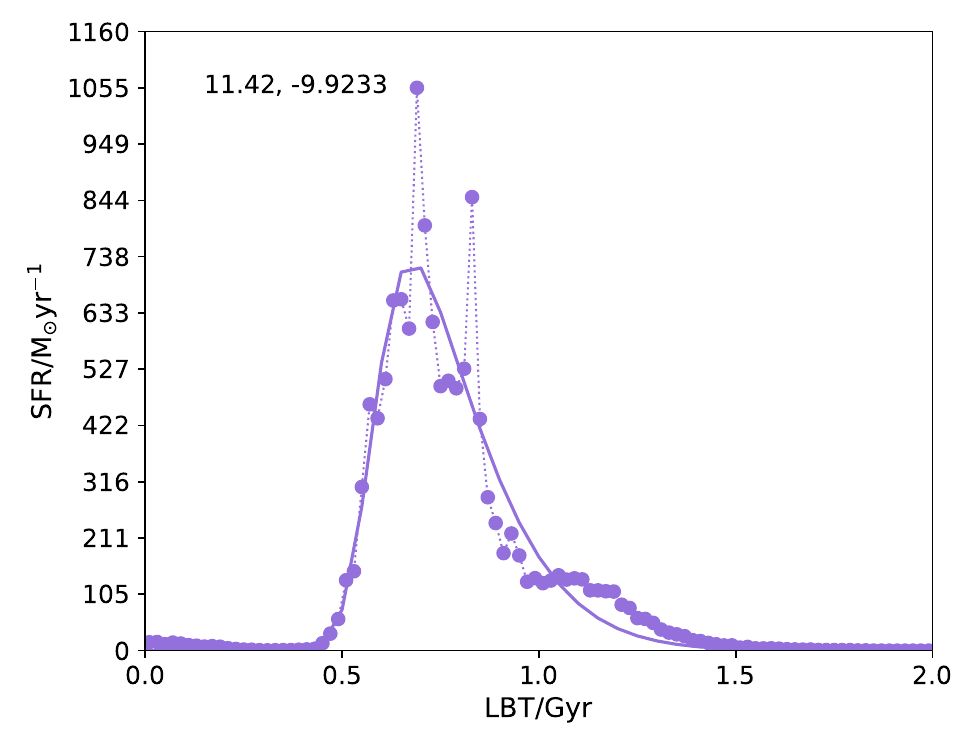}
    \includegraphics[trim=5mm 5mm 1.5mm 3mm, clip,width=0.3\textwidth]{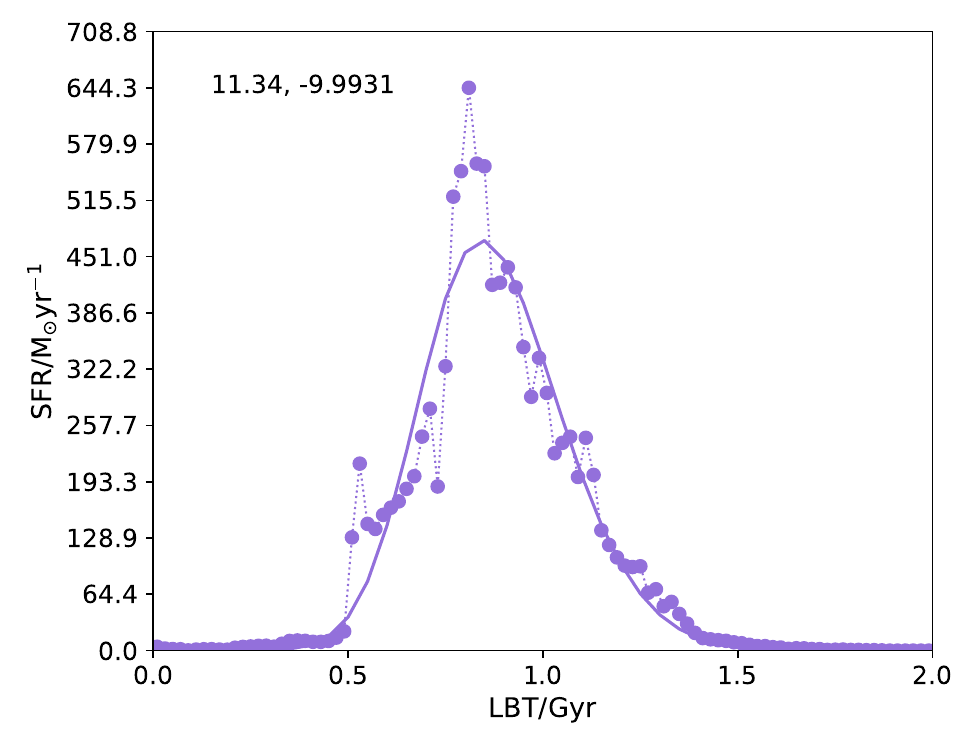}
    \includegraphics[trim=5mm 5mm 1.5mm 3mm, clip,width=0.3\textwidth]{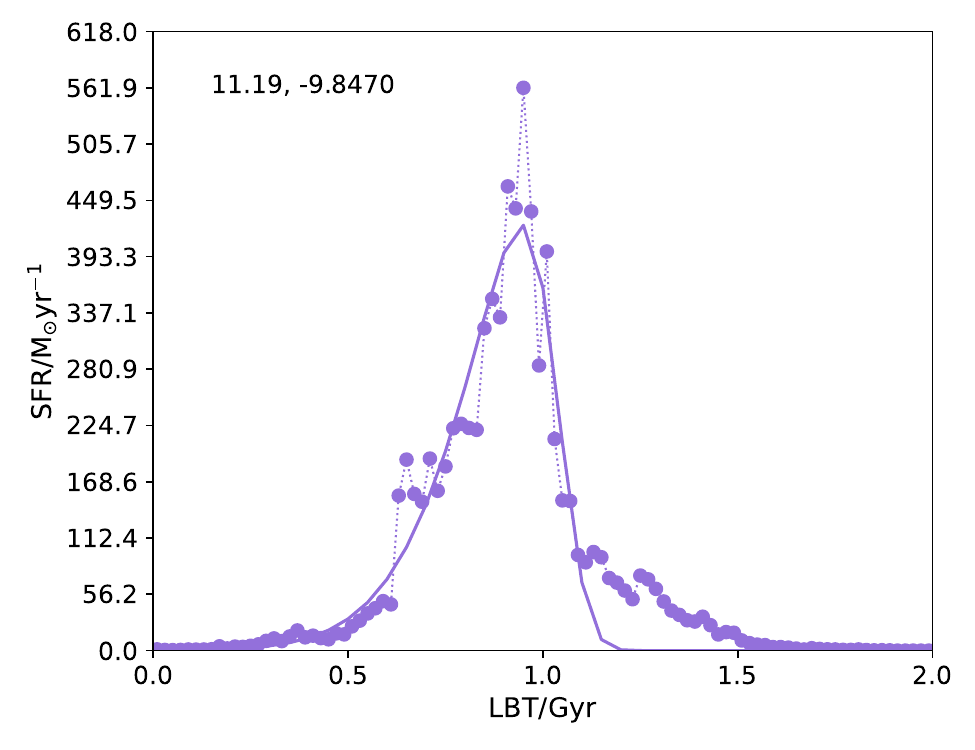}
    \includegraphics[trim=5mm 5mm 1.5mm 3mm, clip,width=0.3\textwidth]{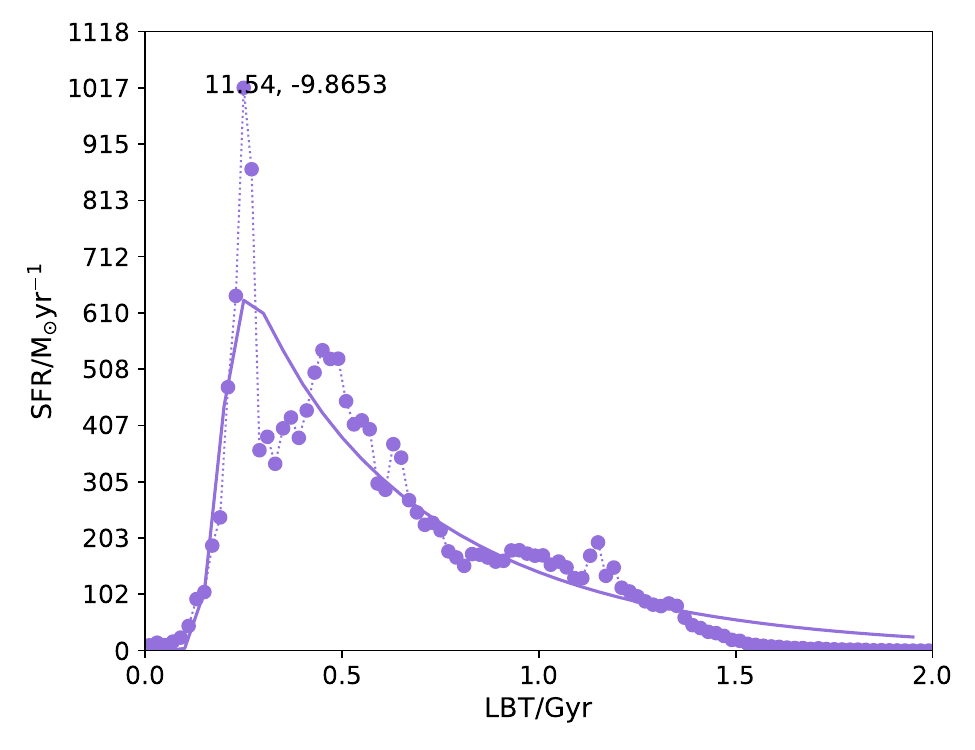}
    \includegraphics[trim=5mm 5mm 1.5mm 3mm, clip,width=0.3\textwidth]{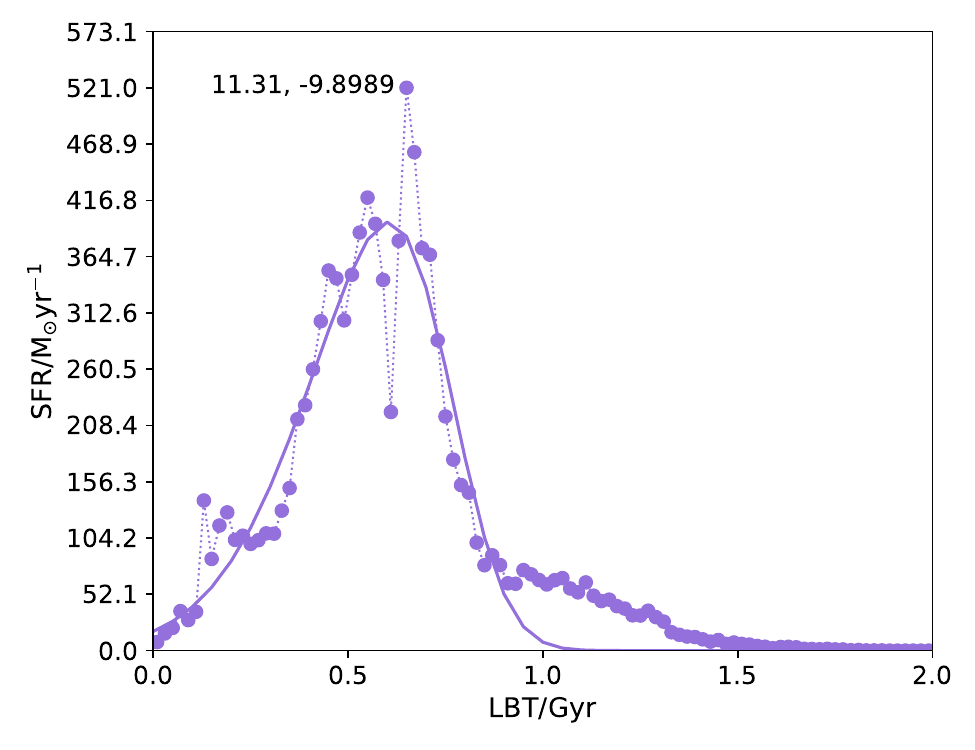}
    \includegraphics[trim=5mm 5mm 1.5mm 3mm, clip,width=0.3\textwidth]{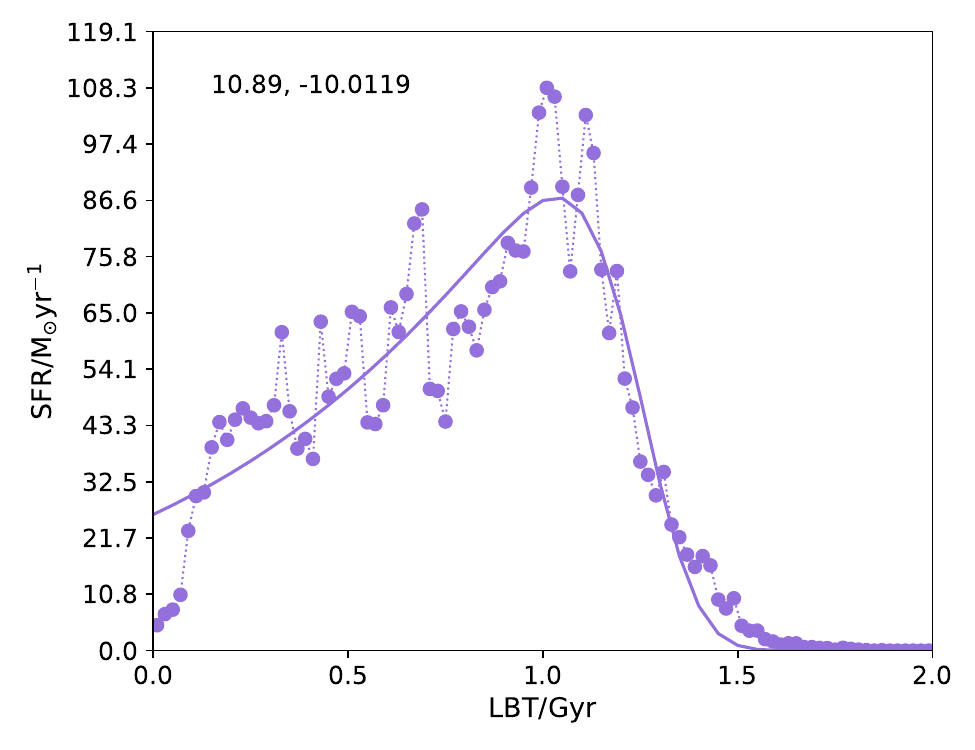}
    \includegraphics[trim=5mm 5mm 1.5mm 3mm, clip,width=0.3\textwidth]{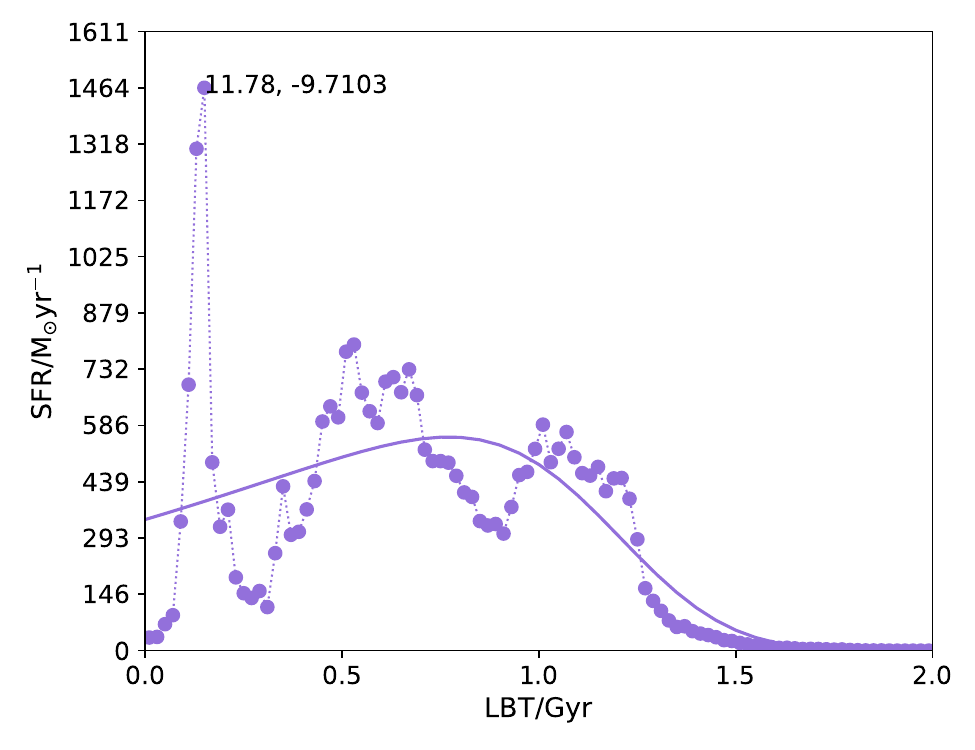}
    \includegraphics[trim=5mm 5mm 1.5mm 3mm, clip,width=0.3\textwidth]{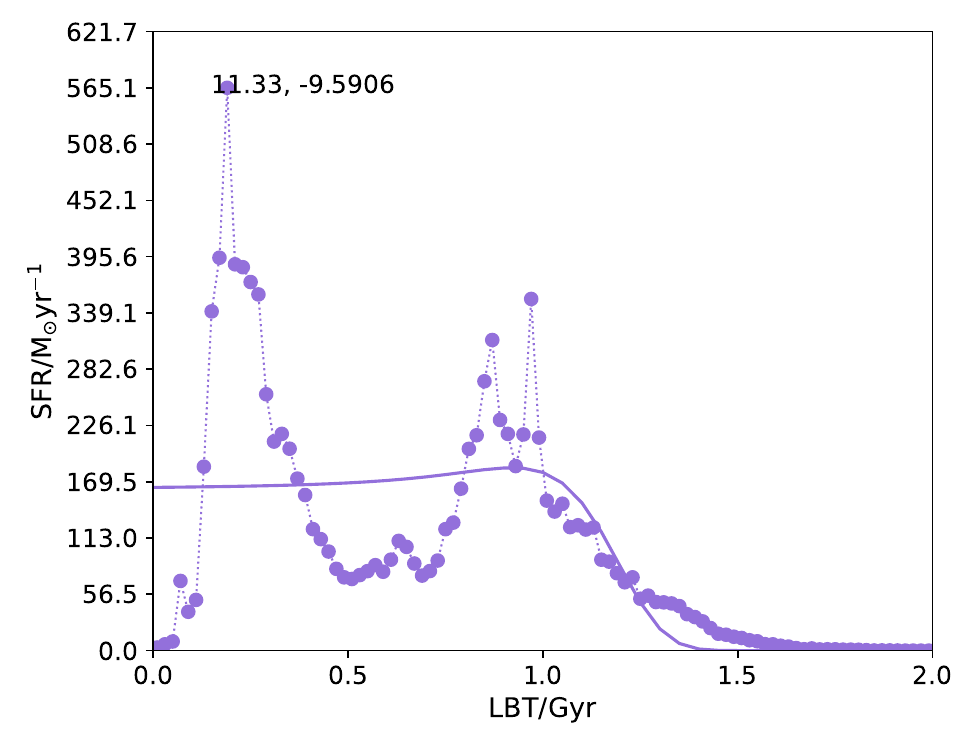}
    \includegraphics[trim=5mm 5mm 1.5mm 3mm, clip,width=0.3\textwidth]{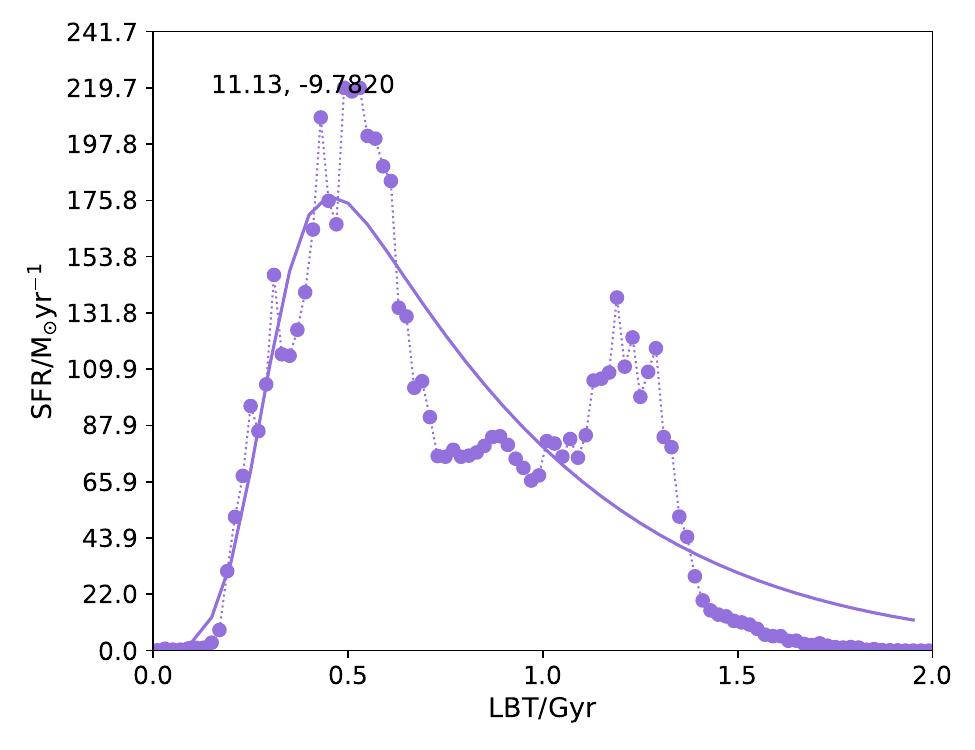}
    \caption{As in Fig.~\ref{SFHfitEAGLE} but for {\sc IllustrisTNG}. The top panel show very good fits, while the middle panels show galaxies with reasonably good fits and small bumps in their SFHs but that we do not consider as being clearly multiple starbursts. The bottom panels show examples of clear multiple starbursts. The goodness threshold used in {\sc IllustrisTNG} to separate uni-model vs multiple starbursts is $-9.83$.}
    \label{SFHfitTNG}
\end{figure*}

\begin{figure*}
    \centering
    \includegraphics[trim=5mm 5mm 1.5mm 3mm, clip,width=0.3\textwidth]{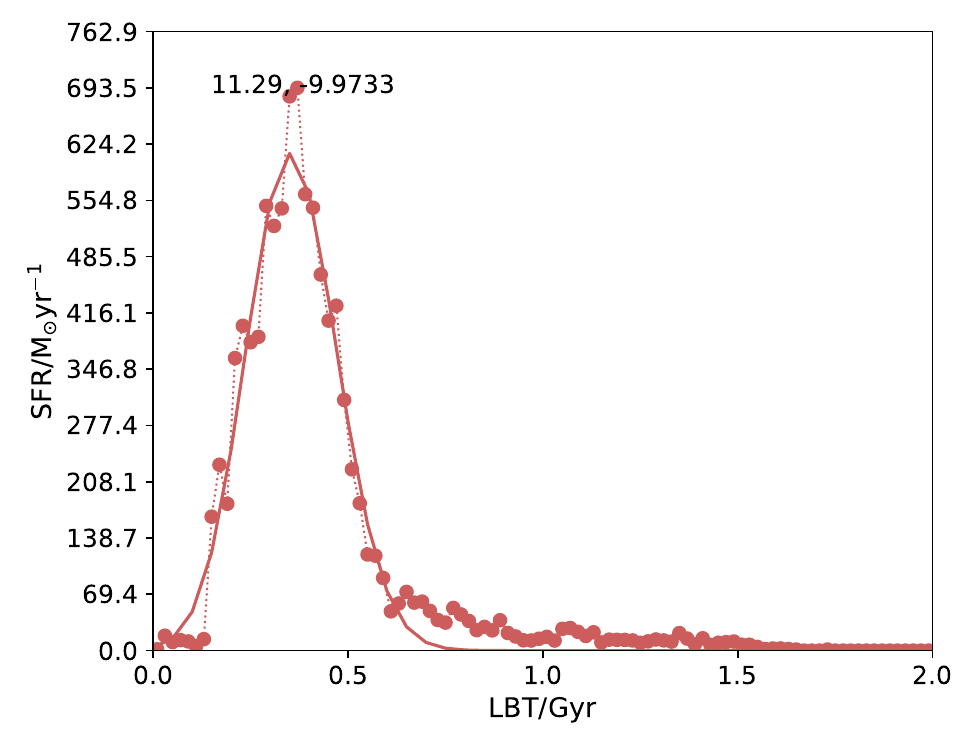}
    \includegraphics[trim=5mm 5mm 1.5mm 3mm, clip,width=0.3\textwidth]{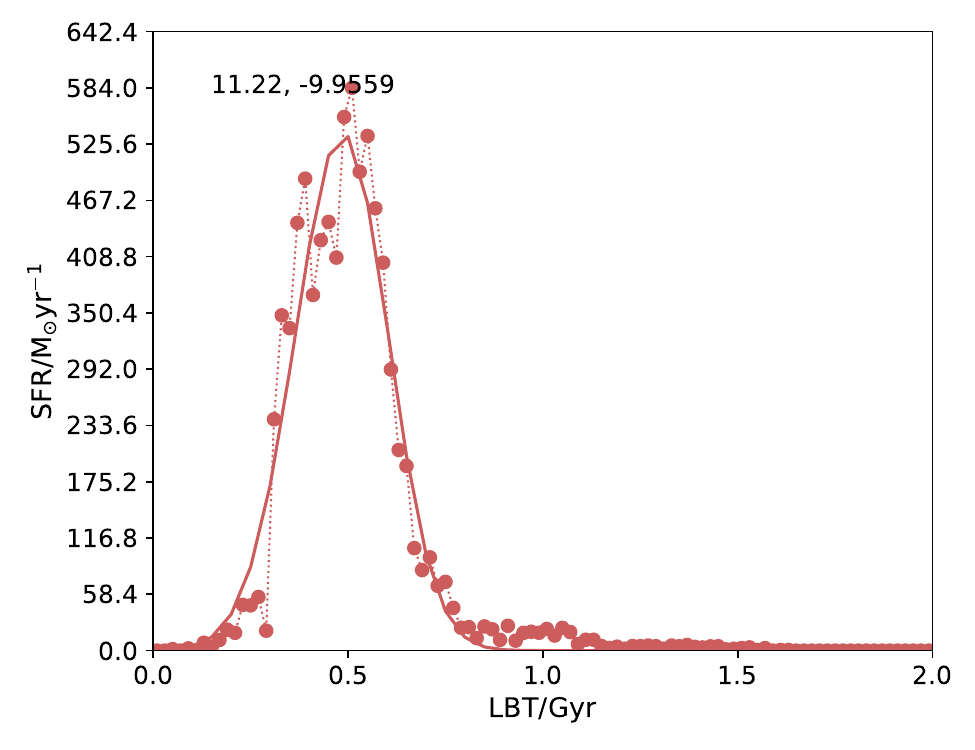}
    \includegraphics[trim=5mm 5mm 1.5mm 3mm, clip,width=0.3\textwidth]{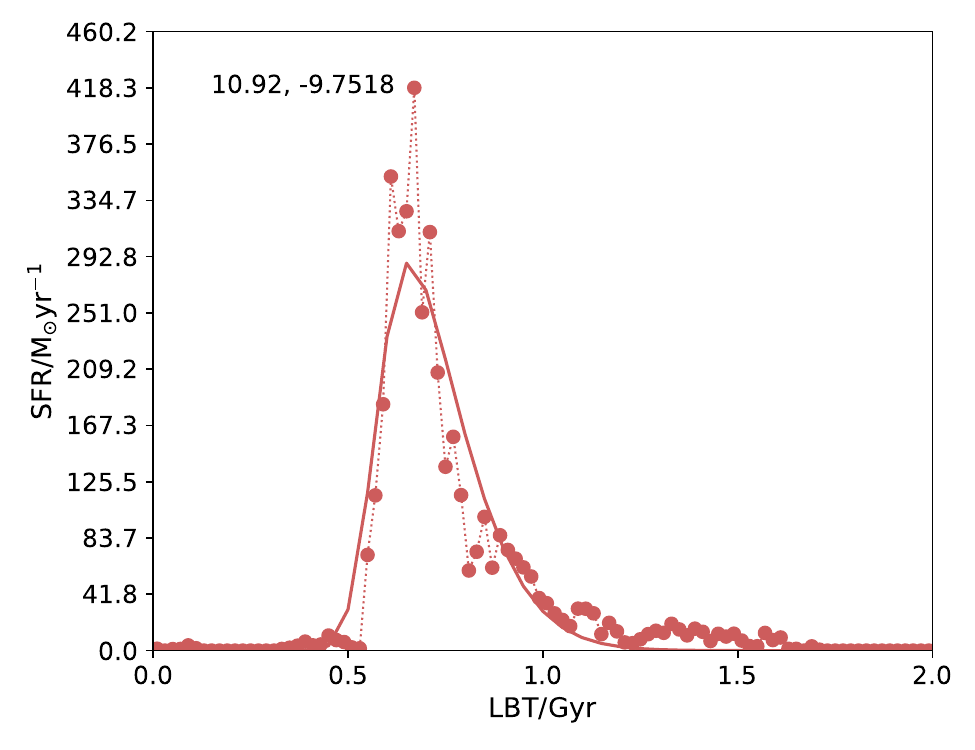}
    \includegraphics[trim=5mm 5mm 1.5mm 3mm, clip,width=0.3\textwidth]{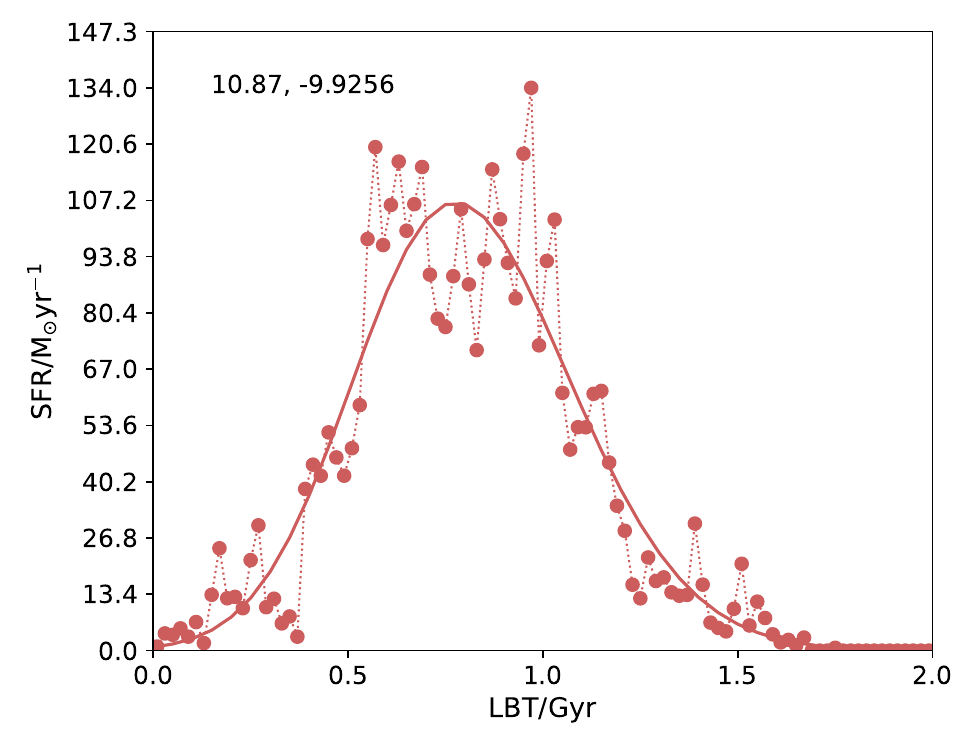}
    \includegraphics[trim=5mm 5mm 1.5mm 3mm, clip,width=0.3\textwidth]{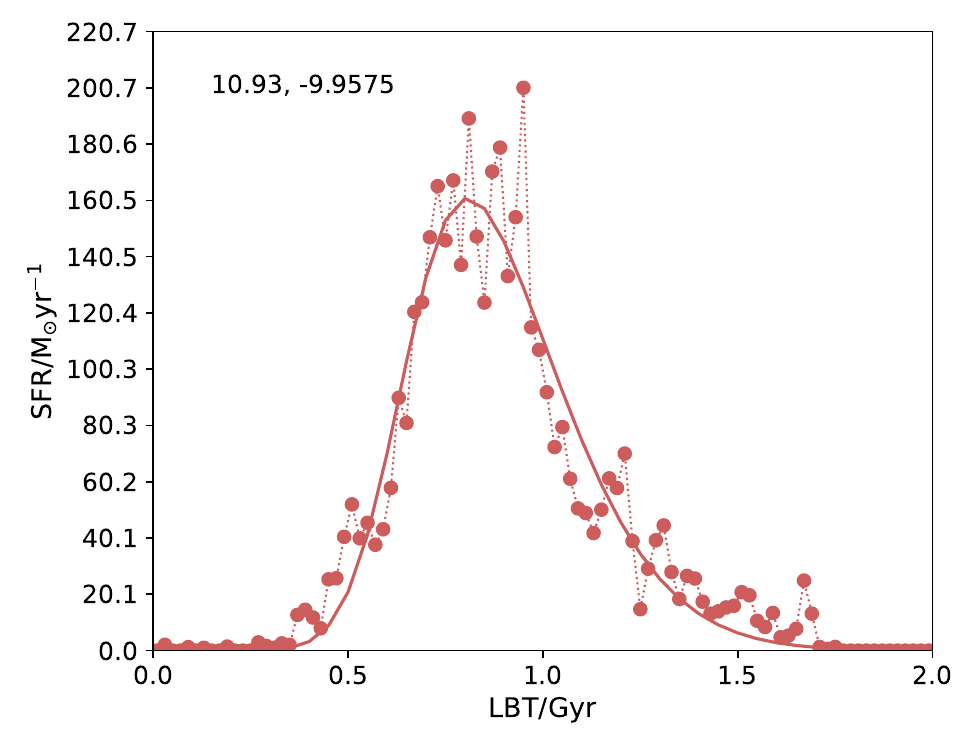}
    \includegraphics[trim=5mm 5mm 1.5mm 3mm, clip,width=0.3\textwidth]{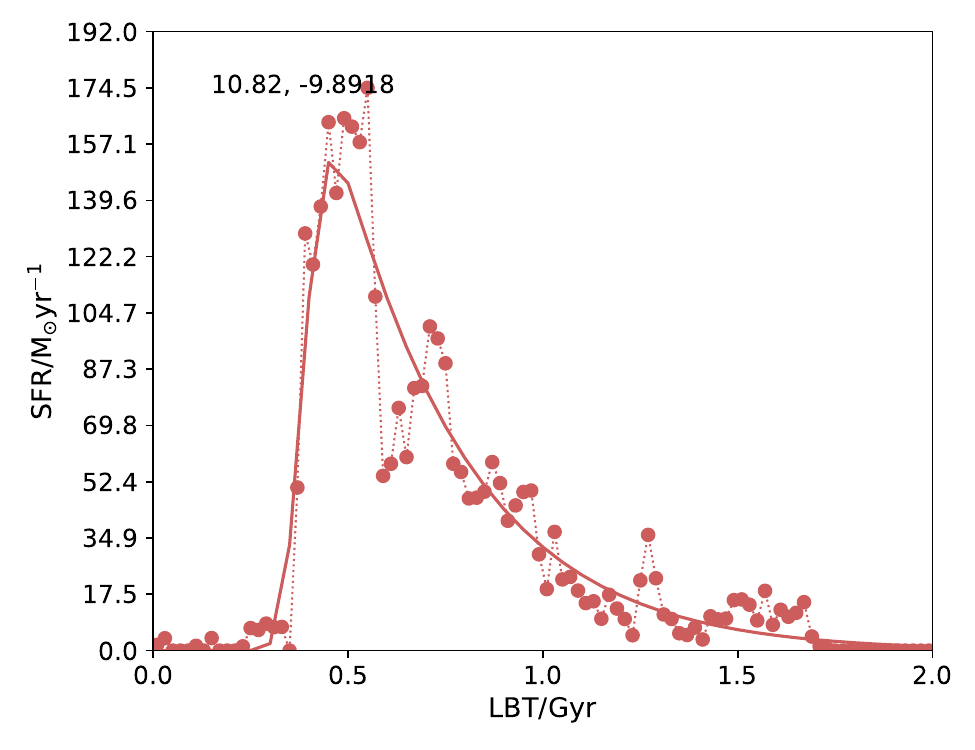}
    \includegraphics[trim=5mm 5mm 1.5mm 3mm, clip,width=0.3\textwidth]{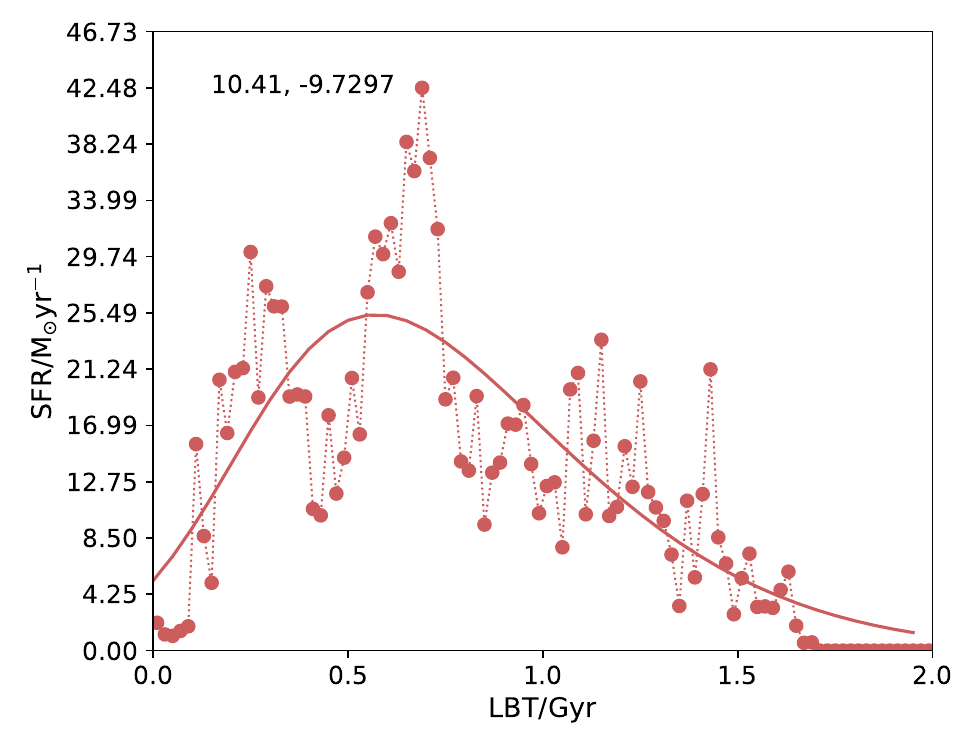}
    \includegraphics[trim=5mm 5mm 1.5mm 3mm, clip,width=0.3\textwidth]{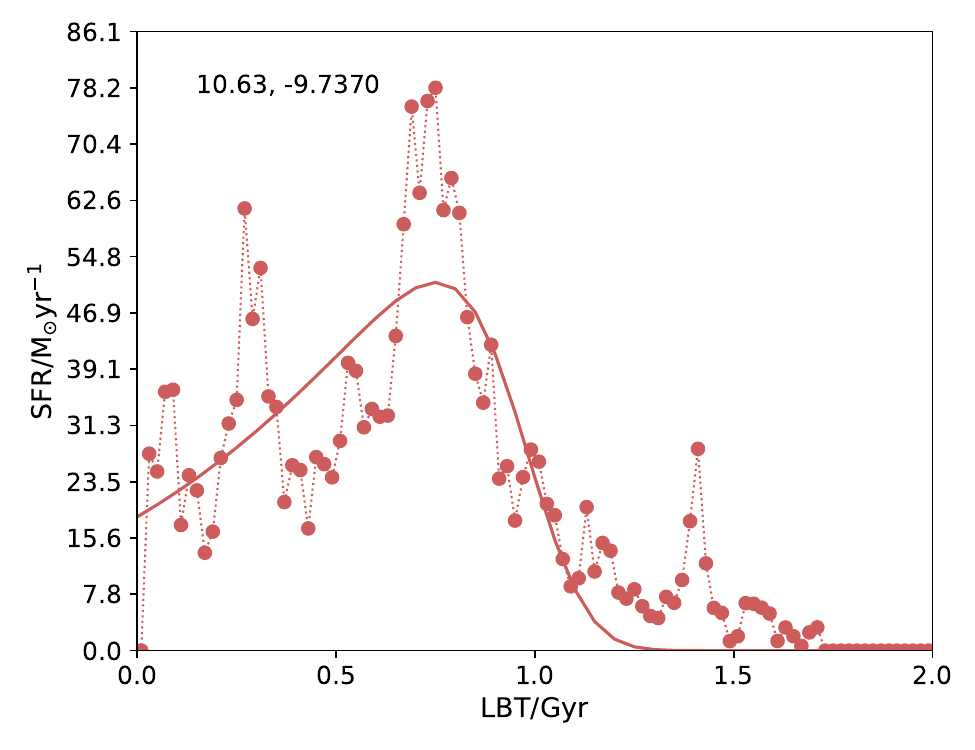}
    \includegraphics[trim=5mm 5mm 1.5mm 3mm, clip,width=0.3\textwidth]{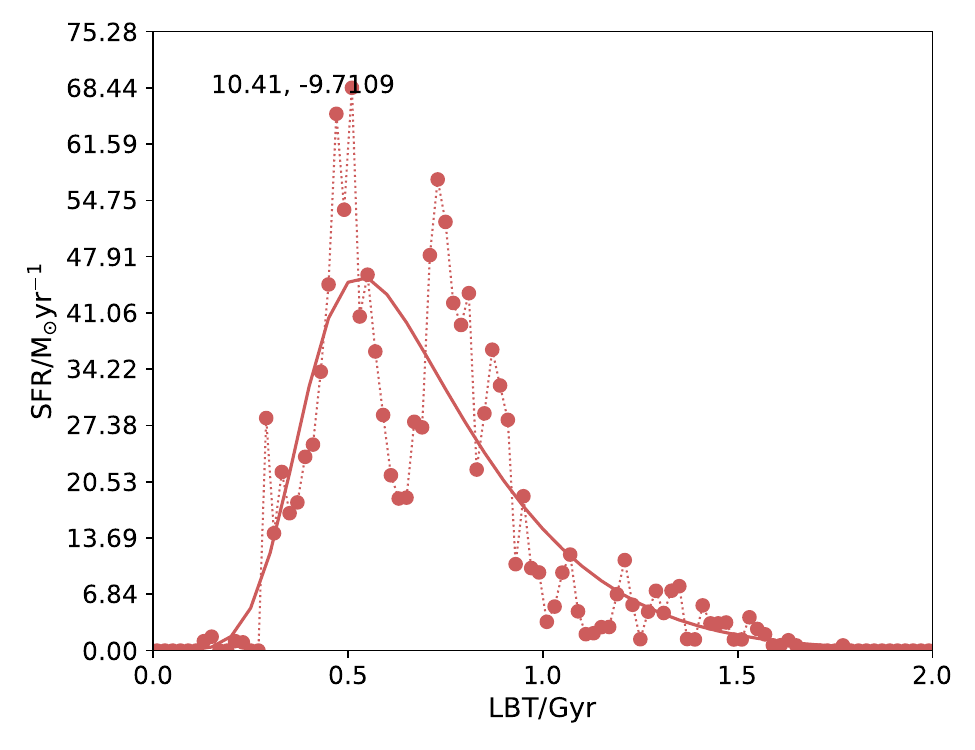}
    \caption{As in Fig.~\ref{SFHfitEAGLE} but for {\sc Simba}. The top panel show very good fits, while the middle panels show galaxies with reasonably good fits and small bumps in their SFHs but that we do not consider as being clear multiple starbursts. The bottom panels show examples of clear multiple starbursts. The goodness threshold used in {\sc Simba} to separate uni-model vs multiple starbursts is $-9.75$.}
    \label{SFHfitSimba}
\end{figure*}

\begin{figure*}
    \centering

    \includegraphics[trim=5mm 5mm 1.5mm 3mm, clip,width=0.3\textwidth]{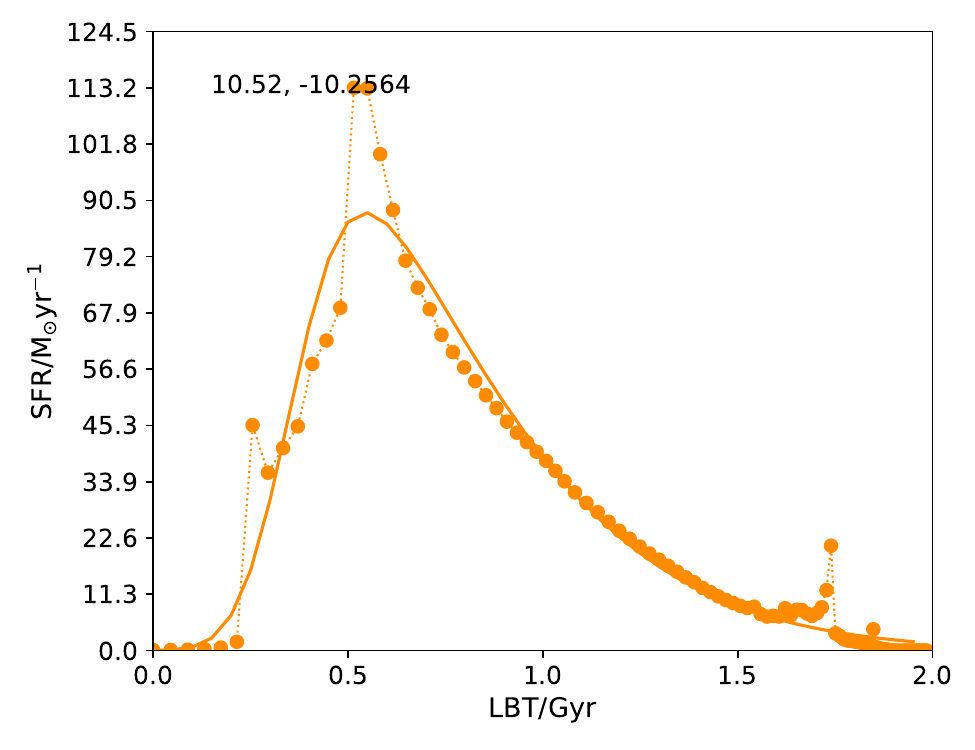}
    \includegraphics[trim=5mm 5mm 1.5mm 3mm, clip,width=0.3\textwidth]{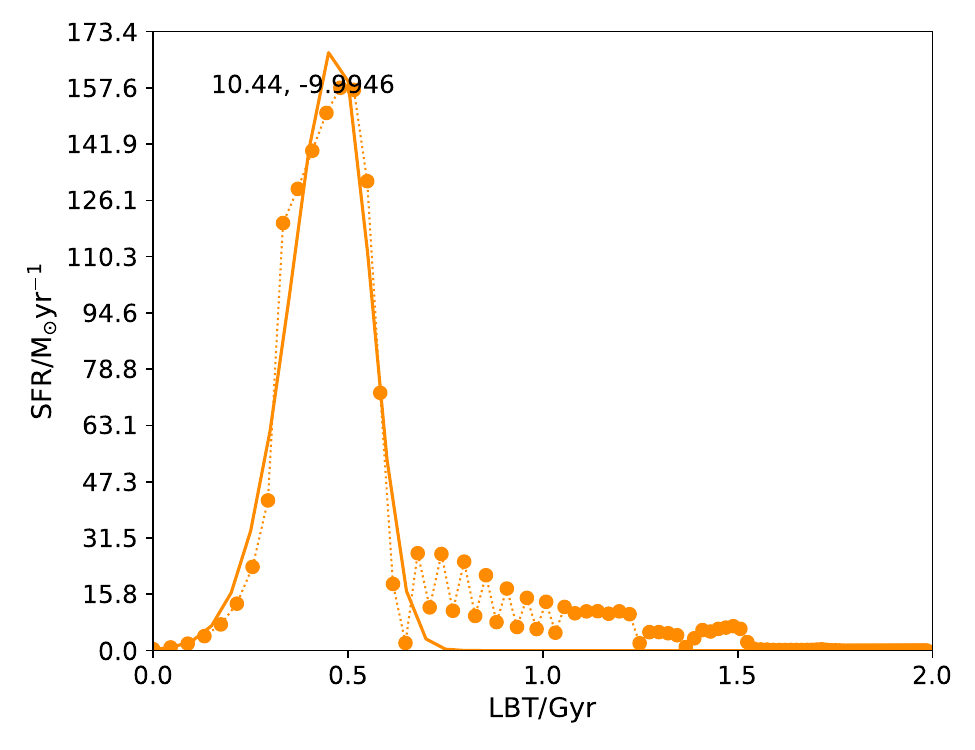}
    \includegraphics[trim=5mm 5mm 1.5mm 3mm, clip,width=0.3\textwidth]{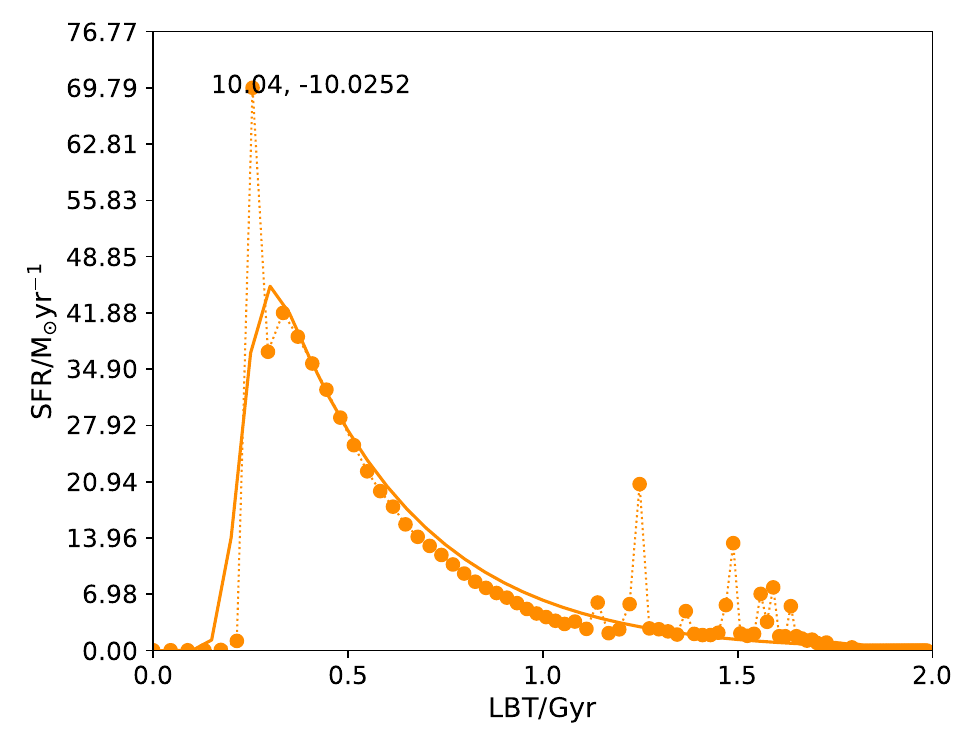}
    \includegraphics[trim=5mm 5mm 1.5mm 3mm, clip,width=0.3\textwidth]{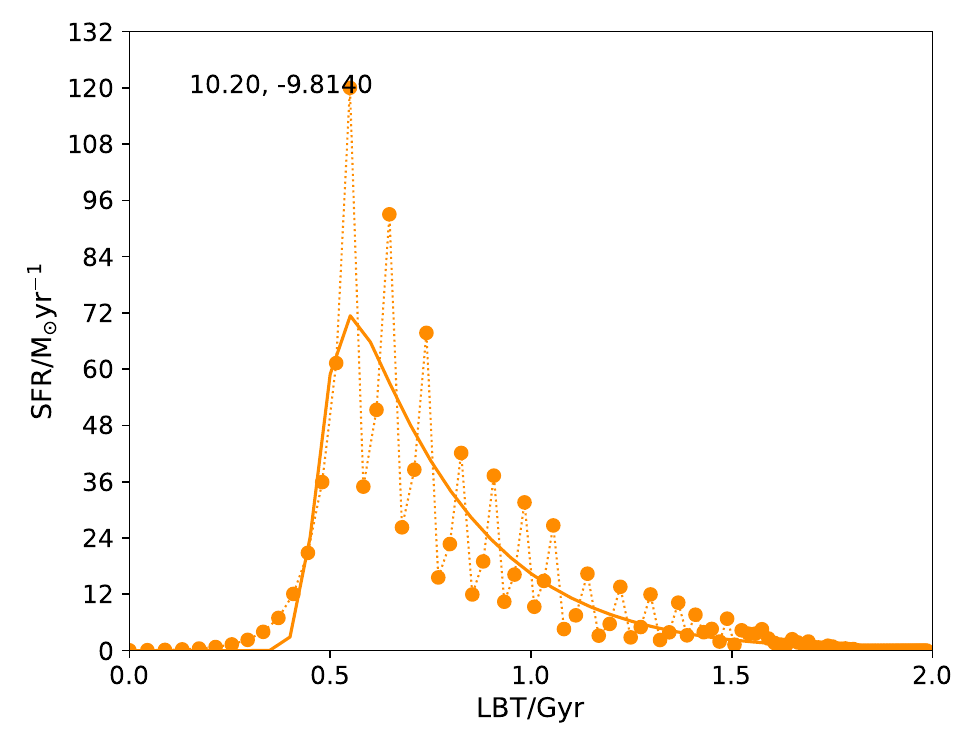}
    \includegraphics[trim=5mm 5mm 1.5mm 3mm, clip,width=0.3\textwidth]{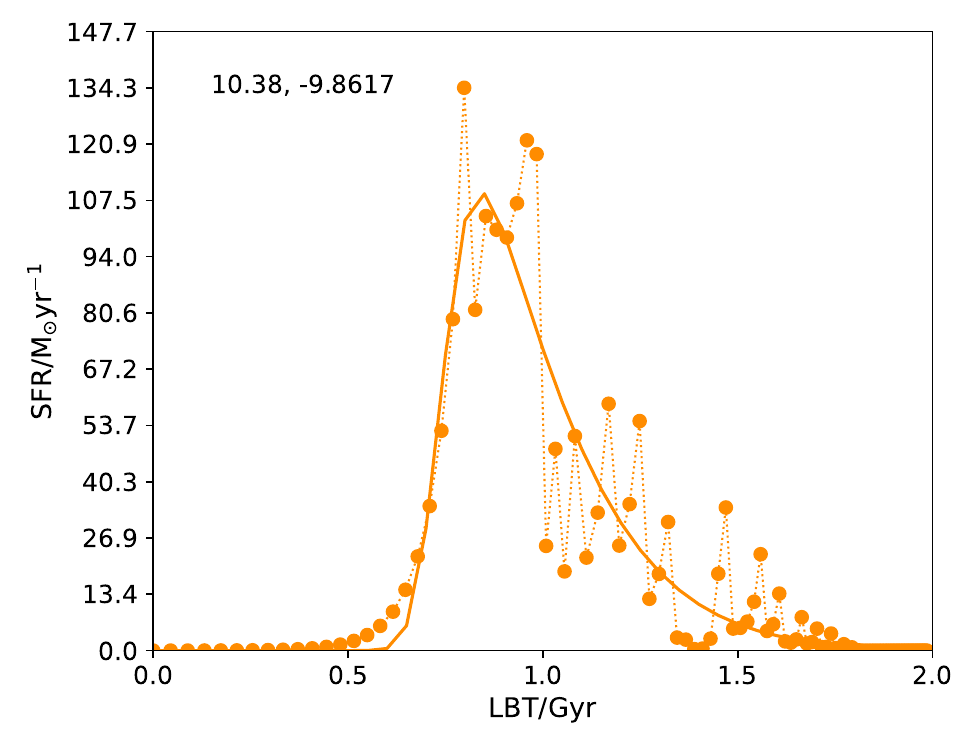}
    \includegraphics[trim=5mm 5mm 1.5mm 3mm, clip,width=0.3\textwidth]{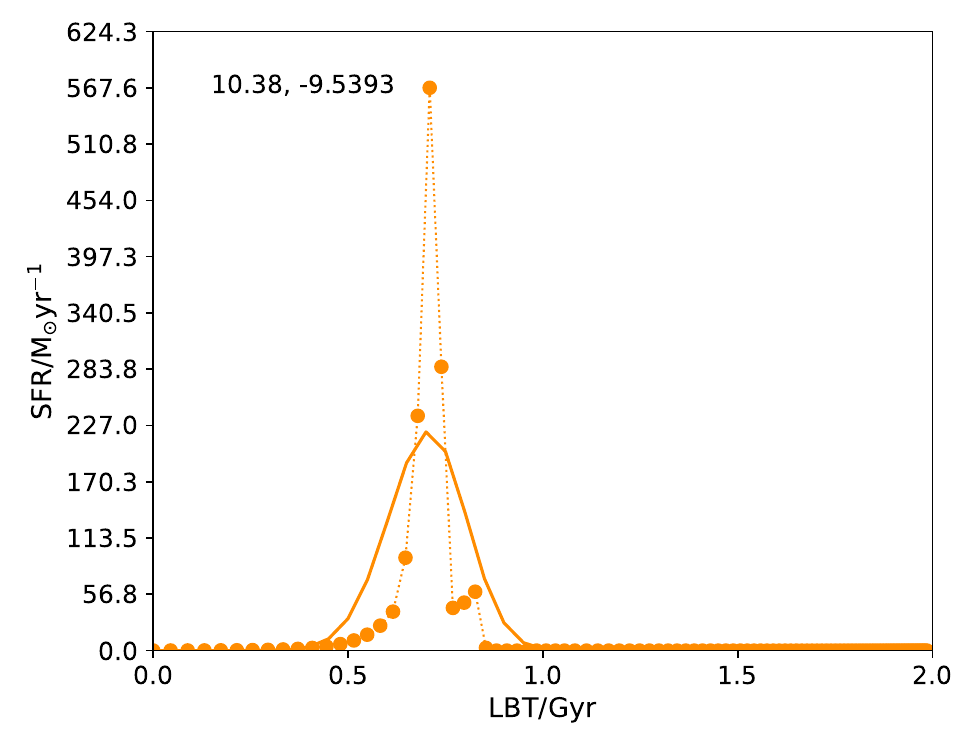}
    \includegraphics[trim=5mm 5mm 1.5mm 3mm, clip,width=0.3\textwidth]{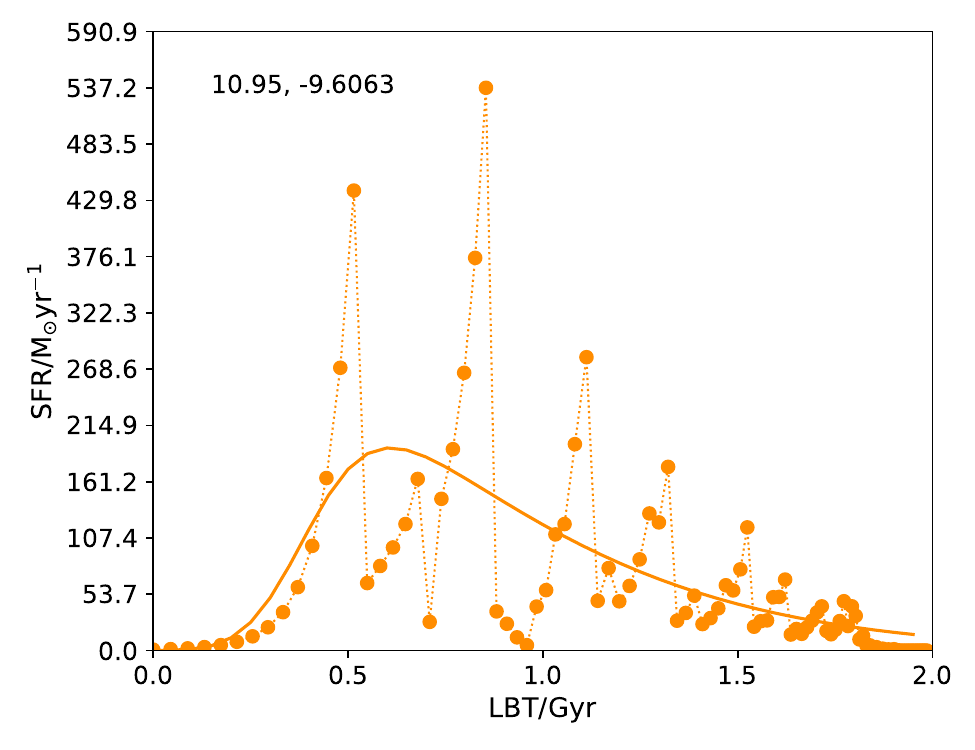}
    \includegraphics[trim=5mm 5mm 1.5mm 3mm, clip,width=0.3\textwidth]{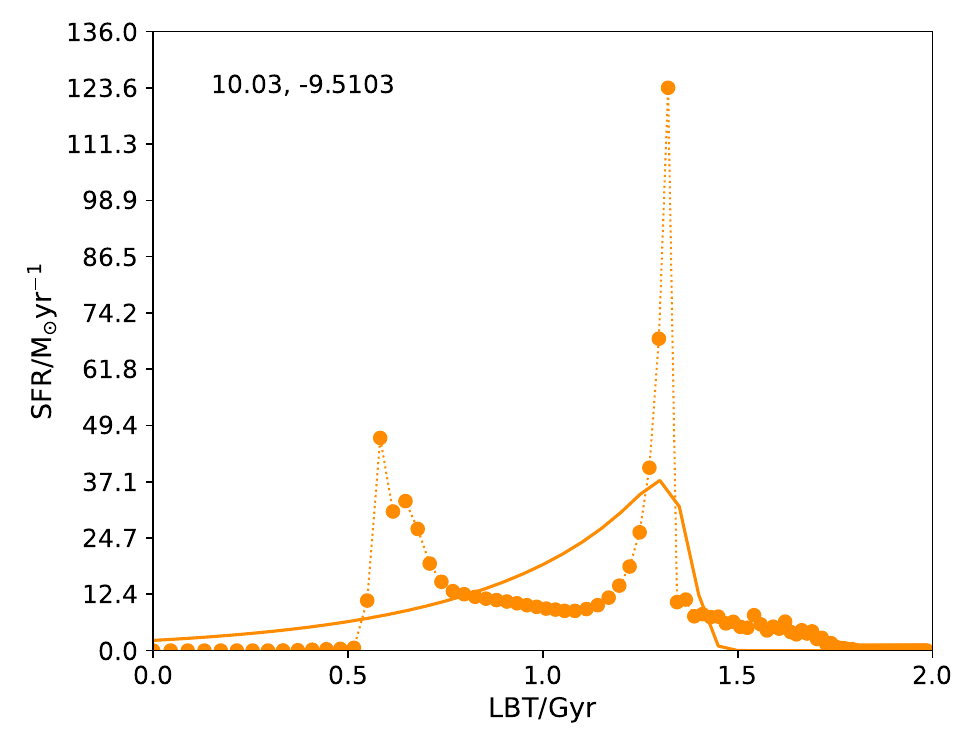}
    \includegraphics[trim=5mm 5mm 1.5mm 3mm, clip,width=0.3\textwidth]{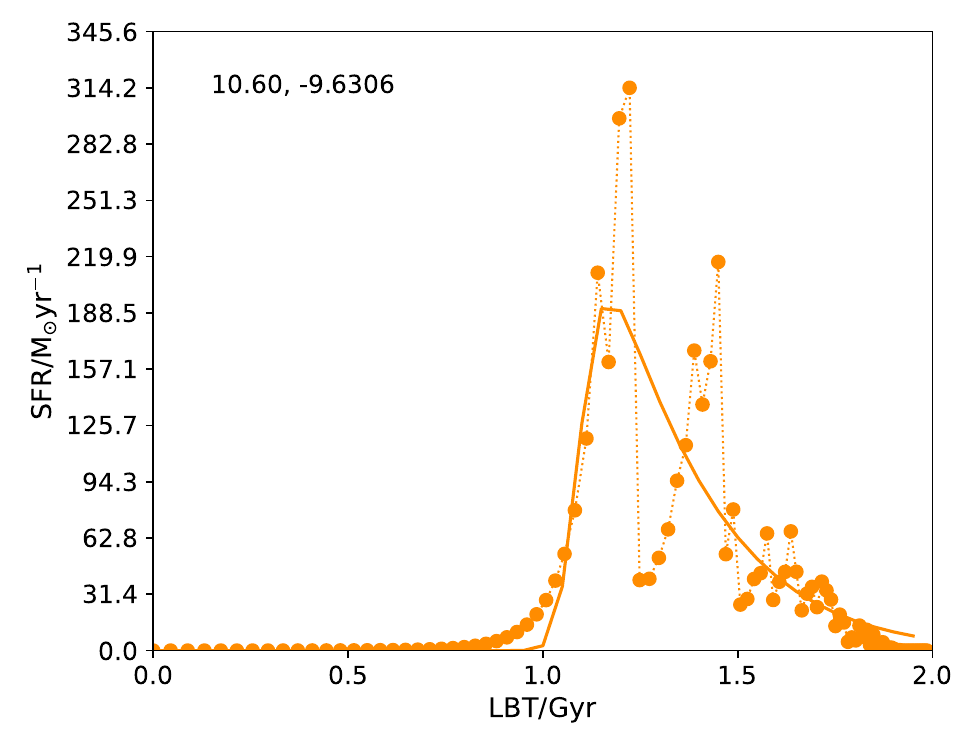}
    \caption{As in Fig.~\ref{SFHfitEAGLE} but for \shark. The top $3$ panels show very good fits. The middle left and middle panels show reasonably good fits, with the panel on the left showing a common ``saw''-like feature which corresponds to a natural stochasticity in the model. We still classify these as being uni-modal SFHs. The middle-right panel shows an example of a uni-modal SFH that is poorly fit. This happens because the burst is extremely short in duration and is not easily modelled with a skewed Gaussian. We reclassify all of those as being uni-modal despite the goodness of fit being poor. The bottom panels show clear examples of multiple starbursts that were found thanks to the goodness of fit being poor. For \shark, we used a goodness of fit threshold of $-9.7$ to separate SFHs between uni-modal and multi-bursts (a bit lower than the other simulations).}
    \label{SFHfitSHARK}
\end{figure*}

\begin{figure*}
    \centering
    \includegraphics[trim=5mm 5mm 1.5mm 3mm, clip,width=0.3\textwidth]{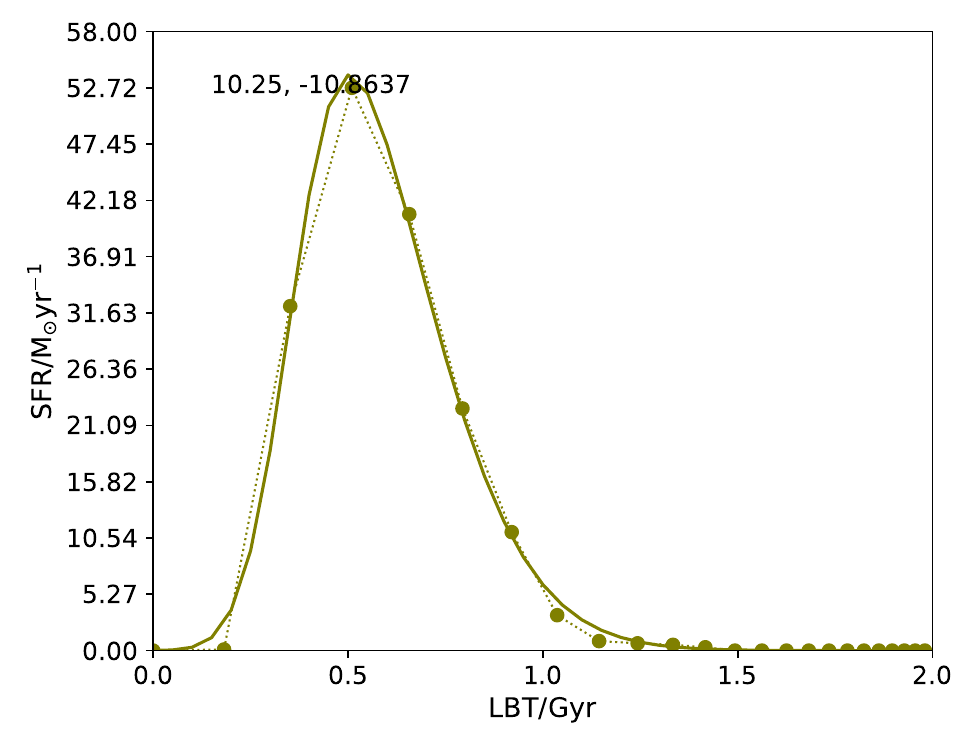}
    \includegraphics[trim=5mm 5mm 1.5mm 3mm, clip,width=0.3\textwidth]{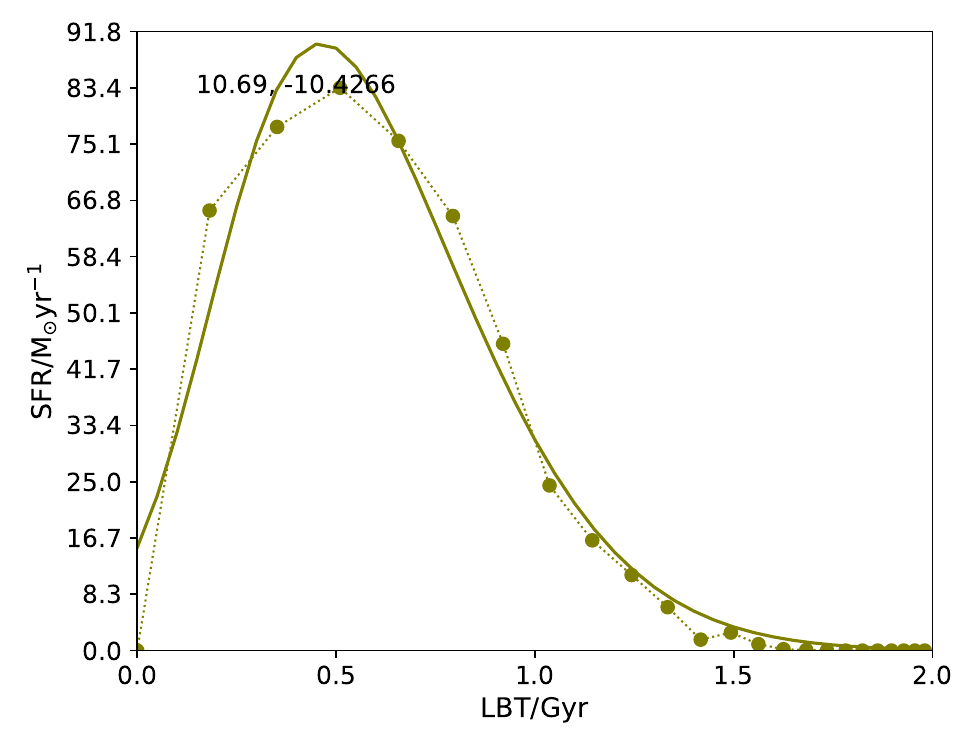}
    \includegraphics[trim=5mm 5mm 1.5mm 3mm, clip,width=0.3\textwidth]{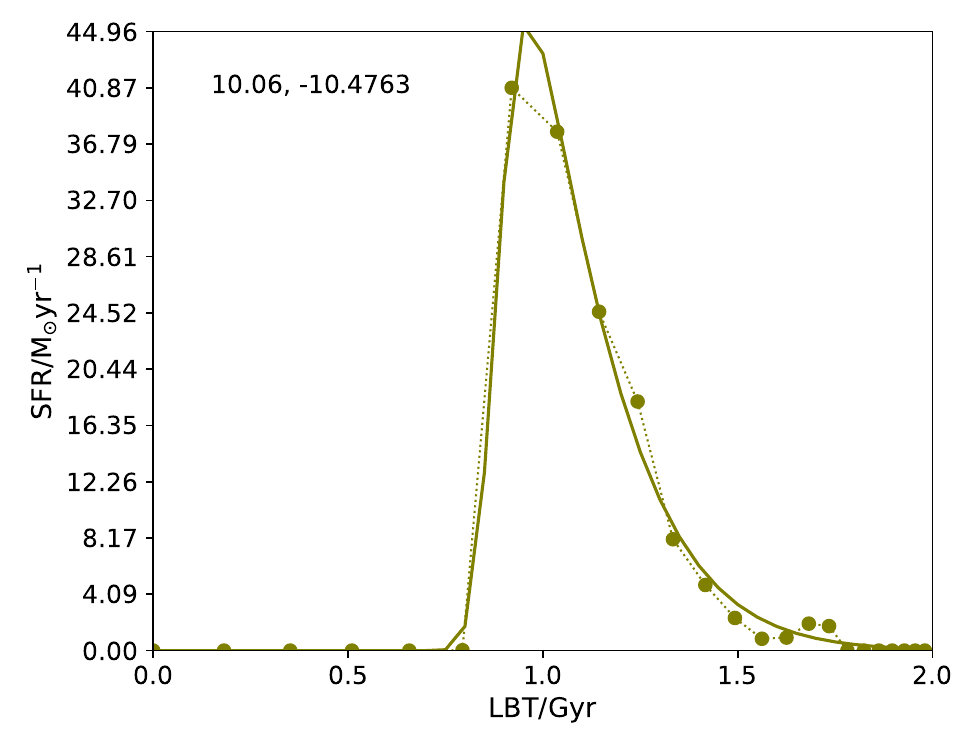}
    \includegraphics[trim=5mm 5mm 1.5mm 3mm, clip,width=0.3\textwidth]{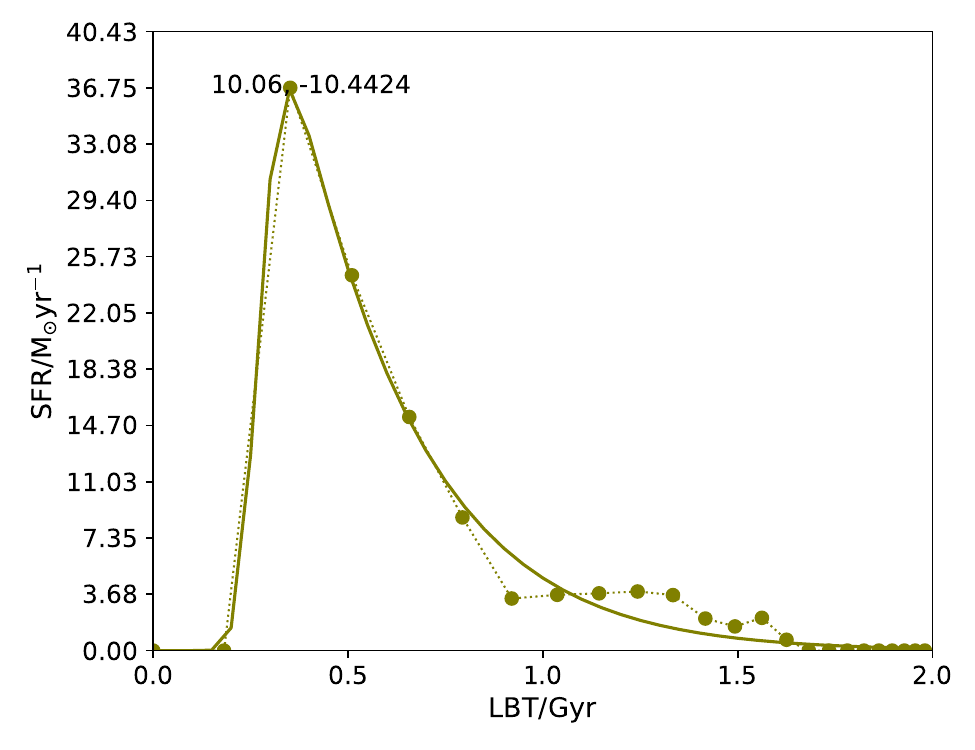}
    \includegraphics[trim=5mm 5mm 1.5mm 3mm, clip,width=0.3\textwidth]{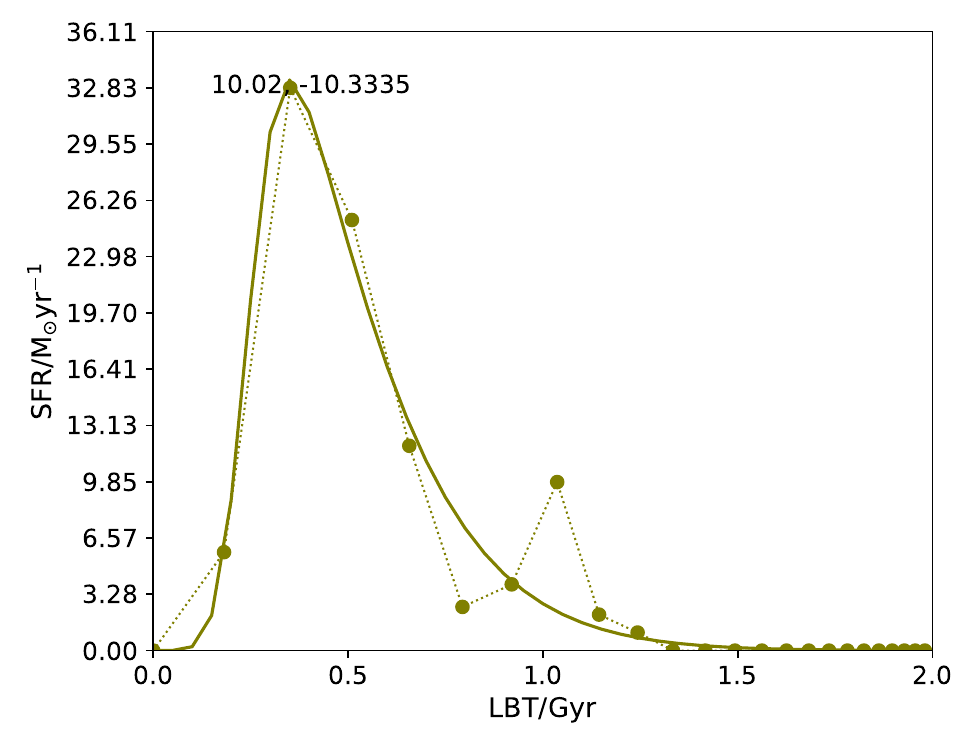}
    \includegraphics[trim=5mm 5mm 1.5mm 3mm, clip,width=0.3\textwidth]{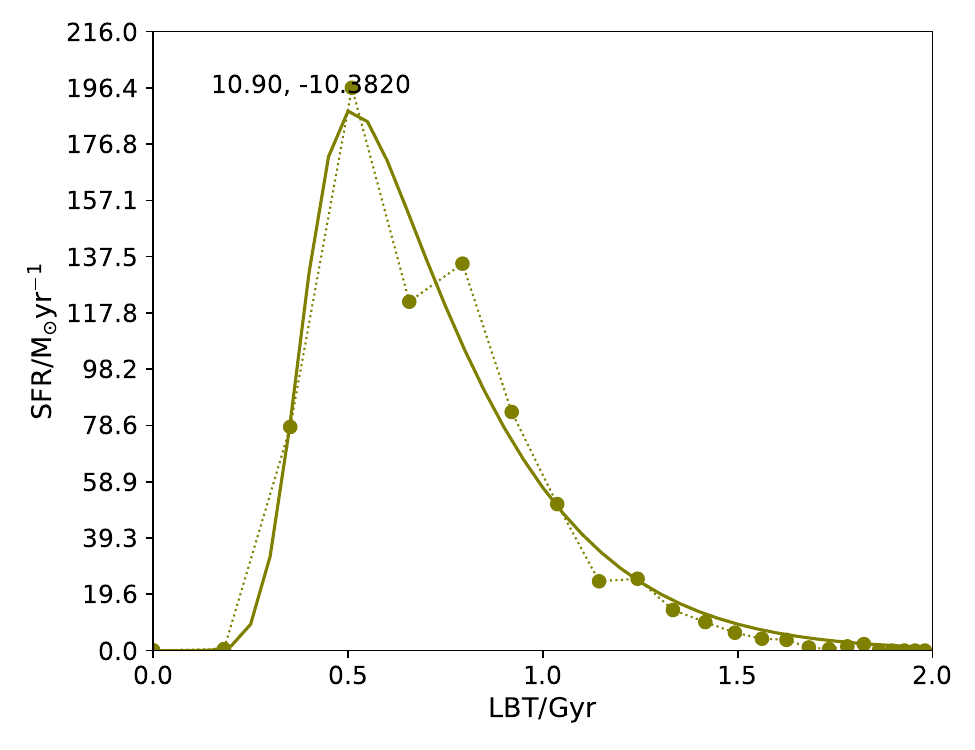}
    \includegraphics[trim=5mm 5mm 1.5mm 3mm, clip,width=0.3\textwidth]{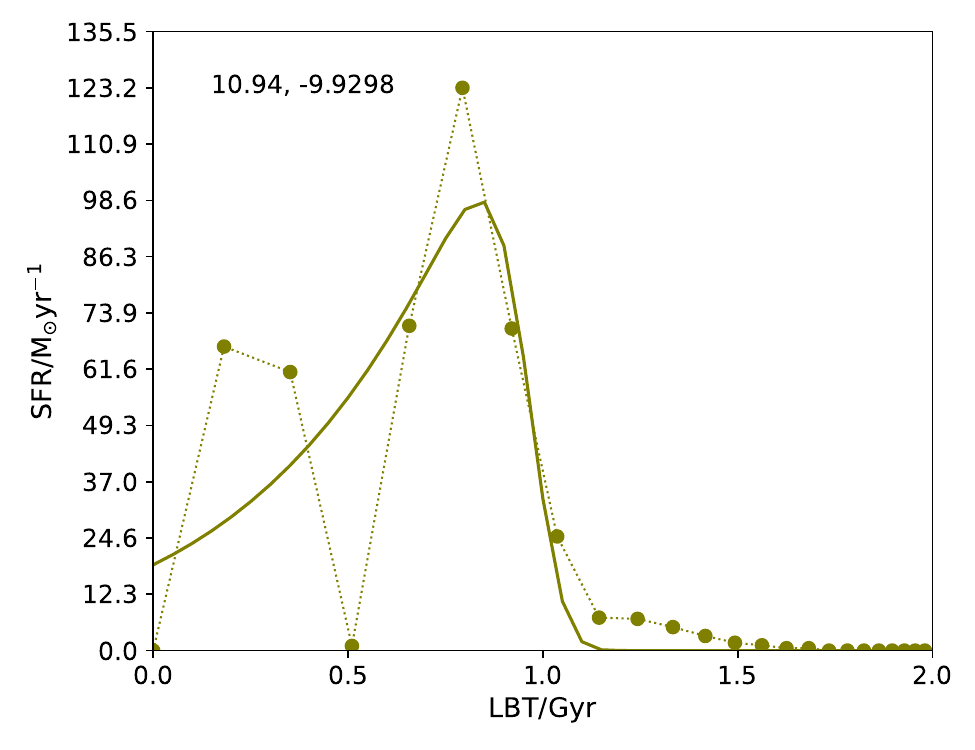}
    \includegraphics[trim=5mm 5mm 1.5mm 3mm, clip,width=0.3\textwidth]{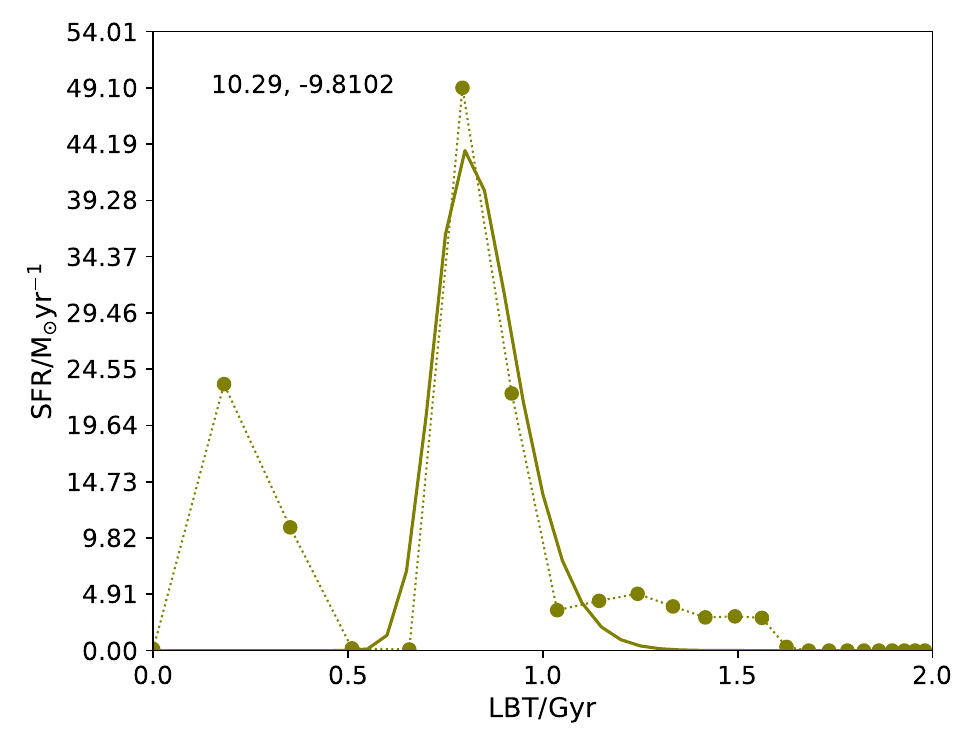}
    \includegraphics[trim=5mm 5mm 1.5mm 3mm, clip,width=0.3\textwidth]{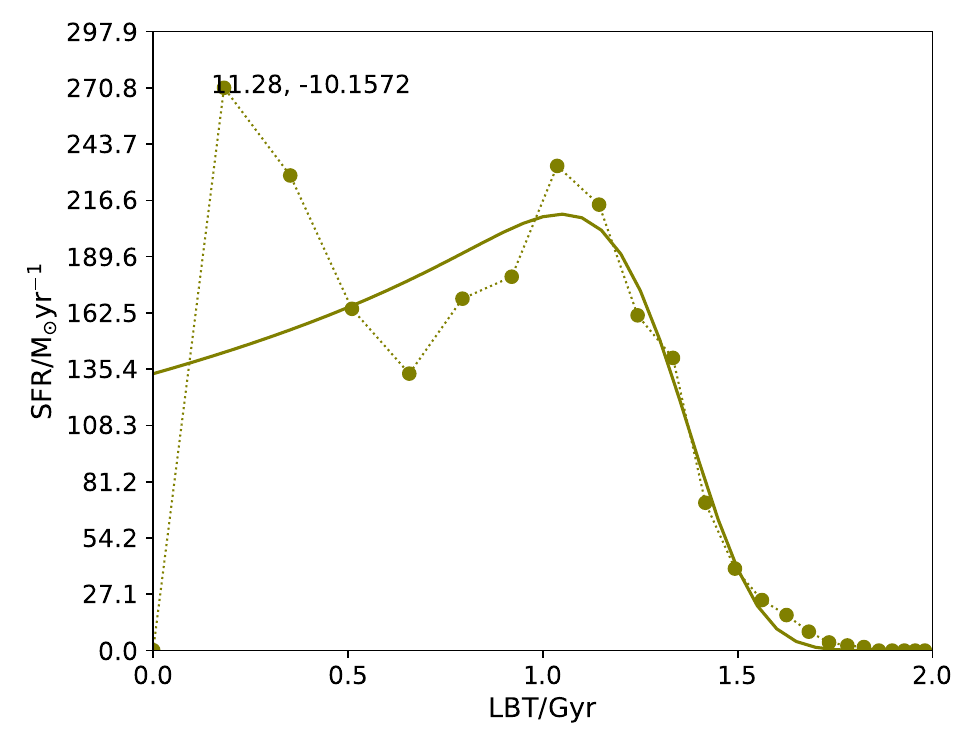}
    \caption{As in Fig.~\ref{SFHfitEAGLE} but for {\sc GAEA}. The top panel show very good fits, while the middle panels show galaxies with reasonably good fits and small bumps in their SFHs but that we do not consider as being clear multiple starbursts. The bottom panels show examples of clear multiple starbursts. The goodness threshold used in {\sc GAEA} to separate uni-model vs multiple starbursts is $-9.9$ (a bit lower than the other simulations).}
    \label{SFHfitGAEA}
\end{figure*}

\begin{figure*}
    \centering
    \includegraphics[trim=5mm 5mm 1.5mm 3mm, clip,width=0.3\textwidth]{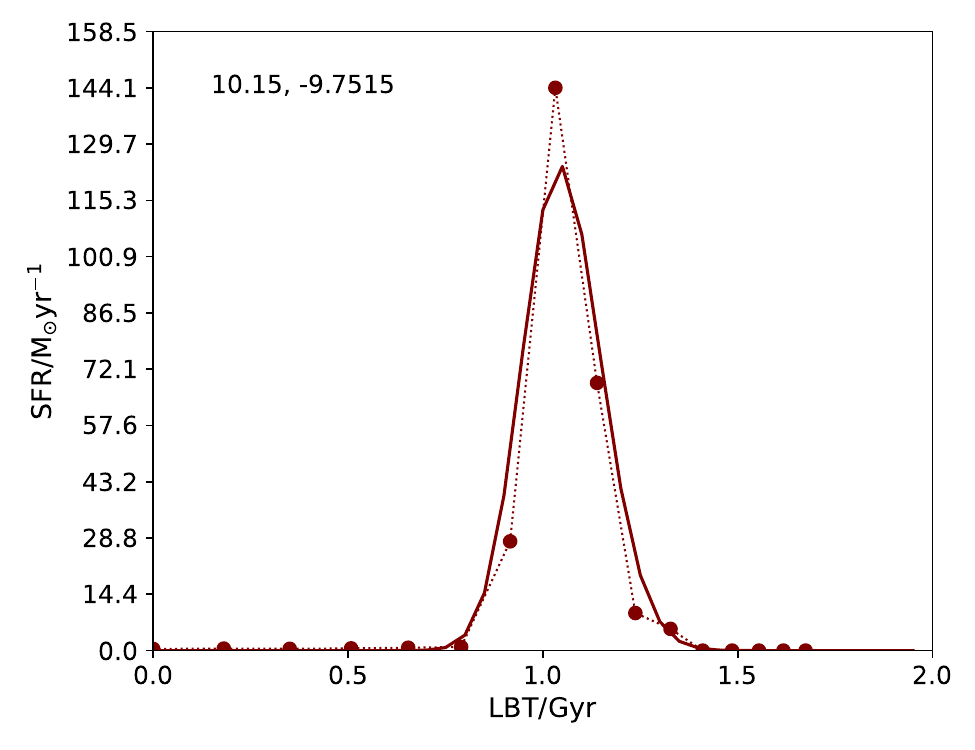}
    \includegraphics[trim=5mm 5mm 1.5mm 3mm, clip,width=0.3\textwidth]{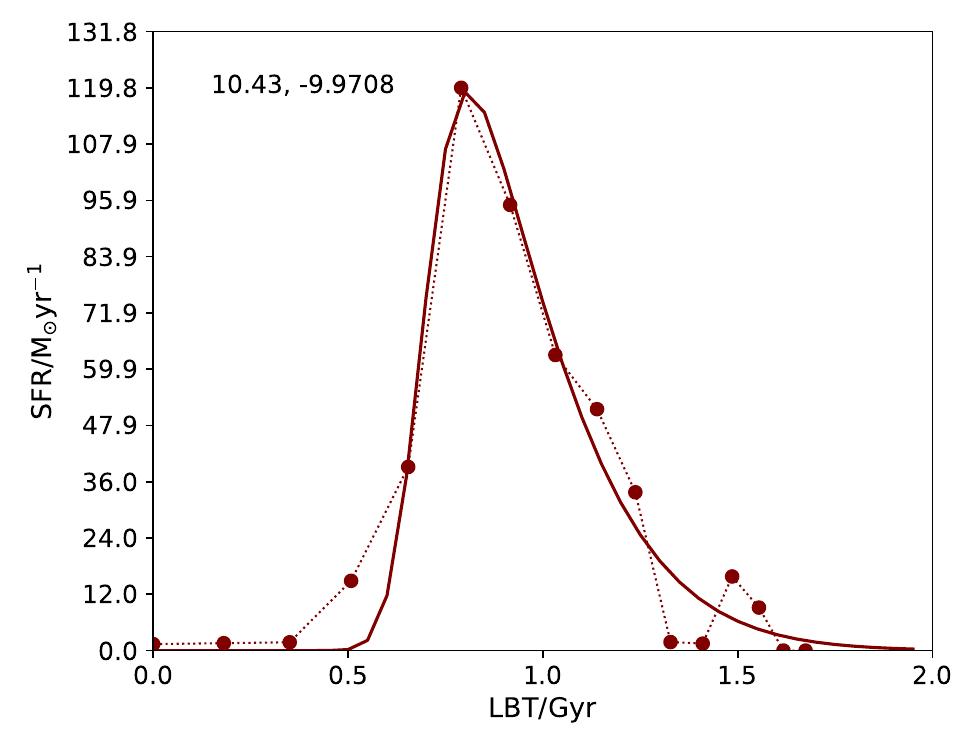}
    \includegraphics[trim=5mm 5mm 1.5mm 3mm, clip,width=0.3\textwidth]{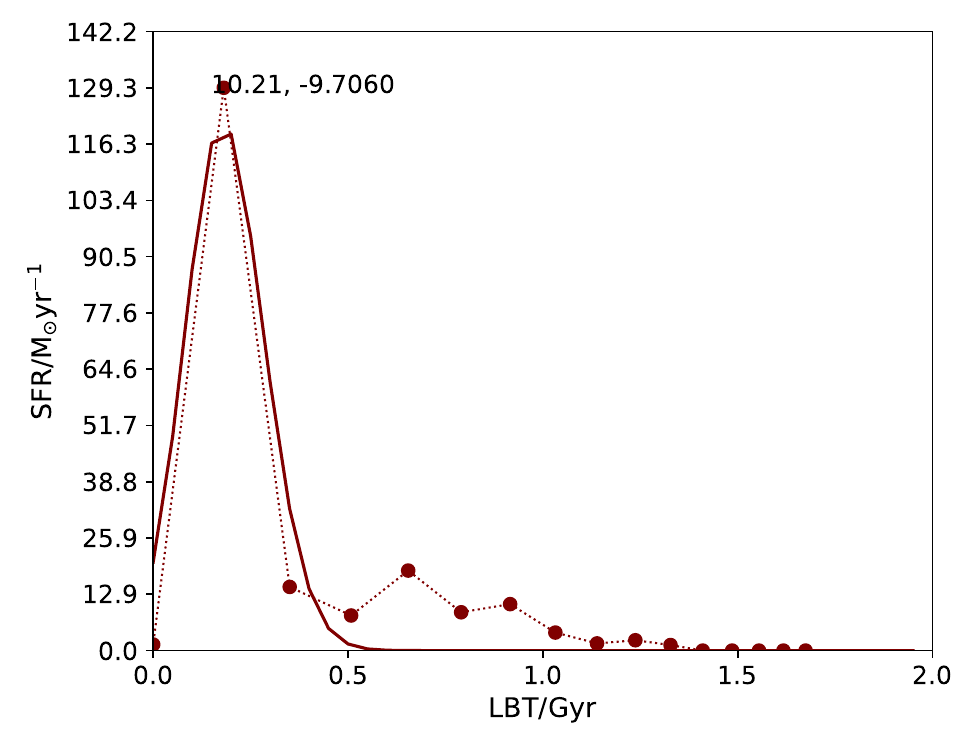}
    \includegraphics[trim=5mm 5mm 1.5mm 3mm, clip,width=0.3\textwidth]{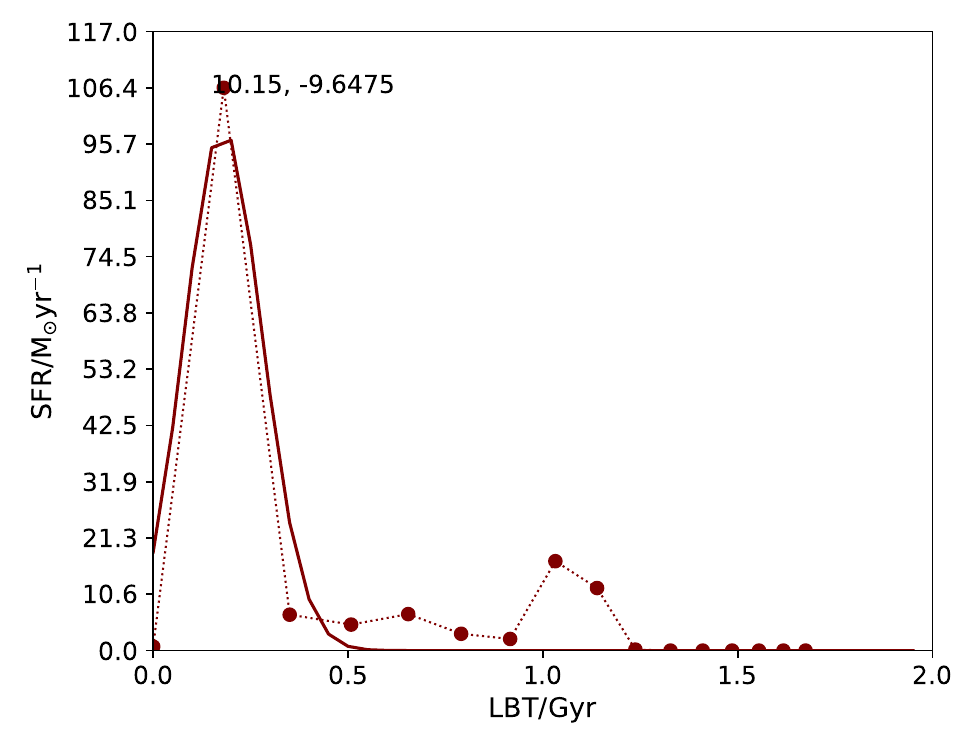}
    \includegraphics[trim=5mm 5mm 1.5mm 3mm, clip,width=0.3\textwidth]{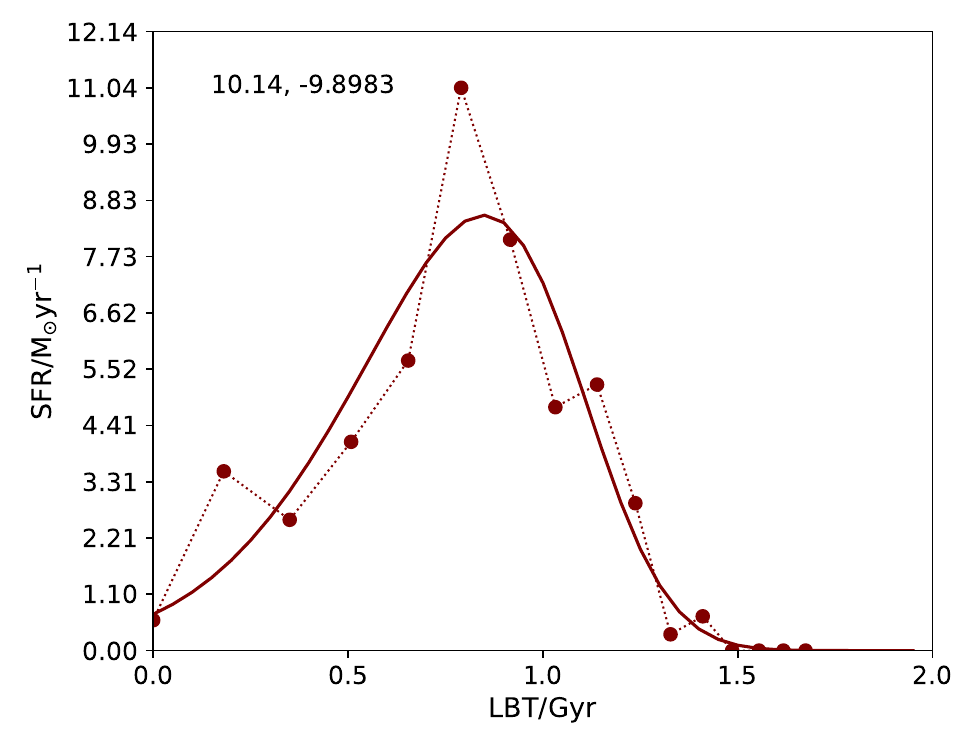}
    \includegraphics[trim=5mm 5mm 1.5mm 3mm, clip,width=0.3\textwidth]{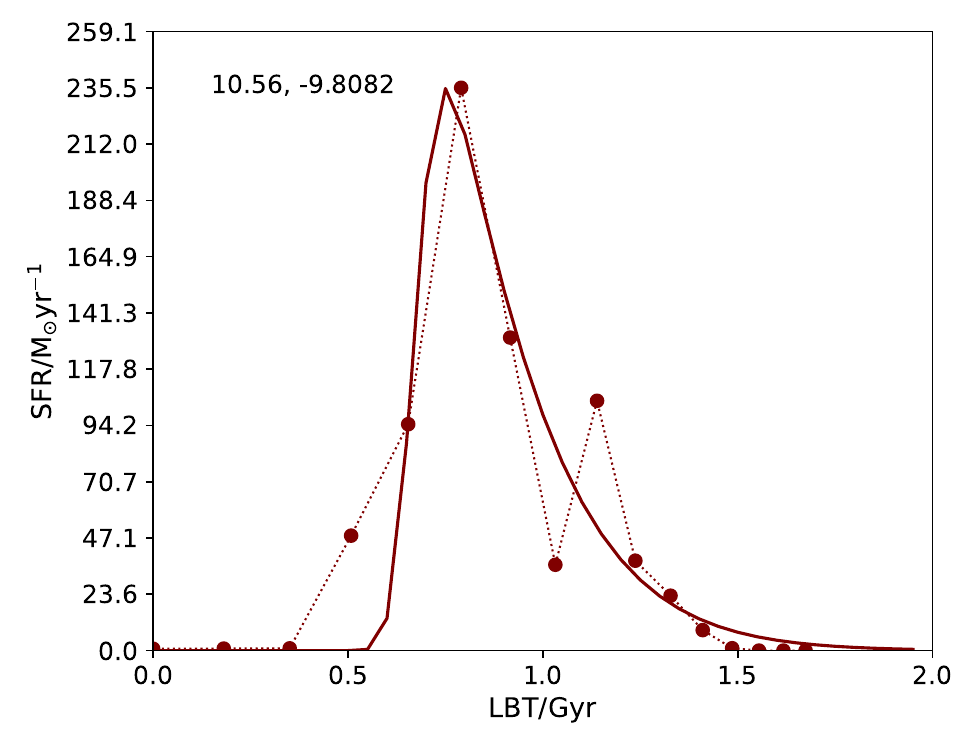}
    \includegraphics[trim=5mm 5mm 1.5mm 3mm, clip,width=0.3\textwidth]{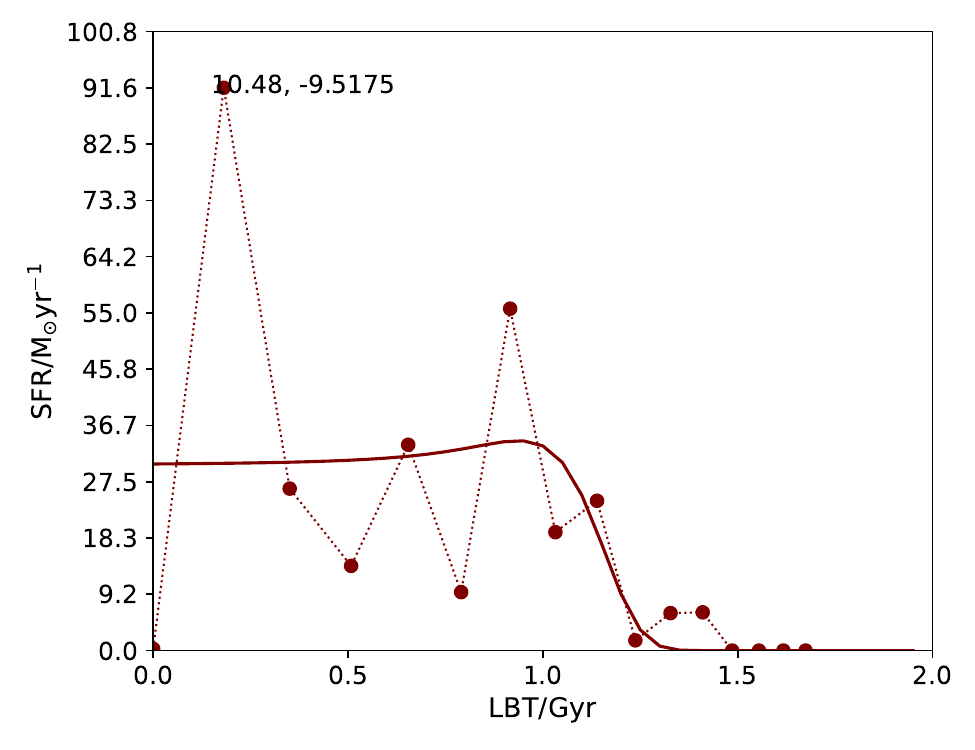}
    \includegraphics[trim=5mm 5mm 1.5mm 3mm, clip,width=0.3\textwidth]{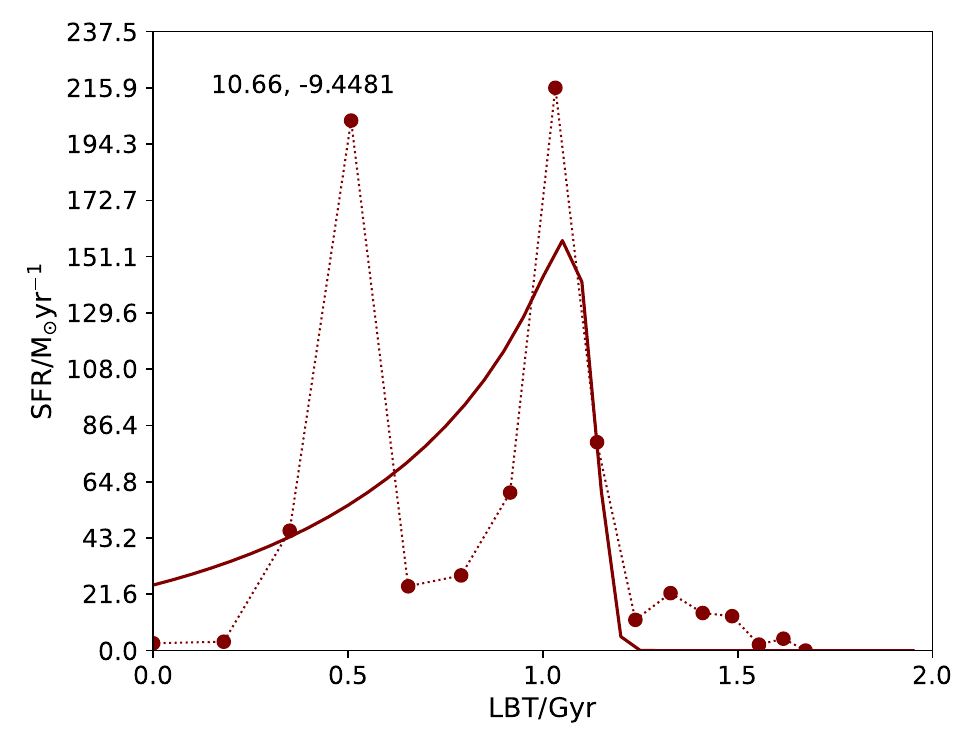}
    \includegraphics[trim=5mm 5mm 1.5mm 3mm, clip,width=0.3\textwidth]{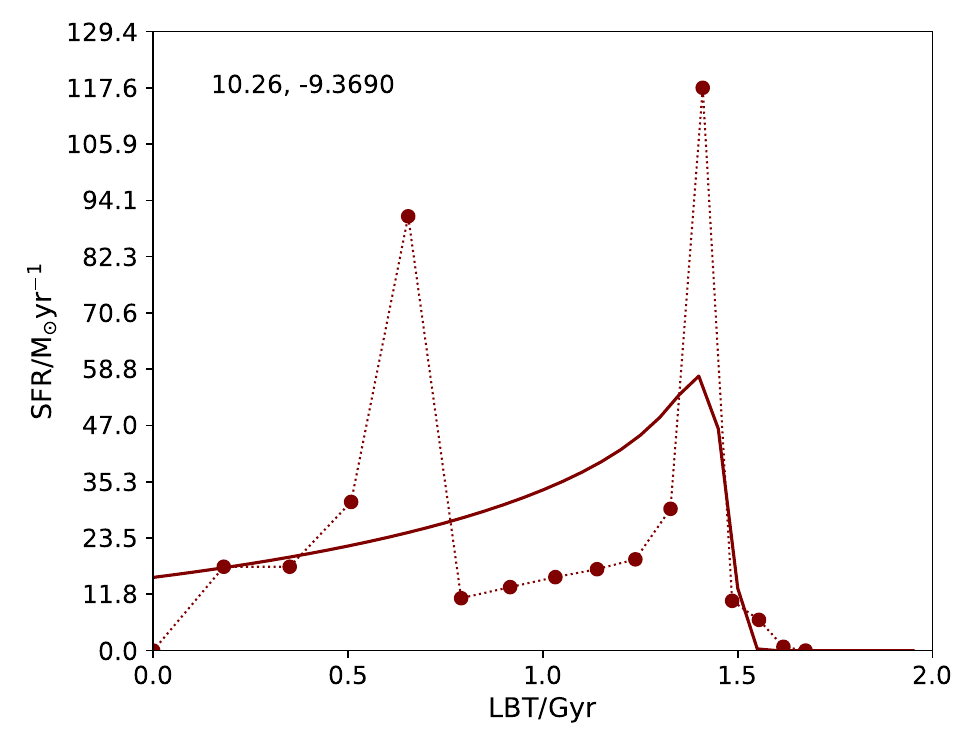}
    \caption{As in Fig.~\ref{SFHfitEAGLE} but for {\sc Galform}. The top panel show very good fits, while the middle panels show galaxies with reasonably good fits and small bumps in their SFHs but that we do not consider as being clear multiple starbursts. The bottom panels show examples of clear multiple starbursts. The goodness threshold used in {\sc Galform} to separate uni-model vs multiple starbursts is $-9.6$ (a bit higher than the other simulations).}
    \label{SFHfitGALFORM}
\end{figure*}

We employ {\sc Highlander} to fit the above function to each SFH, which uses a combination of genetic algorithms and Markov Chain Monte Carlo (MCMC), thus, providing a global minimum and likelihood. 
With the best fit, we compute a ``goodness'' function:

\begin{equation}
    {\rm goodness} = {\rm log}_{10}\left(\frac{\Sigma_{i} |\rm data-model|}{\,N_{\rm d.o.f.}\,\,M^f_{\star}}\right),
\end{equation}

\noindent where $N_{\rm d.o.f.}$ is the number of degrees of freedom (in this case the number of timesteps with which the SFH is sampled),  and $M^f_{\star}$ is the total stellar mass a galaxy has formed (which is different to the actual galaxy stellar mass at $z=3$ due to stellar evolution). Note that the normalisation is required to compare between galaxies, and the quantity is logged to avoid large numbers. We use this measure to isolate galaxy candidates that have had multiple starburst episodes. 
We find that internally to each simulation, ${\rm goodness}$ is a good parameter to isolate those galaxies, even though the exact threshold marking the transition from uni-modal distributions to multiple peaks changes between each simulation. We find that a threshold value between $-9.9$ and $-9.6$ work for all the simulations, and we tune the exact threshold to each simulation (values are included in the captions of Figs.~\ref{SFHfitEAGLE},~\ref{SFHfitTNG},~\ref{SFHfitSimba},~\ref{SFHfitSHARK},~\ref{SFHfitGAEA},~and~\ref{SFHfitGALFORM}). 

Note that here we are interested in multiple starburst episodes that are unrelated with each other. As an example, in \shark\ (Fig.~\ref{SFHfitSHARK}), there are galaxies going through multiple starbursts episodes but that appear correlated (i.e. a galaxy is having multiple starbursts periodically due to disk instabilities, while the average SFH over longer timescales behaves close to being uni-modal). We find that poor fits are generally related to either multiple starbursts (our case of interest) and to very short starbursts that have a sharp decline (those are common in {\sc Eagle} and \shark, but uncommon in the other simulations). The poor fits of those cases are reclassify as being uni-modal. In Figs.~\ref{SFHfitEAGLE},~\ref{SFHfitTNG},~\ref{SFHfitSimba},~\ref{SFHfitSHARK},~\ref{SFHfitGAEA},~and~\ref{SFHfitGALFORM} we show examples of good and bad fits in {\sc Eagle}, {\sc IllustrisTNG}, {\sc Simba}, \shark, {\sc GAEA}, and {\sc Galform}, respectively, and point out which ones are considered as having clear multi-starburst episodes. Table~\ref{SBsFrac} in \S~\ref{multiSBs} shows the fraction of galaxies in each simulation that display multiple starbursts. 

